\documentclass[trackchanges,twocolumn]{aastex701}

\usepackage{CJKutf8}
\usepackage{booktabs}
\usepackage{amsmath}
\usepackage{enumitem}
\usepackage{xspace}
\usepackage{multirow}

\newcommand{\et}{\texttt{EncoTESS}\xspace}
\newcommand{\ageet}{\texttt{ageET}\xspace}
\newcommand{\agep}{\texttt{ageP}\xspace}
\newcommand{\agepfh}{\texttt{agePfH}\xspace}
\newcommand{\ageetxh}{\texttt{ageETXH}\xspace}

\defcitealias{ChronoFlow...2025ApJ...986...59V}{V25}

\begin{document}

\title{\et: Age-Sensitive Encodings from Raw TESS Light Curves}

\author[0009-0009-4567-9946]{Phil R. Van-Lane}
\affiliation{David A. Dunlap Department of Astronomy \& Astrophysics, University of Toronto, 50 St. George St, Toronto, ON M5S 3H4, Canada}
\affiliation{Dunlap Institute for Astronomy \& Astrophysics, University of Toronto, 50 St. George St, Toronto, ON M5S 3H4, Canada}
\affiliation{Department of Astronomy \& Astrophysics, University of California San Diego, 3183 Matthews Ln, La Jolla, CA 92093, USA}
\email{phil.vanlane@mail.utoronto.ca}

\author[0000-0003-2573-9832]{Joshua S. Speagle \begin{CJK*}{UTF8}{gbsn}(沈佳士)\end{CJK*}}
\affiliation{David A. Dunlap Department of Astronomy \& Astrophysics, University of Toronto, 50 St. George St, Toronto, ON M5S 3H4, Canada}
\affiliation{Department of Statistical Sciences, University of Toronto, 9th Floor, 700 University Ave, Toronto, ON M5S 3G3, Canada}
\affiliation{Dunlap Institute for Astronomy \& Astrophysics, University of Toronto, 50 St. George St, Toronto, ON M5S 3H4, Canada}
\affiliation{Data Sciences Institute, University of Toronto, 10th Floor, 700 University Ave, Toronto, ON M7A 2S4, Canada}
\email{j.speagle@utoronto.ca}

\author[0000-0001-5383-9393]{Ryan Cloutier}
\affiliation{Department of Physics \& Astronomy, McMaster University, 1280 Main St West, Hamilton, ON L8S 4L8, Canada}
\email{ryan.cloutier@mcmaster.ca}

\author[0000-0002-9807-5435]{Christopher A. Theissen}
\affiliation{Department of Astronomy \& Astrophysics, University of California San Diego, 3183 Matthews Ln, La Jolla, CA 92093, USA}
\email{ctheissen@ucsd.edu}

\author[0000-0003-3734-8177]{Gwendolyn M. Eadie}
\affiliation{David A. Dunlap Department of Astronomy \& Astrophysics, University of Toronto, 50 St. George St, Toronto, ON M5S 3H4, Canada}
\affiliation{Department of Statistical Sciences, University of Toronto, 9th Floor, 700 University Ave, Toronto, ON M5S 3G3, Canada}
\affiliation{Data Sciences Institute, University of Toronto, 10th Floor, 700 University Ave, Toronto, ON M7A 2S4, Canada}
\email{gwen.eadie@utoronto.ca}

\author[0000-0003-3734-8177]{Ilay Kamai}
\affiliation{Physics Department, Technion: Israel Institute of Technology, Haifa 32000, Israel}
\email{ilay.kamai@campus.technion.ac.il}

\begin{abstract}

Main sequence stars of spectral types late F through M exhibit systematic variability in photometric light curves, particularly when they are young. Rotational modulation of starspots manifests as quasi-sinusoidal variability, which enables the measurement of stellar rotation periods ($P_{rot}$). Variability can also be stochastic, as in stellar flaring. However, since measurements of stochastic processes are highly dependent on the time of observation, they are typically noisier. Considering that different manifestations of variability have unique observational nuances, models that naturally unify these different types of variability are incredibly useful for stellar characterization. Towards this goal, we have developed \et: a Time Series Foundation Model (TSFM) trained on a subset of \textit{TESS} 2-min light curves. \et is specifically designed to handle the observational noise, heteroskedastic measurements, irregular sampling, and large data gaps common to \textit{TESS} data. It is also $\approx 1\%$ of the size of a typical literature TSFM, so can be run easily on a modern laptop. \et encodes light curves into a fixed-size latent parameter space, which can be used to infer physical stellar properties and recovers light curve summary statistics well. \et{} outperforms $P_{rot}$ and variability amplitude as age indicators for stars that have not converged onto the slow rotator sequence yet; broadly these include K and M stars $\lesssim 100$ Myr, and M stars $\lesssim 1$ Gyr. We focus on age inference as an application of \et in this work, but other downstream tasks such as stellar classification could also be explored. The architecture of \et{} enables its future extension to \textit{TESS} light curves of all cadences, and additional surveys such as \textit{Kepler} and the upcoming \textit{PLATO} mission. The core \et framework and library of encodings produced for the stars used in this work are publicly available at \href{https://github.com/philvanlane/encotess}{https://github.com/philvanlane/encotess}.

\end{abstract}

%% cross-references to sections, equations, tables, and Figures.

\section{Introduction} 
\label{sec:introduction}

Photometric variability, a term used to describe the fluctuations in an object's brightness over time, is a fundamentally intuitive property that can be measured from sequential observations.

In the modern era, long-time-baseline observing programs have yielded groundbreaking systematic variability analyses, such as the HK project at the Mount Wilson Observatory \citep[][]{Wilson...1968ApJ...153..221W}, which characterized the magnetic activity cycles of solar-type stars. The recent space-based surveys \textit{Kepler} (\citealt{Kepler...2010Sci...327..977B}), \textit{K2} (\citealt{K2...Howell...2014PASP..126..398H}), and \textit{Transiting Exoplanet Survey Satellite} (\textit{TESS}; \citealt{TESS...2015JATIS...1a4003R}) have produced large catalogs of short cadence photometric time series data, which are useful for studying variability on timescales of minutes to hours. These data are predominantly formatted as ``light curves'' (LCs), which consist of a flux measurement, typically with associated observational error, at each timestamp of observation. Visually, LCs often appear as quasi-sinusoidal due to different types of periodic and stochastic variability of both astrophysical and instrumental origin.

In this work we focus on main sequence (MS) stars with spectral types of late F to M, whose variability is driven by surface magnetic fields that arise due to convective stellar envelopes \citep[][]{Kraft...spindown...1967ApJ...150..551K}. In particular we investigate how variability can be applied to constrain their ages, which are typically difficult to estimate using other methods.

Since these stars evolve minimally in color-magnitude space over their MS lifetimes, isochrone fitting is very challenging if they are not known members of a coeval population \citep[][]{Soderblom...ages...2010ARA&A..48..581S}. Asteroseismology is useful for dating large evolved stars, but the oscillation amplitude of asteroseismic signatures decreases towards cooler stars on the MS. This makes pulsation signals difficult to detect for FGKM stars; oscillations have been detected in only hundreds of F and G type stars, $\sim10$ K type stars, and no M type stars to date \citep[][]{Chaplin...Asteroseismology...K...2026A&A...710A.361C}. Lithium depletion can be useful at young ages, but its reliability decreases for stars $\gtrsim$ 1 Gyr \citep[e.g.][]{Jeffries...LiDepAges...2023MNRAS.523..802J}.

However, rotation period ($P_{rot}$) \textit{can} be used as an age indicator for these stars; this practice is known as \textit{gyrochronology} \citep[][]{Barnes...2003ApJ...586..464B}. The short observational cadences of \textit{Kepler}, \textit{K2}, and \textit{TESS} have increased the number of literature $P_{rot}$ measurements by orders of magnitude \citep[e.g.][]{TARS...Boyle...2026arXiv260305586B,Santos...Prot...2021ApJS..255...17S,Santos...Prot...2019ApJS..244...21S,McQuillan...Prot...2014ApJS..211...24M} over previous ground-based catalogs. This increase in data volume, along with recent innovations in data-driven and machine learning (ML) methods, have enabled a new class of empirical gyrochronology models  \citep[][]{Bouma...gyrointerp...2023ApJ...947L...3B,Lu...Day&Age...2024AJ....167..159L,ChronoFlow...2025ApJ...986...59V} that estimate stellar ages probabilistically from $P_{rot}$ and mass or some proxy for mass such as temperature or color.

Although gyrochronology works reasonably well for FGK stars that have converged onto the ``slow rotator'' sequence, there is still intrinsic scatter in the $P_{rot}$/mass/age relationship, particularly for young stars. There is also multimodality in the spindown behavior of M dwarfs, and these patterns hinder the accuracy and precision of gyrochronology for the youngest, reddest stars \citep[e.g.][]{Barnes...2003ApJ...586..464B,Pass...MM...2022ApJ...936..109P,Pass...2024ApJ...966..231P,ChronoFlow...2025ApJ...986...59V}. This motivates the need for additional indicators to improve age constraints, and LC-derived metrics would be particularly useful as they would allow us to leverage the wealth of LCs already produced by \textit{Kepler}, \textit{K2}, and \textit{TESS}.

One example of another LC-derived age indicator is stellar flaring rate, which is often parameterized by the number of flares, total flare energy, or time spent flaring. Flaring rates change systematically with age at a population level \citep[][]{Davenport...flaring...2019ApJ...871..241D,Feinstein...2020AJ....160..219F,Medina...2022ApJ...935..104M,Feinstein...2024AJ....168...60F,Mamonova...2025AA...700A..53M,Tran...flaring...2026arXiv260220402T}, but since flares are stochastic events it takes an impractically long temporal baseline to constrain flaring rates for individual stars.

Another example of a LC-derived age indicator is the photometric magnetic activity index $S_{ph}$, which \citet{Mathur...Sph...2023ApJ...952..131M} measured from Kepler LCs and showed is correlated with age. Notably, \citet{Messina...PhotAmp...2021A&A...645A.144M} showed that among stars with similar colors and $P_{rot}$, photometric variability amplitude can provide additional age information and can therefore be used as a complementary indicator to $P_{rot}$.

While measurements such as $P_{rot}$, flaring, and $S_{ph}$ are useful age indicators, combining these handpicked features in a single model is difficult to optimize analytically since they are correlated. Furthermore, they are summary statistics calculated from $\mathcal{O}(10^5-10^6)$ data points in each LC, and their values depend on how they are measured. For example, several different methods are widely used in literature to calculate $P_{rot}$, such as Lomb-Scargle Periodograms \citep[LSPs;][]{Lomb...1976Ap&SS..39..447L,Scargle...1982ApJ...263..835S}, autocorrelation functions \citep[ACFs; e.g.][]{McQuillan...ACF...2013ApJ...775L..11M}, wavelet transforms \citep[e.g.][]{Garcia...Wavelet...2014A&A...572A..34G}, and Gaussian Processes \citep[GPs; e.g.][]{Angus...GP.Prot...2018MNRAS.474.2094A}, which each have their own drawbacks that contribute to uncertainty \citep[][]{Angus...GP.Prot...2018MNRAS.474.2094A}. There is also inherent loss of information that comes from deriving any single summary statistic from an entire LC. This motivates the need for a more holistic, data-driven approach to analyzing LC data.

Given recent progress in computing and artificial intelligence (AI), it is now feasible to implement such an approach. In this work, we develop \et: a Time Series Foundation Model \citep[TSFM;][and references therein]{IsmailFawaz...TimeSeriesClassification...2020arXiv201000567I,Ye...TSFM...2024arXiv240502358Y} to encode \textit{TESS} LCs. Unlike task-specific models, TSFMs are trained under self supervision to learn generalized patterns of time series data. The goal is typically to characterize the processes driving the observations. TSFMs are now routinely used across fields and industries to analyze different families of time series data such as stock prices, temperature fluctuations, and the housing market.

Several TSFMs have already been developed for astronomy, including \texttt{Astromer 1}\footnote{Originally named \texttt{ASTROMER} by \citet{DonosoOliva...Astromer...2023A&A...670A..54D} but then referred to as \texttt{Astromer 1} by \citet{DonosoOliva...Astromer2...2026A&A...707A.170D}.} and  \texttt{Astromer 2} \citep[][]{DonosoOliva...Astromer...2023A&A...670A..54D,DonosoOliva...Astromer2...2026A&A...707A.170D}, \texttt{DESA} \citep[][]{DESA...2025ApJ...994..110K}, and \texttt{FALCO} \citep[][]{Zuo...FALCO...2026AJ....171...10Z}. Readers may also be familiar with the prominent astronomy foundation model \texttt{AION-1} \citep[][]{AION1...2025arXiv251017960P}, which combines imaging, spectroscopic, and tabular data, but importantly it \textit{does not} handle time series data. Our work also draws inspiration from other models that are not strictly TSFMs, but that implement state-based approaches to modeling astronomical processes \citep[][]{Esquivel...HMMs...2025ApJ...979..141E,Herrerra...flaring...2026arXiv260622601H,Zimmerman...HMM...2024MNRAS.534.2142Z}.

While previous astronomy TSFMs are similar conceptually to ours, there are several key considerations we aim to address with \et:

\begin{enumerate}
    \item \textit{Training dataset}. Most literature TSFMs are trained on \textit{Kepler} LCs, which have a 4 year baseline but come from a small patch of sky ($\approx$115 deg$^2$). Since \textit{TESS} is an all sky survey, it has observed many more M dwarfs\footnote{ \citet{Eschen...TESS...M...2024MNRAS.531.5053E} quote 117,475 M dwarfs with a \textit{TESS} magnitude $> 13.5$ mag based on the TESS Input Catalog (TIC) Candidate Target List (CTL) v8.01 \citep[][]{Stassun...TESS.CTL...2019AJ....158..138S}, compared to the 4,218 \textit{Kepler} targets classified as M dwarfs by \citet{Gaidos...Kepler..M...2016MNRAS.457.2877G}.}, which are important to this work since they are particularly difficult to date using $P_{rot}$. Since \textit{TESS} photometry is $\approx$10$\times$ noisier than \textit{Kepler} as measured by transit timescale noise \citep[][]{VanCleve...PhotTTPrecision...2016PASP..128g5002V}, and the continuous viewing period is only $\approx$27 days per sector, \textit{Kepler}-trained models are unlikely to transfer well to \textit{TESS} data, so we need a model directly trained on the noisier, shorter \textit{TESS} LCs.

    \item \textit{Observational Sampling.} We need to handle irregularly spaced, heteroskedastic data, and be robust to different temporal resolutions and variable sequence length. This flexibility is crucial for handling the downlink time in \textit{TESS} LCs, which manifest as a data gap of $\approx$1 day in each sector, and for extensibility to different sampling cadences. Most literature models are not explicitly designed with such nuances in mind. 

    \item \textit{Architecture.} Most TSFMs use a \textit{transformer} architecture, which is computationally expensive to scale to long LC sequences. We want a model that can run on consumer hardware efficiently, so we implement a Recurrent Neural Network (RNN) based approach.
\end{enumerate}
Notably, there are 2 literature TSFMs that have been trained on \textit{TESS} LCs that are worth describing here.

\citet{Poznanski...TESSLCs...2026arXiv260505324P} uses a \textit{Quantile Graph} (QG) to convert 2-min \textit{TESS} LCs into fixed-length feature vectors. However, in pre-processing they divide the LCs into approximately continuous segments of 4096 data points ($\approx$6 days) and ignore small gaps within those segments. This methodology is not scalable to other cadences, and it ignores the longer term LC context. They also detrend each LC in pre-processing, which can remove signal from longer term variability such as slow rotation.

\citet{Ding...StarCLR...2026ApJ..1003..141D} developed \texttt{StarCLR}, which is pretrained on \textit{TESS} LCs at multiple cadences, and crucially does use observational timestamps as input so is gap aware. However, they fix LC sequences to 8192 data points in pre-processing, and only pretrain on LCs that already have evidence of periodic variability based on a prior LSP analysis. Since one of our goals is to consider all types of variability and not rely on periodicity, this filter is a crucial restriction.

Additionally, neither \citet{Poznanski...TESSLCs...2026arXiv260505324P} nor \citet{Ding...StarCLR...2026ApJ..1003..141D} incorporate flux measurement uncertainties into their models.

For these reasons, we opted to develop our own TSFM instead of adopting one from literature. \S\ref{sec:data} describes our training dataset, \S\ref{sec:model} presents our model and training architecture, and in \S\ref{sec:flux_recon} we evaluate its ability to reconstruct LCs. In \S\ref{sec:variability_stats} we explore how the latent encodings are correlated with physical properties, and test its recovery of traditional summary statistics. We interpret what \et actually learns from LCs in \S\ref{sec:interpretability}. \S\ref{sec:age_inference} explores age inference as an application, and shows that \et outperforms models that use only $P_{rot}$ and/or variability amplitude as age indicators. We investigate whether \et can predict field star ages in \S\ref{sec:field_ages}. We discuss systematic effects and future directions in \S\ref{sec:future}, and conclude in \S\ref{sec:conclusion}.

\section{Data}
\label{sec:data}

\begin{table}[ht]
\centering
\begin{tabular}{lrr}
\toprule
Reference & \# Stars & \# LCs \\
\midrule
\citet{ChronoFlow...2025ApJ...986...59V}      & 2,495 & 7,166  \\
\citet{Feinstein...2024AJ....168...60F}  & 6,753 & 25,198 \\
\citet{Kiman...2021AJ....161..277K}      & 753  & 2,964  \\
\citet{LWRD...Engle...2024ApJ...960...62E}           & 54   & 205   \\
\citet{Magaudda...Mdwarfs...2020AA...638A..20M}   & 154  & 372   \\
\citet{Mamonova...2025AA...700A..53M}   & 320  & 1,985  \\
MOCAdb \citep[][]{MOCAdb...2026arXiv260215695G}         & 436  & 1,006   \\
\citet{Newton...nearbyMdwarfs...2016ApJ...821...93N}     & 1,615 & 6,571  \\
\citet{Newton...SouthernMdwarfs...2018AJ....156..217N}     & 485  & 1,864  \\
\citet{Pass...MM...2022ApJ...936..109P}       & 24   & 75    \\
\citet{Pass...2023AJ....166...16P}       & 115  & 515   \\
\citet{Pass...2024ApJ...966..231P}       & 119  & 362   \\
\citet{TIMEtable...2023MNRAS.520.5283G}    & 173  & 826   \\
\citet{Kim...thickdisk...2022MNRAS.510.4308K}    & 6,047  & 8,784   \\
NASA Exoplanet Archive    & 3,952  & 19,990   \\
\midrule
\textbf{TOTAL\footnote{De-duplicated across sources.}}          & \textbf{21,507} & \textbf{69,345} \\
\bottomrule
\end{tabular}
\caption{Number of stars and total LCs (\textit{TESS} sectors) in the \et{} training catalog, broken down by literature source.}
\label{tab:lc_sources}
\end{table}

\begin{figure}
  \centering
  \includegraphics[width=\columnwidth]{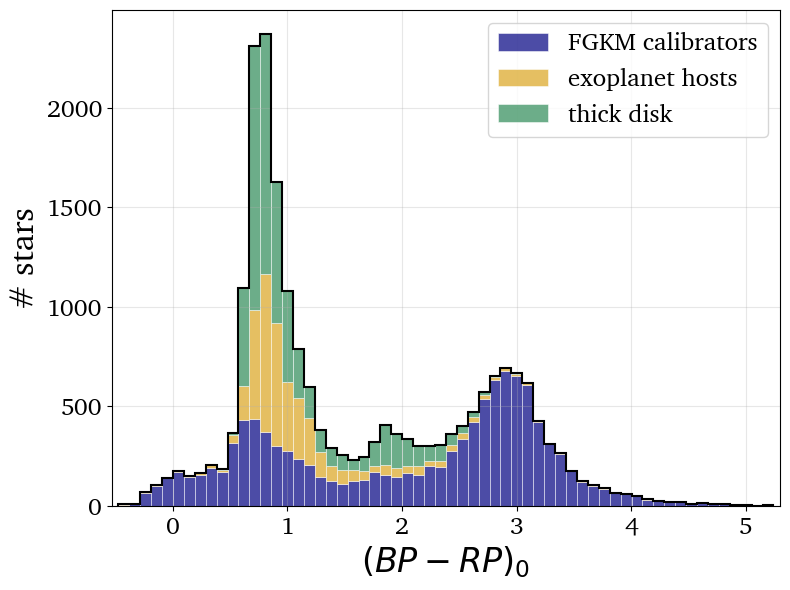}
  \caption{Distribution of Gaia DR3 $(BP-RP)_0$ colors by source category. Exoplanet hosts and thick disk stars are highly concentrated around $(BP-RP)_0 \approx1$, whereas the other targets are more evenly distributed are are responsible for the other peak at $(BP-RP)_0 \approx3$.}
  \label{fig:bprp0_dist}
\end{figure}

To develop \et, we used 2-minute cadence \textit{TESS} LCs from the Science Processing Operations Center pipeline \citep[SPOC;][]{Jenkins...SPOC...2016SPIE.9913E..3EJ}; specifically the Presearch Data Conditioning Simple Aperture Photometry flux measurements \citep[\texttt{PDCSAP};][]{Smith...PDCSAP...2012PASP..124.1000S,Stumpe...PDCSAP...2012PASP..124..985S,Stumpe...PDCSAP...2014PASP..126..100S}. While we restrict this work to 2-min cadence LCs, the \et{} architecture is such that other cadences can be used in the future without redesign (see \S\ref{sec:future}).

We selected targets based on literature catalogs, from three general categories. The first is exoplanet hosts, which includes all stars from the NASA Exoplanet Archive filtered to records with the \texttt{default} flag, having a documented \texttt{TIC\_ID} with associated 2-min LCs, and a populated \texttt{gaia\_dr3\_id} field. The second category is thick disk stars, which we included to bolster the sample of field age LCs used in training. This sample consists of all stars in the \citet{Kim...thickdisk...2022MNRAS.510.4308K} catalog that have associated 2-min LCs.

The third category consists of FGKM MS stars that exhibit variability or have independently calibrated ages. These FGKM MS targets are sourced from literature catalogs of open clusters and associations \citep[][]{ChronoFlow...2025ApJ...986...59V,Feinstein...2024AJ....168...60F,Kiman...2021AJ....161..277K,Mamonova...2025AA...700A..53M,MOCAdb...2026arXiv260215695G}, studies of magnetically active and/or rotating nearby M dwarfs \citep[][]{Magaudda...Mdwarfs...2020AA...638A..20M,Engle...LWRD...2024ApJ...960...62E,TIMEtable...2023MNRAS.520.5283G,Newton...nearbyMdwarfs...2016ApJ...821...93N,Newton...SouthernMdwarfs...2018AJ....156..217N}, and M dwarfs in wide binaries \citep[][]{Pass...MM...2022ApJ...936..109P,Pass...2023AJ....166...16P,Pass...2024ApJ...966..231P}.

Table~\ref{tab:lc_sources} summarizes our LC dataset. It is grouped by literature source, and provides the total number of stars and total number of sectors with 2-min LCs from each. Since some targets were present in multiple sources, the total number of stars and LCs are less than the sum of all individual sources. Our final dataset consists of 69,345 individual LCs from sectors 1--101, which represent 21,507 unique stars.

Figure \ref{fig:bprp0_dist} presents the Gaia DR3 $(BP-RP)_0$ distribution of the entire dataset, broken down by literature source category. The distribution is essentially bimodal; there is a sharp peak at $(BP-RP)_0 \approx1$ from exoplanet hosts and thick disk stars, and a broader peak at $(BP-RP)_0\approx3$ which is almost entirely from the other FGKM calibrators.

\subsection{Light Curve Preprocessing}
\label{subsec:data_lc_filters}

We used \texttt{Lightkurve} \citep[][]{Lightkurve...2018ascl.soft12013L} to download all LCs, and applied the filter \texttt{quality\_bitmask="default"} to remove unreliable flux measurements.\footnote{Version 2.5.1 of \texttt{Lightkurve} was used, in which the default bitmask is 17087. This means that flux measurements with any of the bit value flags 1, 2, 4, 8, 16, 32, 128, 512, or 16,384 as described in Table 28 of \citet{tess_sdpdd} are excluded from our data.} The raw data we extracted from each LC were the observed \texttt{PDCSAP} flux $f$, the measurement uncertainty (\texttt{flux\_err}) $\sigma_f$, and the time of observation $t$. Where $f$ was recorded in the LC but no $\sigma_f$, we used a default $\sigma_f$ of the median value from that LC. We also reset the times of each LC so they started at $t_0=0$.

We normalized each LC to a median flux $f_{med}$ of 0 and a half interquartile range $f_{H}$ of 1, so that the normalized measurements ($\hat{f}$, $\hat{\sigma}_{f}$) were:
\begin{equation}
\label{eq:flux_norm}
    \hat{f} = \frac{f - f_{med}}{f_{H}};\;\;\;\;\; \hat{\sigma}_f = \frac{\sigma_f}{f_{H}}
\end{equation}
These metrics were used because traditional z-score normalization based on mean and standard deviation was highly sensitive to extreme outliers, which are sometimes present at the very start and/or end of sectors. 

The three observational channels we saved from each LC were therefore:
\begin{equation}
    [\hat{f},\hat{\sigma}_{f},t]    
\end{equation}

\subsection{TESS and Light Curve Metadata}
\label{subsec:tess_metadata}

In addition to the three per-observation channels, we saved the \textit{TESS} magnitude $T_{mag}$ of the target, as well as the sector number $S$, camera number $N_{cam}$, and CCD number $N_{ccd}$ of the LC, which are provided by the \texttt{Lightkurve} objects. We also saved the two flux normalization parameters $f_{med}$ and $f_H$, as inputs for \et. This is for two reasons: (i) they track overall activity levels and general variability, and (ii) they are correlated with noise levels and instrumental systematics, so including them explicitly helps \et{} disentangle systematic and astrophysical effects.

Finally, we included the cadence of observation $\Delta t$ as metadata; while this does not inform our model since it is 120s for all LCs in this work, we include it as a semantic step towards extending \et's applicability to all \textit{TESS} LCs.

So, for every LC we save a 7-dimensional array of metadata:
\begin{equation}
    [T_{mag},S,N_{cam},N_{ccd},f_{med},f_{H},\Delta t]    
\end{equation}

\subsection{Gaia Crossmatch}
\label{subsec:gaia_metadata}

For all LCs, we crossmatched TIC IDs with Gaia Data Release 3 \citep[DR3;][]{GaiaDR3...2023A&A...674A...1G} IDs using a combination of the \texttt{GAIA} field in the MAST TIC catalog (which corresponds to Gaia DR2 ID) and the \texttt{dr3.dr2\_neighbourhood} table in the Gaia DR3 database. This was required in some cases to identify the TIC IDs of sources from the literature catalogs, but we always performed this crossmatch regardless to identify any sources that did not have measured parallax, $G$, $BP$, or $RP$ flux measurements in DR3, which we then excluded from this work.

We applied zeropoint corrections to the Gaia DR3 \texttt{parallax} using \texttt{gaiadr3\_zeropoint} (\citealt{Lindegren...zeropoint...2021A&A...649A...4L}. Gaia DR3 $G$, $BP$, and $RP$ were de-reddened using the dust maps from \citet{Edenhofer...2024A&A...685A..82E} and \citet{Bayestar19...dustmap...2019ApJ...887...93G}, with the sampling and conversion process described by \citet{ChronoFlow...2025ApJ...986...59V} (hereafter \citetalias{ChronoFlow...2025ApJ...986...59V}). The one distinction worth noting is that \citetalias{ChronoFlow...2025ApJ...986...59V} samples distances for each star were based on their Gaia DR3 parallax. In this work we sampled distances based on \citet{BailerJones...dist...2021AJ....161..147B} where possible, and fell back to parallax otherwise.

\citet{Edenhofer...2024A&A...685A..82E} is used as the primary dust map in this work, and we supplement with \citet{Bayestar19...dustmap...2019ApJ...887...93G} for stars that \citet{Edenhofer...2024A&A...685A..82E} does not cover. We discuss the implications of this choice in \S\ref{sec:future}.

From this procedure we extract 6 metadata parameters for each star: de-reddened apparent Gaia magnitude $G_0$, de-reddened color $(BP-RP)_0$, parallax $\pi$, and the associated uncertainties for each (notated $\sigma_{G0}$, $\sigma_{BR0}$, and $\sigma_{\pi}$ respectively).

\section{Model architecture}
\label{sec:model}

Here we elaborate on the choice (described briefly in \S\ref{sec:introduction}) to develop \et{} as a new TSFM instead of using a literature model. First, since TSFMs are designed to learn generalizable features of time series data, they naturally encode fundamental properties of the underlying processes instead of only information relevant to a specific task. This provides the additional benefit of characterizing the driving processes in a task-independent framework. TSFMs also often achieve results on downstream tasks comparable to bespoke models \citep[e.g.,][]{TimesFM...2023arXiv231010688D,MoiraiMoE...2024arXiv241010469L}. Therefore, while age inference is one of the primary goal of this work, we developed \et{} as a TSFM so it could potentially be used for other downstream tasks such as classification of spectral types or inference of other stellar properties.

\begin{figure*}
  \centering
  \includegraphics[width=0.9\textwidth]{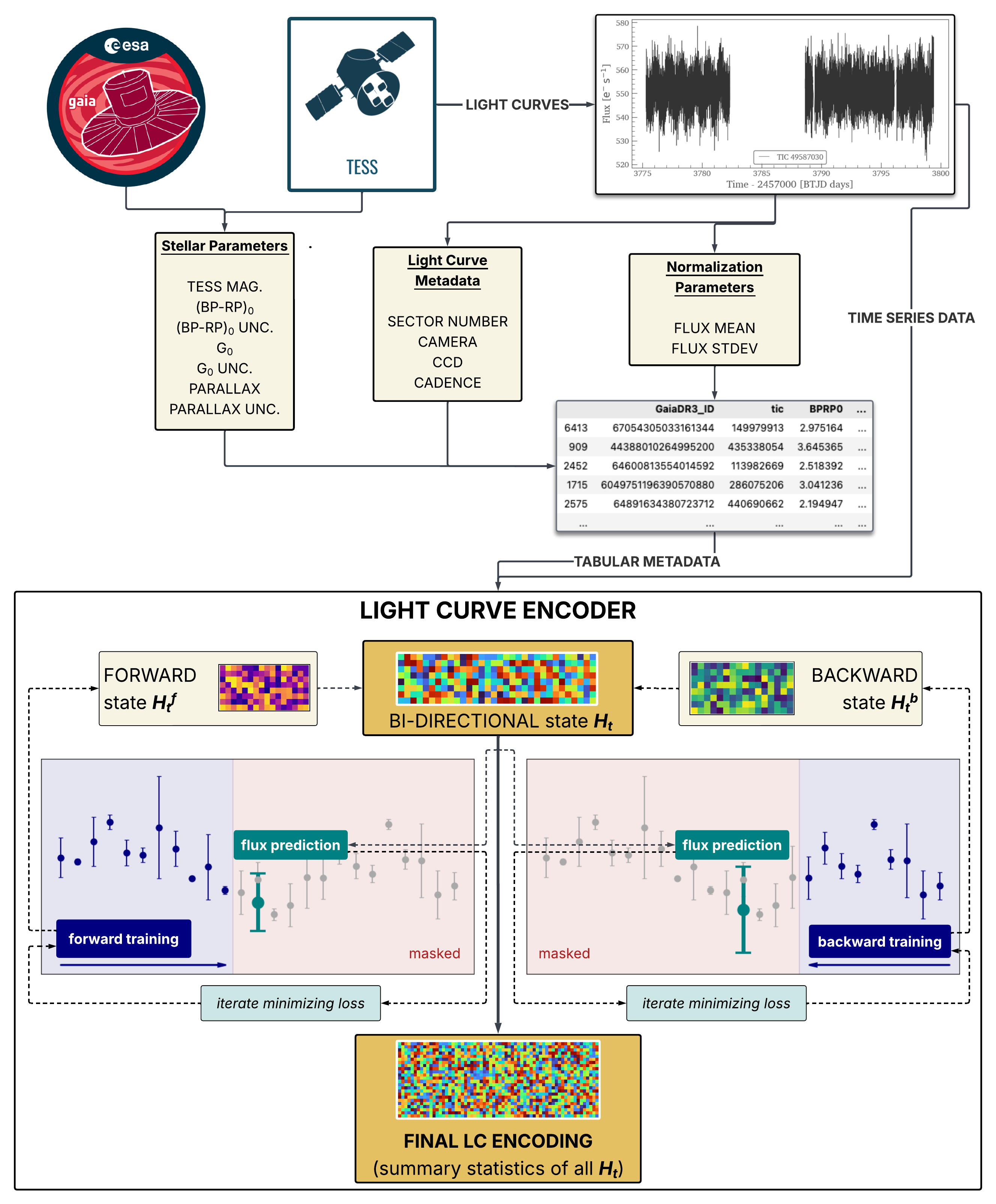}
  \caption{A schematic view of the data flow and training process of \et. The training dataset consists of raw LCs (timestamp, flux, and flux measurement uncertainty values for each point), and their corresponding metadata consisting of stellar and sector parameters from \textit{Gaia DR3} and \textit{TESS}. Training is via reconstruction: subsets of an LC are encoded using both forward and backward processing into hidden states ($H^f_t$ and $H^b_t$ respectively), which are used to predict the flux at some offset. \et{} learns to construct these hidden states to maximize the likelihood of its predictions. This is repeated at variable prediction offsets. After training is complete, the hidden states corresponding to every observation in a LC are aggregated into a 1536-dim latent representation of each LC. In this schematic, solid lines represent one-time data processing and dashed lines represent computations that are executed iteratively during training. The colored grids represent vectors, where each pixel is a dimension and the color represents the magnitude of that dimension. The dimensions shown here are accurate: each $H^f_t$ and $H^b_t$ is 64-dim, so each $H_t$ is a 128-dim concatenation of $H^f_t$ and $H^b_t$, and the final LC encoding is a 1536-dim vector constructed by pooling all $H_t$ vectors in 12 different ways (see Appendix~\ref{app:model_output_computation} for details).}
  \label{fig:schematic}
\end{figure*}

In the following subsections we briefly describe the \et{} architecture in four logical segments, but refer readers who wish to understand the full mechanics to Appendix~\ref{app:detailed_model_architecture}. Figure~\ref{fig:schematic} also presents an abstracted overview of \et, including the input data, training procedure, and outputs.

\subsection{Input Data}
\label{subsec:model_inputs}

Our input data for an LC of length $L$ is a $(3\times L)$-dimensional matrix of normalized time series data:
\begin{equation}
    [\hat{f},\hat{\sigma}_{f},t]    
\end{equation}
and a 13-dim metadata vector:
\begin{equation}
\begin{aligned}
 [&T_{mag},S,N_{cam},N_{ccd},f_{med},f_{H},\Delta t\\
    &G_0,\sigma_{G0},(BP-RP)_0,\sigma_{BR0},\pi,\sigma_{\pi}]
\end{aligned}
\end{equation}

We treat time as an observable instead of interpolating the data to a uniform grid because we want \et{} to learn the concept of absolute time. This is for two main reasons: (i) the model should understand the difference between an abrupt astrophysical jump in flux vs a gradual change in flux over a gap in measurements (e.g. \textit{TESS} downlink time), and (ii) this is required for future integration of LCs with different observational cadences. \et{} needs to understand the difference between 30 $\times$ 2-min cadence observations spanning an hour, and 30 $\times$ 20-sec cadence observations spanning 10 minutes, and be able to project both sets of observations into a shared interpretation.

To the metadata fields, we apply simple normalizations to ensure all fields are approximately of order unity:
\begin{equation}
\begin{aligned}
    \label{eq:metadata}
    M_S = \{\;\;\;&\frac{G_0}{100},\;\log_{10}(\sigma_{G0}),\;(BP-RP)_0,\;\log_{10}(\sigma_{BR0}),\\
    &\log_{10}(\pi),\;\log_{10}(\sigma_{\pi}),\;\log_{10}(T_{mag}),\;\frac{S}{100},\\
    &N_{cam},\;N_{ccd},\;\log_{10}(f_{med}),\;\log_{10}(f_{H})\;\;\;\}\\
\end{aligned}
\end{equation}
Appendix~\ref{app:model_input_enc} describes in detail how the LC and metadata parameters are encoded in the \et{} implementation.

We note here that including additional ``views'' of time series data as input, such as power spectra and/or ACFs, can help foundation models learn \citep[e.g.][]{DESA...2025ApJ...994..110K}. To determine whether this would help \et, we benchmarked the performance of several versions of \et{} which included frequency power spectra and ACFs calculated from each LC as additional input channels. We found that these additional channels did not improve performance (see Appendix \ref{app:extra_channels} for details).

\subsection{Recurrent Neural Network Architecture}
\label{subsec:model_rnn}

Transformers are often considered to be the state-of-the-art for foundation models because of their ability to model short and long scale interactions between features via their ``attention'' mechanisms. However, the number of compute operations in transformers scales with sequence length $L$ as $\sim\mathcal{O}(L^2)$. This makes processing long sequences a computational concern.

In contrast, RNNs do not have a built in attention mechanism, and instead model long scale dependencies by carrying information through the time series sequentially. This means that while there isn't the same capability to model direct interaction between every feature as in transformers, the number of compute operations scales as $\sim\mathcal{O}(L)$ instead of $\sim\mathcal{O}(L^2)$. This is an important computational advantage when considering long and densely sampled LCs.

There are two other distinct advantages of RNNs. First, transformers require fixed feature size (i.e., sequence length), but RNNs do not. Second, RNNs typically use significantly fewer parameters than transformers. 

While another advantage of transformers is their ability to process sequences in parallel, in contrast with tradition RNNs that process sequentially, \citet{RNNs...2024arXiv241001201F} developed an RNN architecture termed a \texttt{minGRU} which implements parallel processing and therefore removes that key bottleneck.

For these reasons, we have developed \et{} as an RNN, and implement the \texttt{minGRU} architecture developed by \citet{RNNs...2024arXiv241001201F}. The final \et{} architecture consists of only $\approx64,000$ parameters. This is tiny compared to most TSFMs, which are typically $\mathcal{O}(10^6)$ parameters at a minimum.

A new family of \textit{State Space Models} \citep[SSMs; e.g.,][]{Mamba...2023arXiv231200752G} have also become more popular over the last several years, which have computational costs similar to RNNs but more flexible architecture that enables more efficient training on modern hardware. While it may prove useful to implement an SSM in future work, for \et{} we used the simpler RNN approach.

We describe the RNN implementation in detail in Appendix~\ref{app:model_rnn_framework}, but the key concept is that \et learns to represent the LC at any point $t$ using a 64-dim ``forward hidden state'' $H^f_t$ and a 64-dim ``backward hidden state'' $H^b_t$. $H^f_t$ represents the context of the LC up until time $t$, and $H^b_t$ represents the context after time $t$. The hidden states depend on both the observational and metadata channels. Combining them yields a 128-dim hidden state $H_t$, which therefore represents the full LC context at any point $t$ including the stellar and observational information from the metadata too.

\subsection{Training Procedure}
\label{subsec:model_training}

Training is executed via unlabeled self supervised reconstruction. In practice, this means that we train \et{} to predict the flux value at some time based on the LC context and metadata. This method stems from the idea that to reasonably replicate a sequence of observations from partial information, a model must be able to intelligently describe the underlying processes, and therefore training in this way will force the model to learn rich representations of the stars. The simplest version of this training task is predicting the observation at some timestamp $t_y$ given observations from $t_0$ to $t_{y-1}$.

To train \et{} we iterate this procedure over variable offsets (prediction ``horizons'') extending from 2 minutes to 4 days, to encourage \et{} to capture long terms trends. We also apply masking to parts of each LC and metadata during training to improve \et's robustness against missing data. The prediction component itself is a \textit{Conditional Normalizing Flow} (CNF), which enables full probabilistic flux likelihood estimation instead of a point estimate with Gaussian uncertainty. These components are described in detail in Appendix~\ref{app:detailed_model_training}.

\subsection{Model Output}
\label{subsec:model_output}

Our RNN framework learns to represent each point in a LC as the 128-dim vector $H_t$. A typical LC is therefore represented with $\mathcal{O}(10^6)$ data points, with the true size depending on the number of data points in the LC. This representation is not useful for downstream applications due to its large and variable size, so we apply different types of aggregation to reduce this to a fixed 1536-dim vector that we call $\Theta$ for each LC. Details of this aggregation are provided in Appendix~\ref{app:model_output_computation}. Importantly, while we mask portions of the input during training, we use the full metadata and set of observations in this work when processing LCs to compute $\Theta$.

A useful exercise is estimating the ``intrinsic dimensionality'' of a parameter space such as $\Theta$ to estimate the \textit{minimum} number of dimensions required to fully describe it. We used \texttt{TwoNN} \citep[][]{Facco...TwoNN...2017NatSR...712140F} to probe this and measured an intrinsic dimensionality of 13.

Therefore, many of these 1536 dimensions are highly redundant, so for visualization and downstream tasks it can be useful and/or necessary to reduce the  dimensionality further. We use three different dimensionality reduction techniques in this work, which are summarized here and explained in further detail in Appendix~\ref{app:low_dim}.

\begin{enumerate}
    \item \textit{Uniform Manifold Approximation and Projection} \citep[\texttt{UMAP};][]{UMAP...McInnes2018} is used to compress $\Theta$ into 2 dimensions for visualization. The dimensions themselves are not meaningful so we do not perform any quantitative analysis with them, but we use \texttt{UMAP} for plots in the rest of this work.
    \item \textit{Principal Component Analysis} \citep[PCA;][]{PCA...Hotelling} is an unsupervised method that re-orients $\Theta$ into a new 1536-dim parameter space that we call $\Theta'$, by maximizing the variance along its axes. These are called ``principal components'' (PCs), and PCA is designed such that information is concentrated in the earliest PCs. Therefore taking a small subset of PCs is much more useful than a small set of random features from $\Theta$. For example, 44\% of the variance in $\Theta$ is explained by PC1 alone. In this work we use different subsets of the PCA space depending on task complexity, but often use the top 4 PCs (which we call $\Theta'_4$), top 8 ($\Theta'_8$), or top 16 ($\Theta'_{16}$).
    \item We use \textit{Partial Least Squares} (PLS) to construct a 1536-dim space analogous to $\Theta'$, however PLS is a supervised method that we design to maximize the covariance with  age ($\tau$) along its axes. We call this space $\Theta^{\tau}$. It therefore has its age-relevant information concentrated in its top components by design, so we use the top 3 components of $\Theta^{\tau}$ (we call this subspace $\Theta^{\tau}_3$) for age inference tasks in this work.
\end{enumerate}
\begin{figure*}
  \centering
  \includegraphics[width=\textwidth]{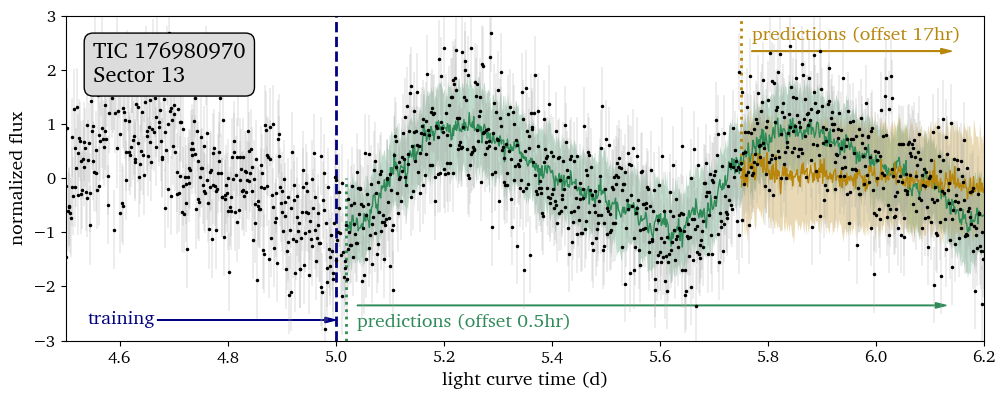}
  \caption{The confidence of flux predictions is highly dependent on the prediction horizon. Here we compare forward model predictions where the horizon is (i) 0.5 hours, and (ii) 17 hours. Shaded regions represent the 1$\sigma$ confidence interval of predictions. Black points with the grey error bars are true measured values, normalized as per Equation~\ref{eq:flux_norm}.}
  \label{fig:post_width_offset_comp}
\end{figure*}
\begin{figure*}
  \centering
  \includegraphics[width=\textwidth]{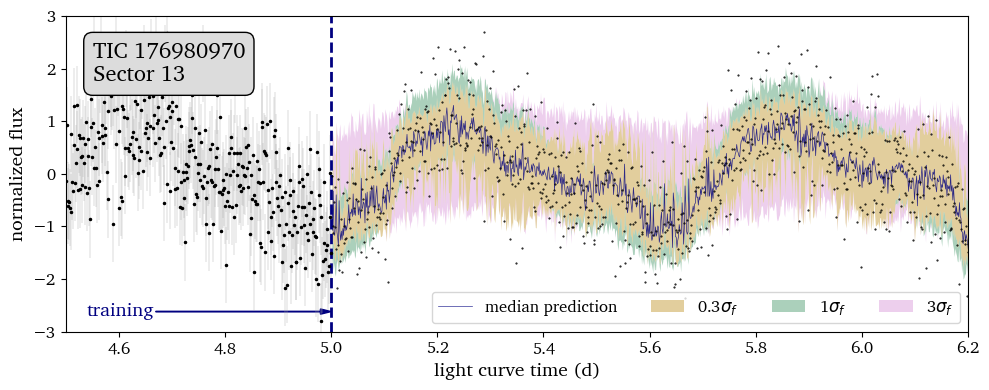}
  \caption{The confidence of flux predictions also depends on the measurement error. Here we compare flux predictions using the real measurement error against reconstructions based on models with artificially dampened (0.3 $\times$ true) and inflated (3 $\times$ true) \textit{TESS} flux uncertainty. Shaded regions represent the 1$\sigma$ confidence interval of $p(\hat{f})$ in each scenario. Black points with the grey error bars are true measured values, normalized as per Equation~\ref{eq:flux_norm}. Error bars are not shown in the prediction regime to avoid clutter. While the credible regions do not always symmetrically widen at every point with inflated $\hat{\sigma}_f$, the uncertainty increases overall.}
  \label{fig:post_width_flux_err_comp}
\end{figure*}
In the rest of this work we generally use the 2-dim \texttt{UMAP} for visualizations, the PCA space $\Theta'$ for interpretability and correlation analyses, and the 3-dim $\Theta^{\tau}_3$ for age inference.

\section{Flux Reconstruction}
\label{sec:flux_recon}

Here we examine how well the final model is able to accomplish its training task: flux reconstruction. We explore its performance at predicting the flux at ``unobserved'' points in an LC, and also what factors affect its predictions. Practically, there are several contexts where this could be applied, including infilling lower cadence observations with higher cadence predictions, interpolating through large data gaps such as the \texttt{TESS} downlink time, or predicting future observations.

\subsection{Context Effects}
\label{subsec:flux_recon_context}

We first evaluate \et's performance at different prediction ``horizons'' $h_p$, which is the time interval between data included in training context and the prediction target time. For example, a prediction horizon of one hour in the forward direction means that at a given time $t$, the flux measurement $f_t$ is predicted based on a training scope that includes all observed flux measurements from $f_0$ to $f_{t-1hr}$. We apply this horizon equally in both the forward and backward directions. As a consequence, flux measurements within $1h_p$ of the beginning of the LC have no forward training context, and flux measurements within $1h_p$ of the end of the LC have no backward training context, other than metadata. In Figure~\ref{fig:post_width_offset_comp}, we plot $p(\hat{f})$ using two different prediction horizons: 0.5 hours and 17 hours. These correspond to average offsets of 16 data points and 1024 data points, respectively, in a 2-minute cadence LC. It is clear that at shorter horizons, \et{} is more confident and able to track variability better. At longer horizons, its confidence is weaker and its predictions trend closer to the long-term median flux. This is not surprising, given that uncertainty of specific flux measurements will accumulate over time in a noisy light curve.

\begin{figure}
  \centering
  \includegraphics[width=\columnwidth]{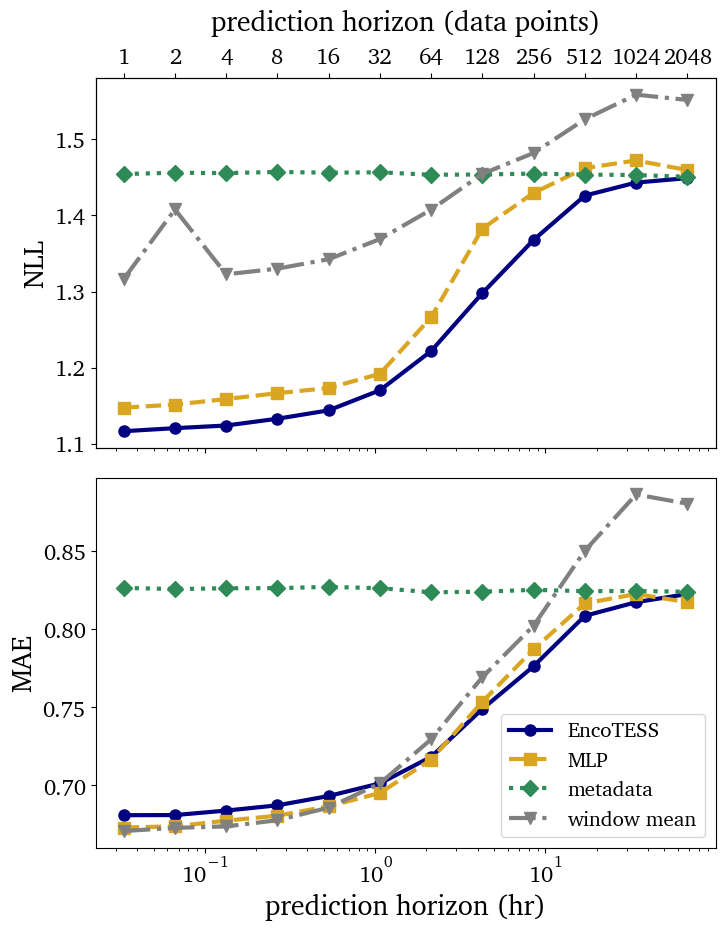}
  \caption{Flux reconstruction performance for the four model variants, using negative log likelihood (NLL) and median absolute errors (MAE) as metrics. Individual NLL values were clipped to $\approx10\sigma$ for stability. The RNN architecture and CNF prediction head used in \et{} provide value, particularly when looking at prediction horizons of a $\sim$few hours to days and when considering the entire likelihood shape, although the overall \textit{accuracy} of \et and the MLP are comparable.}
  \label{fig:prediction_baseline_comp}
\end{figure}

Next, we test the impact of varying observational uncertainty. Figure~\ref{fig:post_width_flux_err_comp} compares $p(\hat{f})$ for $\hat{\sigma}_f$ scalings between 0.3$\times$ and 3$\times$ the real $\hat{\sigma}_f$, which are used in the CNF prediction head. We see that there appears to be a minimum prediction uncertainty, characterized by the $p(\hat{f})$ width, that the model learns even for low $\hat{\sigma}_f$. This is representative of the typical noise floor of the LCs in our dataset. As expected, above that floor the $p(\hat{f})$ width increases with $\hat{\sigma}_f$, and the $p(\hat{f})$ median begins to regress towards the median of the entire LC.

\subsection{Performance Compared to Baselines}
\label{subsec:flux_recon_baseline_comp}

It is also important to characterize how well \et{} can reconstruct flux compared to more naive models, to determine whether its architecture is providing value. To that end, we tested \et{} against three baseline models: (i) a naive deterministic baseline model, (ii) a model based only on stellar metadata, and (iii) an MLP-based model.

The naive baseline model uses a windowed mean to predict the flux at any time based on the 32 observations preceding the target time and the 32 observations following the target time. Uncertainty for this model is estimated as the root mean square error (RMSE) of predictions, which is a metric used only to estimate this model's prediction uncertainty and not to compare between models.

The metadata-only baseline uses a CNF prediction head conditioned only on the 13 LC metadata fields and the target prediction time $\tilde{t}_y$, so cannot use the preceding or following LC context.

The learned MLP baseline uses the 32 preceding and following observations, combined with the same stellar metadata fields used in \et, in an MLP to predict the flux along with a learned Gaussian uncertainty.

For reference, \et has 92,857 learned parameters; this includes the prediction head which is why this number is larger than the parameter count of the core model quoted elsewhere. The metadata baseline has 14,386 learned parameters and the MLP baseline has 69,658 learned parameters. The windowed mean model is fully deterministic model so has 0 learned parameters.

These baselines were used because they help us assess the value added by different components of the \et{} architecture:

\begin{itemize}
    \item The windowed mean model is effectively the most naive linear method of flux prediction, so it provides a minimum performance baseline.
    \item The metadata model is \et{} without $H_t$, so it provides a baseline for how much additional value the hidden states provide.
    \item The MLP model uses only short-term LC context and no CNF prediction head, so it provides a baseline for how effective the RNN architecture is at capturing long term trends, and for how much value the CNF prediction head provides.
\end{itemize}

Figure~\ref{fig:prediction_baseline_comp} compares \et{} against the three baseline models using two different prediction metrics: negative log likelihood (NLL) and median absolute error (MAE). It is useful to consider both, since MAE measures how accurately the median $p(\hat{f})$ estimates the true flux while the NLL also considers the shape and width of $p(\hat{f})$. The absolute MAE and NLL values are not important, but lower corresponds to a better fit.

By both metrics, we see that the metadata model is invariant with prediction horizon, which is expected given that it has no LC context. Also, at the longest horizons shown in this plot ($\approx100$~hr), the predictive capabilities of the MLP and \et asymptote to approximately the level of the metadata model. This indicates that at long horizons, those models rely heavily on stellar properties to predict flux, and the LC context captured in $H_t$ is not useful.

\begin{figure*}
  \centering
  \includegraphics[width=\textwidth]{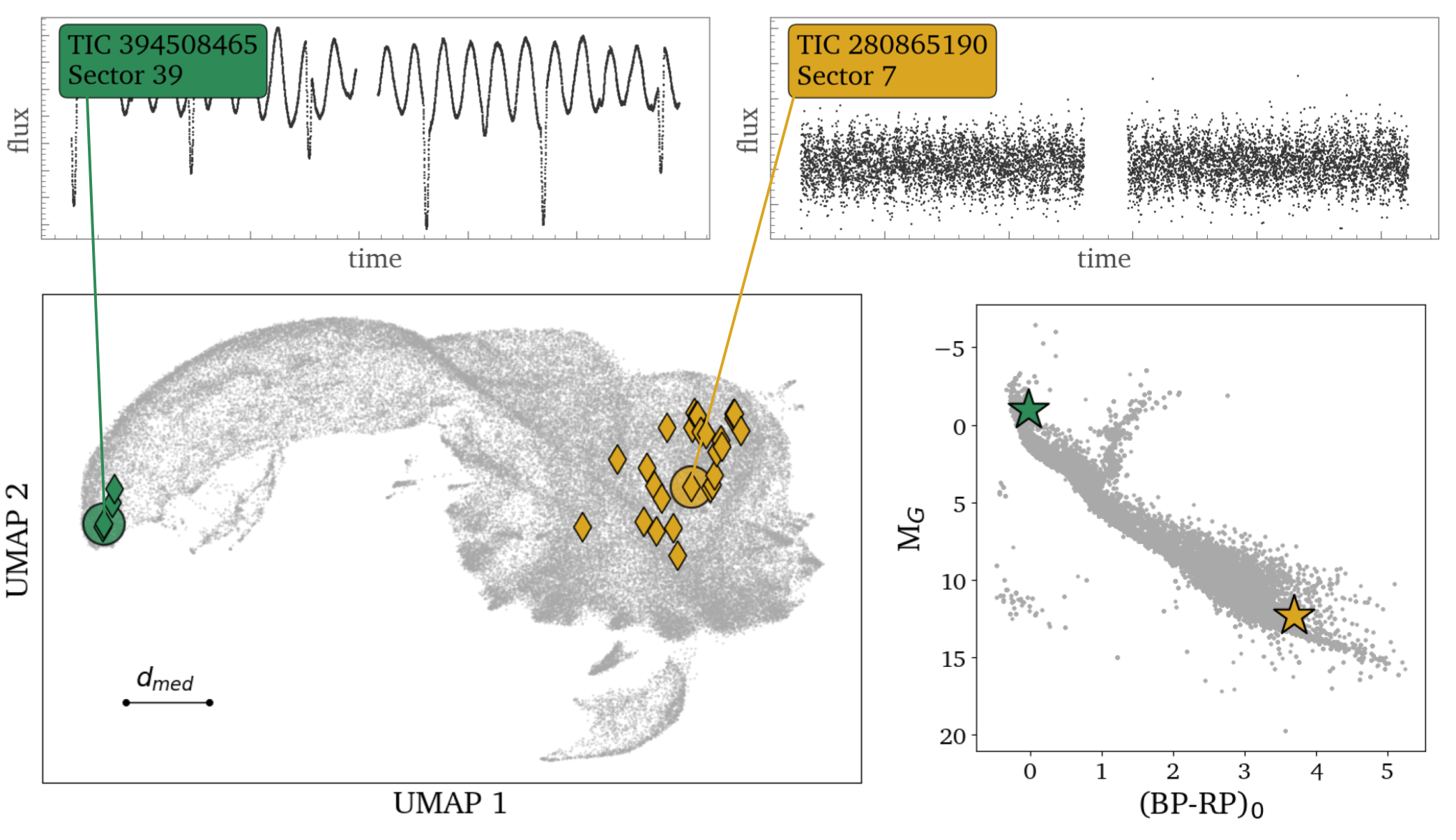}
  \caption{LC encodings ($\Theta$) for two sample stars. The bottom right panel shows each star's position on a color-magnitude diagram. The bottom left panel shows the positions of each $\Theta$, compressed into two dimensions with \texttt{UMAP}. The background points are all other stars in our catalog. Example LCs from each star are shown in the top panel, which correspond to the circled LCs in the \texttt{UMAP} plot. The median distance between points ($d_{\mathrm{med}}$) representing different LCs from the same star is plotted for reference.}
  \label{fig:umap_lcs}
\end{figure*}

Comparing the other three models, by NLL it is clear that the naive windowed mean model performs the worst at all horizons, and \et{} performs the best. In contrast, by median absolute error (MAE), all 3 models are comparable at short term horizons, but the windowed mean performs worse at long horizons. This indicates that disregarding uncertainty, flux prediction by averaging between neighboring points works reasonably well at short timescales, however at longer timescales using this performs worse than a learned global mean, which is what the MLP and \et{} converge to. 

The fact that \et{} outperforms the MLP in NLL more than it does in MAE indicates that the flexibility of the CNF head is an advantage. Since \et{} does still slightly outperform the MLP at long prediction horizons, the RNN architecture provides value for capturing trends on the scale of $\sim$hours to $\sim$days, although its advantage does seem to disappear at $\approx100$ hrs. Since the model size of \et is only slightly larger than that of the MLP, this indicates that there is real value in the architecture, and scaling up the size of \et in the future could improve performance. Furthermore, \et provides a built-in method of extracting the state of the LC at any point in time via the $H_t$.

\section{Variability characterization with light curve encodings}
\label{sec:variability_stats}

As described in \S\ref{subsec:model_output}, we can compute $\Theta$ for every LC in our dataset using the final trained version of \et. Figure~\ref{fig:umap_lcs} plots the 2-dim \texttt{UMAP} compressed versions of $\Theta$ with two stars highlighted as examples. In green (TIC 394508465) is a high mass eclipsing binary (EB), and in gold (TIC 280865190) is a rapidly rotating M dwarf that exhibits frequent flaring. We have 4 LCs from the EB in our dataset, and 24 LCs from the M dwarf. One sample LC from each is plotted in full. For reference, we have also plotted the median pairwise difference between $\Theta$ of LCs from the same stars across our entire catalog, demonstrating that unique LC features of the same star can shift $\Theta$.

While these stars were chosen as examples because they are distinctly different in many traditional ways, we can see how those properties translate to $\Theta$:

\begin{itemize}
    \item They are on opposite ends of a color-magnitude diagram as shown in the lower right panel.
    \item The sample LCs exhibit very different properties; aside from the obvious EB signature, the M dwarf has a much noisier LC and exhibits flaring, which the EB does not.
    \item Even though each LC is encoded independently, the LCs from each star are clustered together. The M dwarf's LCs are scattered more in the \texttt{UMAP} space than the EB's.
\end{itemize}

\begin{figure}
  \centering
  \includegraphics[width=\columnwidth]{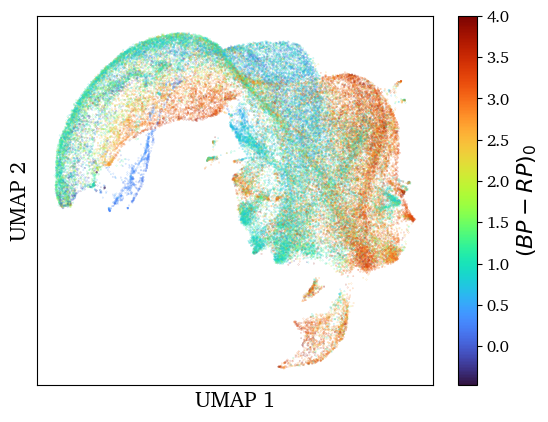}
  \caption{All LC encodings, compressed from the 1536-dim $\Theta$ into 2 dimensions with \texttt{UMAP}. Points are color coded by the Gaia $(BP-RP)_0$ color of the star.}
  \label{fig:umap_bprp0}
\end{figure}

\begin{figure*}
  \centering
  \includegraphics[width=\textwidth]{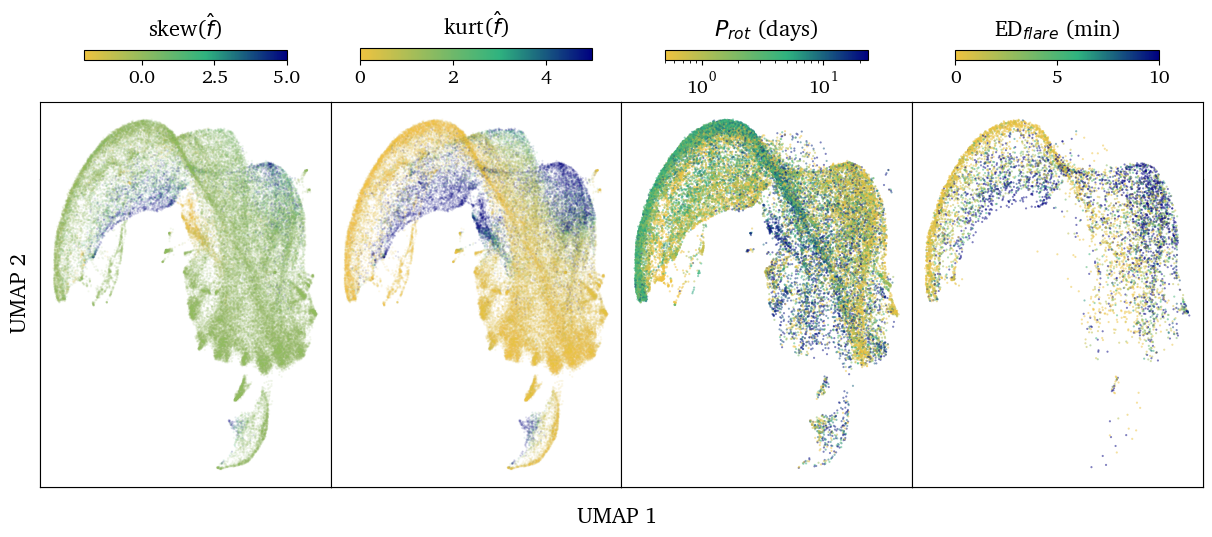}
  \caption{All stars in our dataset, positioned by two dimensional UMAP compression of their latent encodings. Each subplot shows the stars colour-coded by specific measured parameters, which are clearly tractable in all cases through the latent representations.}
  \label{fig:umap_diagnostics}
\end{figure*}

These observations reinforce several characteristics of \et's expected behavior. Different LCs from the same star should be clustered given that they share stellar properties, but the scatter is also informative. While it partly depends on the absolute number of LCs, the noise properties of different sectors may also affect LC location. Furthermore, the stochastic nature of flares means that they may not appear in all of the M dwarf's LCs. Finally, noisier LCs are more susceptible to sector-specific systematics, and although we have tried to mitigate these effects in \et, there is likely some persistence of such artifacts.

This was one specific example, but one could investigate LCs across $\Theta$ to study LC properties and stellar characteristics. For example, we found that most of the EB LCs in our dataset are extremely close to TIC 394508465 in the \texttt{UMAP} space. \et{} could be used in this way to classify different types of variability (e.g., flaring, rotation, eclipsing binaries, exoplanet transits, asteroseismic pulsations) automatically from light curves, similar to \citet{Poznanski...TESSLCs...2026arXiv260505324P} and \citet{Ding...StarCLR...2026ApJ..1003..141D}.

\subsection{Correlation with Physical Properties}

In addition to examining specific light curve features, we can plot trends in different stellar properties and summary statistics across $\Theta$. As an example, in Figure~\ref{fig:umap_bprp0} we have color coded each LC by Gaia $(BP-RP)_0$ of the star, plotted in \texttt{UMAP} space. Since color is one of our metadata input parameters, this partly reflects \et's own conditioning, but nevertheless the clear structure shows that color persists strongly through $\Theta$.

We can also go beyond color to inspect variability indicators that were \textit{not} included as metadata input. Figure~\ref{fig:umap_diagnostics} shows the \texttt{UMAP} LC projections color-coded in each panel by the value of a summary statistic. Skewness measures the asymmetry in the flux distributions of each LC, and kurtosis measures how heavy the tail of the distribution is. Both of these were calculated directly from the LCs. The third panel shows rotation $P_{rot}$ measurements derived by \citet{TARS...Boyle...2026arXiv260305586B}, and the last panel plots flaring equivalent duration ($ED_{\mathrm{flare}}$) as measured by \citet{Feinstein...2024AJ....168...60F}\footnote{This is not the total time that a star is in a flaring state during any given sector, but rather the time that it would take a star in its normal quiescent phase to emit as much energy as was produced by its flares. This is why it is on a timescale of minutes in this plot, not hours or days.}. To get a single $ED_{flare}$ per LC, we simply added up all of the individual flare ED measurements as calculated by \citet{Feinstein...2024AJ....168...60F} per sector and star.

There is clear structure in each plot, demonstrating that \et{} captures these measures of variability in $\Theta$. The quantities are also related in systematic, physical ways. To first order, we would expect stars that are rotating quickly to be younger and more magnetically active, and likely flaring more than slowly rotating stars. Stars that flare more are also more likely to have a larger flux kurtosis, and a larger skewness since flares manifest as positive jumps in flux. These relative trends are reflected in Figure~\ref{fig:umap_diagnostics}, where we see that skewness/kurtosis/ED$_{\mathrm{flare}}$ are generally positively correlated with each other and negatively correlated with $P_{rot}$. Interestingly, a negative skewness can also be an indicator of a transiting exoplanet, which manifests as periodic dips in flux. We investigated this specifically and found that the region of negative skewness around the center of the \texttt{UMAP} also has a high density of transiting exoplanets.

\begin{figure*}
  \centering
  \includegraphics[width=\textwidth]{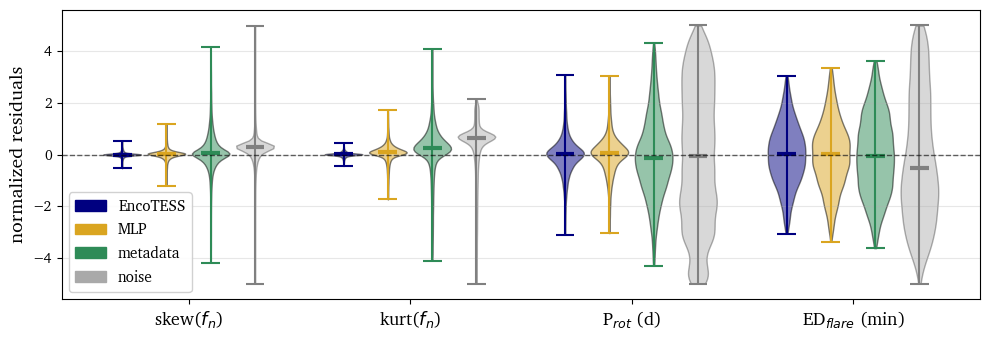}
  \caption{Violin plot comparing the recovery of 4 LC summary statistics from 4 different inputs: (i) The real $\Theta$ computed by \et, (ii) pure noise, (iii) only  metadata, and (iv) an MLP-derived baseline. (iii) and (iv) are analogous to the metadata and MLP variants described in \S\ref{subsec:flux_recon_baseline_comp}, but with a metric recovery target instead of flux reconstruction target. \et{} performs best overall, indicating that our architectural choices provide value. Residuals are transformed and scaled for ease of visualization, but key metrics are provided in native units in Table~\ref{tab:metric_recovery_stats}}
  \label{fig:baseline_tests}
\end{figure*}

\subsection{Recovery of Variability Statistics}
\label{subsec:var_state_recovery_perf_comp}

As a more quantitative test, we used \et to recover the summary statistics shown in Figure~\ref{fig:umap_diagnostics} from $\Theta$, and compared those results to three control models. In all four cases we trained a simple MLP predictor $\mathcal{M}$ to estimate each summary statistic from the inputs. Since the architecture of $\mathcal{M}$ is the same for each variant, this tells us whether $\Theta$ truly is adding value over more naive inputs. The 4 models we compare are: 
\begin{enumerate}[itemsep=0pt]
    \item \et. Input is $\Theta$.
    \item \textit{Pure Noise.} Input is 1536-dim Gaussian noise, normalized to the same scale as $\Theta$. This model represents the performance we would expect if $\Theta$ contained no useful information.
    \item \textit{Metadata Baseline.} Input is the 13 metadata fields used by \et. This model represents how well we can recover the summary statistics from stellar properties, so any improvement on that is due to information captured in $H_t$ from the LCs.
    \item \textit{MLP Baseline.} Input is a 256-dim vector pooled from the output of the last layer of the MLP model described in \S\ref{subsec:flux_recon_baseline_comp}. Any improvement over this can be attributed to the RNN architecture and \et's ability to capture longer-term LC trends.
\end{enumerate}
For each variant, no LCs were held out from pretraining, but recovery metrics were computed using 5-fold cross validation, and reported from the held out fold so there was no leakage between training and test datasets.

\begin{table*}
\centering
\caption{Median and interquartile range (IQR) of the residuals for each model/metric combination from a recovery test, in native units. }
\label{tab:metric_recovery_stats}
\begin{tabular}{lcccccccc}
\toprule
Model & \multicolumn{2}{c}{skew($f_n$)} & \multicolumn{2}{c}{kurt($f_n$)} & \multicolumn{2}{c}{$P_{\rm rot}$ (d)} & \multicolumn{2}{c}{ED$_{\rm flare}$ (min)} \\
\cmidrule(lr){2-3} \cmidrule(lr){4-5} \cmidrule(lr){6-7} \cmidrule(lr){8-9}
   & Med. & IQR & Med. & IQR & Med. & IQR & Med. & IQR \\
\midrule
EncoTESS & \textbf{0.002} & \textbf{0.037} & \textbf{0.013} & \textbf{0.091} & \textbf{0.01} & \textbf{0.67} & \textbf{0.00} & \textbf{3.22} \\
MLP & 0.014 & 0.084 & 0.101 & 0.337 & 0.03 & 0.71 & \textbf{0.00} & 3.62 \\
metadata & 0.027 & 0.310 & 0.273 & 0.847 & -0.09 & 2.20 & -0.01 & 5.10 \\
noise & 0.134 & 0.181 & 0.675 & 0.986 & -0.04 & 5.33 & -0.66 & 8.51 \\
\bottomrule
\end{tabular}
\end{table*}

We compare the models in Figure~\ref{fig:baseline_tests} by showing the distribution of residuals for each metric. These residuals are transformed and scaled for each of visualization, so only the relative spreads in that figure are meaningful, but we report key summary statistics in native units in Table~\ref{tab:metric_recovery_stats}. Specifically, we highlight: (i) the median residual for each model/metric combination, which is an indicator of overall bias, and (ii): the IQR of residuals, which is a measure of accuracy/precision.

For the $P_{rot}$ and $ED_{flare}$ indicators, we see that the \et bias is tied or comparable with the MLP, and \et does marginally better than the MLP by IQR. Aside from that, \et outperforms the other models by both indicators across all metrics, demonstrating that \et's architecture provides a quantifiable advantage over simpler implementations.

In particular, it clearly outperforms the other models in recovering the higher order flux summary statistics (skewness and kurtosis). This points to a distinct advantage of the RNN architecture: it is able to retain memory over the entire LC better than the other models which focus on short-term variability or the ability to persist LC context.

It is not surprising that the performance of the MLP is comparable to \et in recovering $ED_{flare}$, since flares manifest over relatively short segments of the LCs. It is more interesting that the performance is comparable for $P_{rot}$, but as shown in Figures~\ref{fig:post_width_offset_comp} and \ref{fig:prediction_baseline_comp}, \et flux predictions starts to regress to the mean LC behavior over intervals on the order of $\sim$ days, so it doesn't capture long period rotation as well. As suggested in \S\ref{subsec:flux_recon_baseline_comp}, a larger version of \et may be able to capture these trends better.

\section{Interpretability}
\label{sec:interpretability}

Here we investigate our results in the context of 2 specific questions:
\begin{enumerate}
    \item Which LC features or segments are important to \et?
    \item How does $\Theta$ correlate with summary statistics of variability and/or other key features?
    \item What are the age important features that are complementary to $P_{rot}$?
\end{enumerate}
which we address in the following subsections through analysis of the lower-dimensional representations of $\Theta$.

\subsection{Saliency Analysis}
\label{subsec:saliency}

We can apply a deterministic approach to evaluate which parts of LCs are most heavily weighted by \et. This is typically called ``saliency analysis'': the evaluation of which input parameters most strongly influence a model's output. In the context of \et, this means measuring how much each point in an LC contributes to the LC's learned representation. While it is likely rare for any single point in an LC to have a uniquely critical impact, saliency analysis can reveal important structure in the way LCs are encoded.

There are different targets that can be considered in this analysis as measures of the model's ``output'', and multiple different methods to evaluate the contribution of each input. For \et{} in particular, there are three targets that could all provide useful information: (i) the hidden states $H_t$ that represent the LC at any point $t$, (ii) the encoding $\Theta$ of each LC, and (iii) a lower dimensional representation of $\Theta$ such as $\Theta'$ (the PCA vectors as described in \S\ref{subsec:model_output}. Here, we use the first principal component from $\Theta'$, PC1, as our target since it is a single information-rich number that describes any LC.

\begin{figure*}
  \centering
  \includegraphics[width=\textwidth]{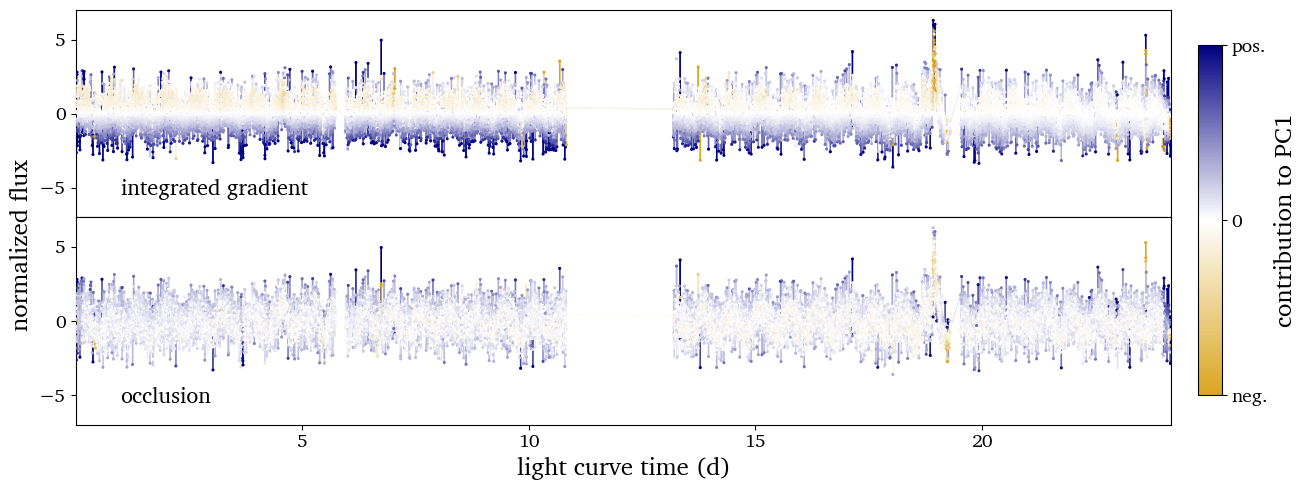}
  \caption{Colour coded contribution of every point in the LC (TIC 280865190, sector 7) to PC1, measured using an integrated gradient (IG) technique and occlusion. Positive and negative contributions move PC1 in different directions, but the magnitude of the contribution can be considered as an ``importance'' score.}
  \label{fig:saliency_M}
\end{figure*}

\begin{figure*}
  \centering
  \includegraphics[width=\textwidth]{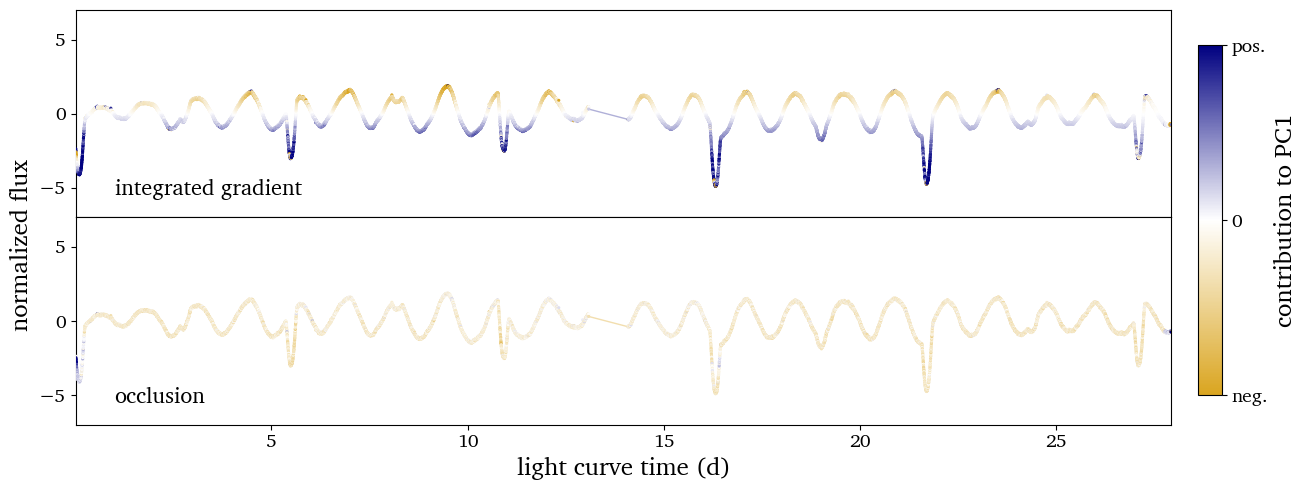}
  \caption{Analogous to \ref{fig:saliency_M}, but for the eclipsing binary TIC 394508465, sector 39.}
  \label{fig:saliency_EB}
\end{figure*}

\begin{figure*}
  \centering
  \includegraphics[width=\textwidth]{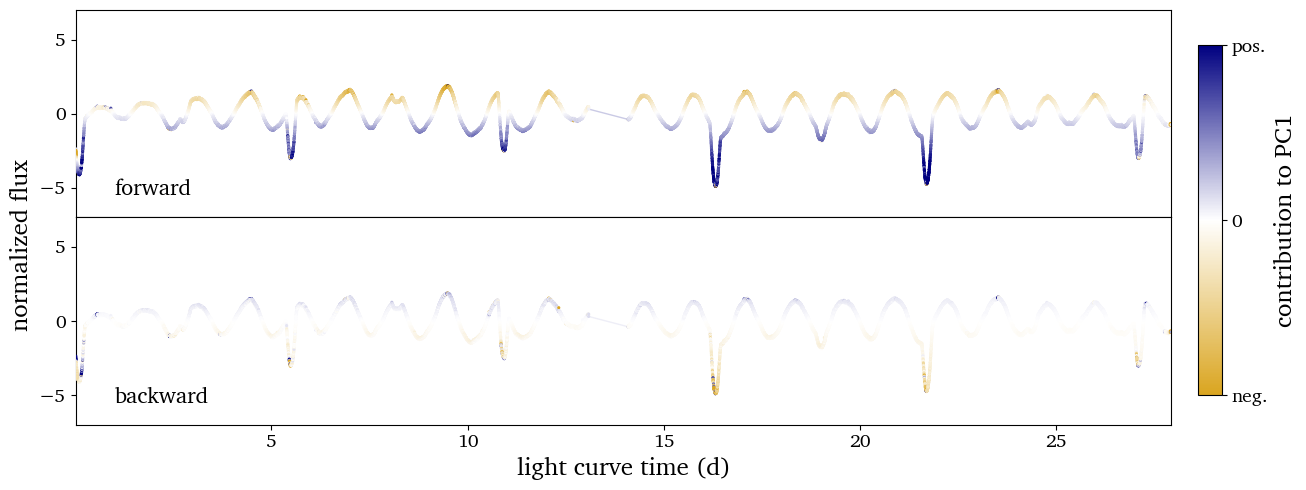}
  \caption{Saliency analysis of the eclipsing binary TIC 394508465, sector 39, comparing the magnitudes of the integrated gradients of the forward and backward hidden states. It is clear that the forward state is more sensitive to individual measurements overall, but the backward state is particularly sensitive to the eclipse features.}
  \label{fig:saliency_fwd_bwd_EB}
\end{figure*}

We also consider two methods of evaluating the contributions of input points: the integrated gradient (IG) and occlusion. Since the gradient of our target PC1 with respect to any input point $x=(f_t,\sigma_f,t)$ is defined, we calculate the IG as;
\begin{equation}
    IG = \int_0^{\hat{f}} \frac{\partial (PC1)}{\partial x}
\end{equation}
This provides a measurement of how sensitive PC1 is to changes in each input point. In contrast, occlusion measures the change in the target PC1 that results from masking any input point. Since this is a different metric than IG, we do not expect both to exhibit the same structure, but we do expect some correlation because both are some measure of the importance of each input point. While the absolute values of changes in PC1 as measured by IG and occlusion are not directly interpretable, the relative differences between points in the LC are.

In Figures~\ref{fig:saliency_M} and~\ref{fig:saliency_EB}, we show traces of the two LCs that were also highlighted in Figure~\ref{fig:umap_lcs} (the M dwarf TIC 280865190 and the eclipsing binary TIC 394508465), color coded by these saliency measurements.

Several interesting features can be seen in these. First, observations that are around the median flux generally contribute very little information (with the exception being the occlusion test for the EB). This effect seems to be accentuated in the noisier M dwarf LC ($LC_M$ hereafter).

Second, the flare towards the end of $LC_M$ has a strong influence on PC1, indicating that flaring is an important component of our latent space. Interestingly, most of the points in the flare push PC1 in the positive direction, but the very tip of the flare pushes it in the negative direction. This apparent inconsistency is likely the result of only considering the single dimension PC1, as the latent space is much richer and the contributions of these points in the flare are unlikely to be truly opposite when the entire $\Theta$ is considered.

Third, both the positive and negative local extrema in $LC_M$ push PC1 in a negative direction, but moderate positive flux values push PC1 in the positive direction. This indicates that \et{} is interpreting these points differently. Perhaps it interprets the local extrema as mostly noise, and that the smaller deviations model the rotation of the star more accurately. This is supported by inspecting $LC_M$ closely; the gold points at moderate positive flux do seem to visually track the rotation of this star very well.

Fourth, we see in the LC of the eclipsing binary ($LC_{EB}$ hereafter), that in this high signal-to-noise regime, there is not the same noise characterization as seen with $LC_M$. Positive flux values all push PC1 in the positive direction, and negative flux values push PC1 in the negative direction.

Finally, \et{} weights the eclipses seen in $LC_{EB}$ heavily, indicating that these contribute significantly to $\Theta$.

These saliency tests help us understand what LC features are contributing to the \et{} encodings in a deterministic way. While we only analyzed two specific LCs here, these tests could be widely applied across LCs to interpret $\Theta$.

\begin{figure*}
  \centering
  \includegraphics[width=\textwidth]{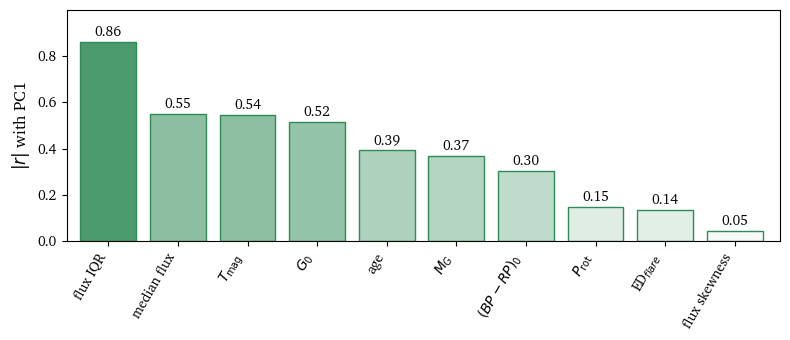}
  \caption{Top 10 light curve features that have the strongest correlation with the PC1 of $\Theta'$. We included metadata features in this analysis as well as age $\tau$ (as provided in the source catalogs) and the variability statistics shown in Figure~\ref{fig:umap_diagnostics}. It is clear that PC1 is primarily driven by the overall level of variability and target brightness, and color/magnitude measurements.}
  \label{fig:pc1_corr_features}
\end{figure*}

\subsubsection{Forward vs backward encoding}
\label{subsubsec:saliency_fwd_bwd}

Saliency analysis can also help us understand the differences in \et's forwards and backwards encoding channels. Naively, since LCs are generally symmetric (with the exception of flares, and in some cases occultations, which have a distinct morphology), we would expect the forwards and backwards channels to behave similarly. Interestingly, this is not true, as shown in Figure~\ref{fig:saliency_fwd_bwd_EB}. Overall, the magnitudes of the IG measurements are much higher when only considering the flow through the forward state. In contrast, while the backward state is generally less sensitive to each point individually, the relatively high sensitivity to the eclipse features stands out. Furthermore, we found that each direction contributes approximately equally (within 5-10\% depending on the metric used) $\Theta'$, indicating that the backward state encodes just as much useful information. This has interesting implications for our knowledge of how \et{} behaves and how TSFMs learn in general: the forward and backward directions naturally encode different types of features, even when provided training sequences that are mostly symmetric. The forward training learns to respond aggressively to local fluctuations, while the backward state is more robust to those, but both contribute important information to the global structure of $\Theta$.

\subsection{Correlation with variability summary statistics}
\label{subsec:var_stat_correlation}

At a broader level, we can inspect which traditional variability statistics of LCs are most strongly correlated with $\Theta$. Here, we again use PC1 as our low-dimensional representation of $\Theta$, and we measure correlation by computing the Pearson $r$ statistic between each feature and PC1. Since PCA is unsupervised, PC1 is a good diagnostic here as it is not biased towards age or any other features by design; it represents the axis where there is the greatest variance between LCs.

We included every metadata field in this test aside from $S$, $N_{cam}$, and $N_{ccd}$. We excluded these because even though they may be correlated with PC1, the numerical values and their ordering are not meaningful. We also included stellar age ($\tau$), absolute magnitude ($M_G$), and the 4 variability summary statistics presented in Figure~\ref{fig:umap_diagnostics}.

Figure~\ref{fig:pc1_corr_features} presents the results. Only the top 10 features ranked by $r$ are shown, but all others had correlations near the noise floor anyway. The strongest correlation is with $f_H$, indicating that the overall amplitude of variability has a large influence on an LCs location in $\Theta$-space. The other strongest correlations are primarily related to the overall brightness of the target and/or its position on a CMD; these are med$(f)$, $M_G$, $T_{mag}$, $G_0$, and $(BP-RP)_0$.

It is notable that there is significant correlation with age in PC1, even though PC1 was not designed specifically as an age diagnostic. The correlation with $P_{rot}$ is relatively minor, and approximately the same strength as with $ED_{\mathrm{flare}}$.

While this is a useful test, it comes with the caveat that this only explores PC1, which we did for ease of interpretability. Many of these features may have stronger correlations with later PCs, so these numbers should not be interpreted as the ceiling of correlation between $\Theta$ and these features.

In Appendix~\ref{app:pc_corner_plot} we show a corner plot of $\Theta_8'$ color coded by age as an example of how multiple PCs can be examined at once, however the makes interpretation more challenging.

\subsection{Complementary Age Information}
\label{subsec:interpretability_prot_orth}

\begin{figure*}
  \centering
  \includegraphics[width=\textwidth]{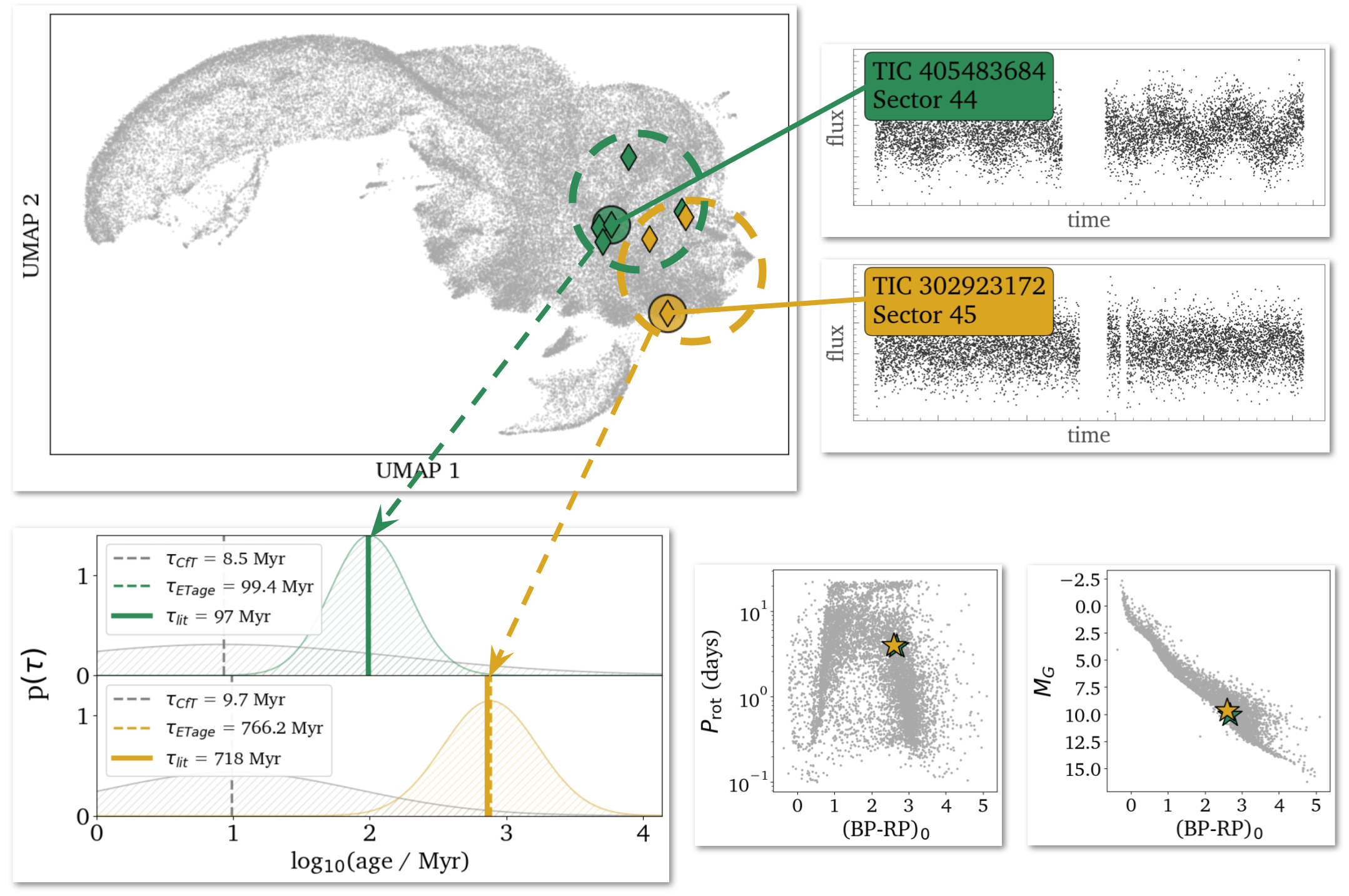}
  \caption{Schematic showing age inference for two stars that are in similar places on a color-magnitude and color-rotation diagram, but have different light curve properties and different encodings. \agep{} is unable to recover accurate ages from rotation alone, but \ageet{} is able to recover very accurate ages using the light curve encodings.}
  \label{fig:age_inf_comp}
\end{figure*}

Here, we investigate what age information we can extract from $\Theta$ that is complementary to $P_{rot}$. We start with a specific example in Figure \ref{fig:age_inf_comp}. The stars are located at very similar points on a CMD and color-$P_{rot}$ diagram, so they have almost identical age estimates from \agep. However, their true ages (from literature) differ by almost an order of magnitude. We also see that despite having similar rotation periods, they have different LC properties: in particular, there is a stronger rotational variability signal exhibited by TIC 405483684. This is reflected in $\Theta$: while there is some overlap in the positions of individual sectors (top left panel), the mean position of the LCs for each star are in different places in the \texttt{UMAP} encoding space. We see in the bottom left panel that \ageet{} therefore estimates different ages for these stars, which are quite close to the literature ages.

\begin{figure*}
  \centering
  \includegraphics[width=\textwidth]{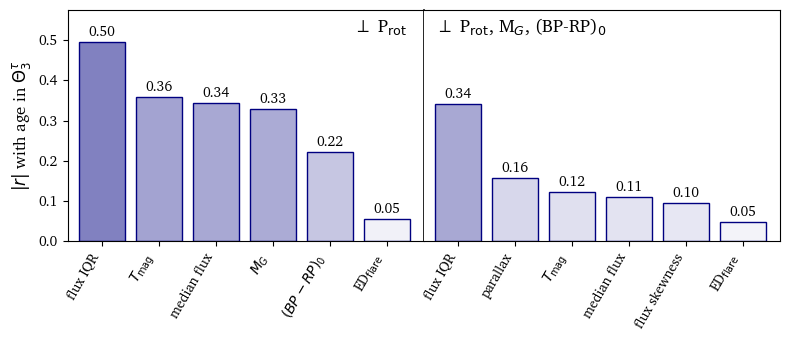}
  \caption{Features other than $P_{rot}$ that are most aligned with age in the \et{} latent space. In the left panel we measure the correlation of each feature with age in the 3-dim $\Theta^{\tau}_3$ space that is linearly orthogonal to $P_{rot}$. We do the same in the right panel, but in a space linearly orthogonal to the primary $P_{rot}$/$M_G$/$(BP-RP)_0$ direction. In each panel we only show the top six features; the correlation with any other features is $r\leq0.02$.}
  \label{fig:prot_orth_features}
\end{figure*}

We can check this in a more systematic way by probing the LC properties that are aligned most strongly with $\Theta$ \textit{but not with $P_{rot}$}. To do this, we first determined the linear direction in $\Theta$ for which covariance with $P_{rot}$ is maximized. Conceptually, this can be thought of as the primary $P_{rot}$ axis in $\Theta$.

We then performed a linear regression to fit:
\begin{equation}
    \Theta = \beta_0 + \beta_1 \cdot P_{rot}
\end{equation}
across all stars, from which we obtain a global $\beta_0$ and $\beta_1$. From these we compute a ``fitted'' $\Theta$ using the true $P_{rot}$ of each star: $\bar{\Theta}$, which we then subtract from $\Theta$ to get the ``residual'' $\Theta_{\mathrm{res}}$, which conceptually is the part of $\Theta$ that $P_{rot}$ cannot account for\footnote{Since LC features may be correlated with each other in complex and highly nonlinear ways, this does not completely scrub $\Theta$ from $P_{rot}$ information. Some non-linear correlation with $P_{rot}$ will survive, and this procedure also likely removes some non-$P_{rot}$ information. We attempted to apply more complex nonlinear methods, but some residual information always remained, so we decided to use this method as an imperfect but useful and interpretable first-order approximation.}.

With $\Theta_{\mathrm{res}}$ we used PLS as described in \S\ref{subsec:model_output} to obtain compute a 3-dim space analogous to $\Theta^{\tau}_3$ that is linearly orthogonal to $P_{rot}$, and within this space we measured the correlation of various metadata features to the most predictive age direction. The provides a a quantification of the amount of $P_{rot}$-independent age information that each can provide. In this exercise we used all ages compiled from the source catalogs in Table~\ref{tab:lc_sources}, excluding ages from the NASA Exoplanet Archive. Importantly, while these are defined metrics that we can measure age correlation with, there are likely additional important features derived from $H_t$ that cannot be explained as easily with summary statistics.

Figure~\ref{fig:prot_orth_features} presents our results. We measure the correlation between tested features and age using the absolute value of the Pearson $r$ coefficient. In the left panel, we see that $f_H$, which is the half interquartile range of flux measurements as described in \S\ref{subsec:model_inputs} and a measure of the total level of variability in a LC, is the single most descriptive feature. The next four features (\textit{TESS} magnitude $T_{mag}$, med$(f)$, absolute magnitude $M_G$ and $(BP-RP)_0$) are all strong correlated with a star's position on a CMD. This is not surprising, as $P_{rot}$ is not a useful age indicator without some proxy for the mass of a star, which CMD position describes and typical gyrochronology relations already include.

To also remove the most CMD-dependent features, we repeated this test, but residualized against the $\Theta$ direction most aligned with the combination of $P_{rot}$, $M_G$, and $(BP-RP)_0$, instead of just $P_{rot}$. Parallax $\pi$, $T_{mag}$ and med($f$) are still important features, indicating that the model may rely on these features as a proxy for absolute magnitude. $f_H$ is still the most age-correlated feature, re-emphasizing that it holds $P_{rot}$-independent value, although its Pearson $r$ decreases (likely due to its strong correlation with $M_G$). Both tests also show that the total equivalent flare duration is the next most important purely physical age-correlated feature. There is minor age correlation with overall flux skewness, but at a level close to the noise floor. Overall flux kurtosis is not helpful and is not shown on these plots.

The fact that $f_H$ is important to \et corroborates the findings of \citet{Mathur...Sph...2023ApJ...952..131M} and \citet{Messina...PhotAmp...2021A&A...645A.144M}, the latter in particular having shown that photometric variability amplitude is correlated with age even in stars with similar colors and rotation periods across five young associations and open clusters. These results confirm that those findings hold true across a larger dataset extending to older ages. The results also agree with \citet{Poznanski...TESSLCs...2026arXiv260505324P} who found that variability amplitude was heavily reflected in their \textit{LC} encodings.

In addition to the overall level of photometric variability, this shows that flaring also provides age information complementary to $P_{rot}$. There are quite possibly more LC features that do not manifest in the summary statistics we have included in this test, but are captured by $\Theta$ via the $H_t$ computation and useful for age inference. Exploring that regime and connecting such LC features to physical processes in the future could be yield useful insights.

\section{Age Inference}
\label{sec:age_inference}

\begin{figure}
  \centering
  \includegraphics[width=\columnwidth]{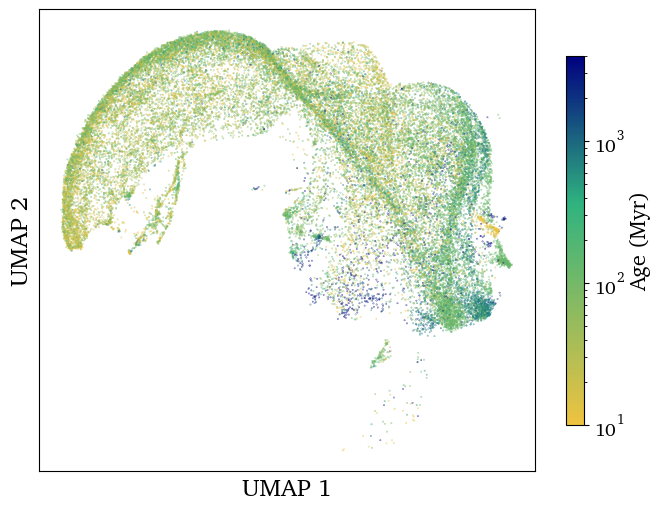}
  \caption{All LC encodings, compressed from 1536 dimensions into 2 dimensions with \texttt{UMAP}. Points are colour coded by the age of the star from the literature catalogs described in \S\ref{sec:data}, and clear structure is visible.}
  \label{fig:umap_age}
\end{figure}

While \et could be conceivably used in many downstream applications, here we specifically examine its age inference capabilities. A qualitative test of $\Theta$'s age information is to plot stellar age over our \texttt{UMAP} encodings, analogous to Figures \ref{fig:umap_bprp0} and \ref{fig:umap_diagnostics}. We do this in Figure~\ref{fig:umap_age}, and see clear structure. Since age was not directly included as an input parameter when training \et, these means that \et naturally encodes age-related observables. As discussed previously, there is vast literature on rotation as an age indicator, and on other observables such as $S_{ph}$ \citep[e.g.][]{Mathur...activity...2025ApJ...982..114M} and flaring rate \citep[e.g.][]{Feinstein...2024AJ....168...60F}, however it is challenging to derive an analytical age prescription that combines these metrics in a complementary way. Furthermore, there are likely additional LC features that could provide even more complementary age information. The \et{} encodings $\Theta$ offer an alternative method for stellar age inference, where we can leverage the information that \et{} has already learned to characterize these stars.

\subsection{Bayesian Inference Framework}
\label{subsec:age_inf_etage}

To clearly distinguish between the core \et{} architecture and this downstream age inference framework, which are decoupled, we will refer to the age inference framework that uses $\Theta$ as \ageet{} going forward.
\ageet is architecturally identical to the \texttt{ChronoFlow} age inference framework presented by \citetalias{ChronoFlow...2025ApJ...986...59V}, with an added step. \texttt{ChronoFlow} trained a CNF to estimate the likelihood:
\begin{equation}
    \label{eq:prot_cf_likelihood}
    p(P_{rot}\;|\;\tau,(BP-RP)_0,\sigma_{BR0})
\end{equation}
where $\sigma_{BR0}$ is the observational uncertainty on $(BP-RP)_0$ and $\tau$ is the stellar age.

In \ageet{}, we replace $P_{rot}$ with $\Theta^{\tau}_3$ as an age-informative representation of $\Theta$ as described in \S\ref{subsec:model_output}. In this context, since we want to predict the age once for every star, we create an aggregate $\Theta$ for each star by taking the maximum value of each of the 1536 dimensions across all $\Theta$ from different LCs of the same star. This is commonly referred to as \textit{max pooling}. We also tested mean and median pooling, but found max pooling to be the most stable.

We use $\Theta^{\tau}_3$ instead of using $\Theta$ directly because training a model to learn
\begin{equation}
    \label{eq:theta_etage_likelihood}
    p(\Theta\;|\;\tau,(BP-RP)_0,\sigma_{BR0})
\end{equation}
where is difficult and unstable due to the high dimensionality of $\Theta$. $\Theta^{\tau}_3$ is already designed to carry maximal age information in minimal dimensions, which is perfect for this task. We also tested training an MLP to learn a low-dimensional age-relevant representation of $\Theta$, however that was unstable and led to overfitting. A more thoughtful implementation of a targeted MLP could help in principle, but we leave that for future work and use the PLS-derived $\Theta^{\tau}_3$ here. Therefore we parameterize the \ageet{} likelihood as:
\begin{equation}
    \label{eq:theta_PCA_etage_likelihood}
    p(\Theta^{\tau}_3\;|\;\tau,(BP-RP)_0,\sigma_{BR0})
\end{equation}

Using $C_0$ as shorthand for $(BP-RP)_0$ and $\sigma_0$ as shorthand for $\sigma_{BR0}$ for readability, we follow the methodology of \citetalias{ChronoFlow...2025ApJ...986...59V} to compute posterior age probability distributions by applying Bayes' theorem and expanding:
\begin{equation}
\label{eq: bayes_post_expanded}
\begin{aligned}
    p&(\tau\;|\;\Theta^{\tau}_3, C_0, \sigma_{0}) = \\
    &\frac{p(\Theta^{\tau}_3\;|\; {C_0},\sigma_{0},\tau)
    \cdot p(C_0\;|\;\sigma_{0},\tau) \cdot p(\tau\;|\;\sigma_{0})}{p(\Theta^{\tau}_3,C_0\;|\;\sigma_{0})}
\end{aligned}
\end{equation}
We also use the same age prior as \citetalias{ChronoFlow...2025ApJ...986...59V}:
\begin{equation} \label{eq: age_prior}
    p(\tau\;|\;\sigma_{0}) \sim \mathcal{U}[1,13800]\;\mathrm{Myr}
\end{equation}
and the same color prior, which is the convolution of a uniform distribution of colors that we would expect to see ignoring measurement error:
\begin{equation}
\label{eq:c0_given_age}
    p(C_0\;|\;\tau) \sim \mathcal{U}[-0.05,3.8]
\end{equation}
with Gaussian measurement error $\mathcal{N}(0,\sigma_0)$. So, our final color prior conditioned on both age and color uncertainty is:
\begin{equation} \label{eq:prior_C0_cond_age_err}
    p(C_0\;|\;\tau,\sigma_{0}) \sim \mathcal{U}[-0.05,3.8] * \mathcal{N}(0,\sigma_{0})
\end{equation} 
Also, the evidence term $p(\Theta^{\tau}_3,C_0\;|\;\sigma_0)$ is constant across our model, and gets normalized out when evaluating the age posterior, so it can be ignored.

By using these closed form priors, we can easily evaluate the likelihood we learn from \ageet over a grid of possible ages to compute our age posteriors $p(\tau\;|\;\Theta^{\tau}_3, C_0, \sigma_{0})$.

\subsection{Comparison to Gyrochronology}
\label{subsec:age_inf_gyro_comp}

\begin{table*}[t]
\centering
\renewcommand{\arraystretch}{1.15}
\setlength{\tabcolsep}{12pt}
\begin{tabular}{lrrrrr}
\toprule
Method & Pearson $r$ & MAE & median res & \%\,$|$res$|$ & Coverage \\
       &     & (dex) & (dex)     & $>0.5$\,dex   & p16-p84 (\%) \\
\midrule
\multicolumn{6}{l}{\emph{All Stars}}\\
\addlinespace[2pt]
\ageet{} ($\Theta^{\tau}_3$)     & \textbf{0.674} & \textbf{0.431} & \textbf{$+$0.028} & \textbf{31.0} & 72.1 \\
\agepfh{} ($P_{rot},f_H$) & 0.601 & 0.496 & $-$0.039 & 34.7 & \textbf{66.5} \\
\agep{} ($P_{rot}$)        & 0.532 & 0.529 & $-$0.064 & 38.7 & 72.8 \\
\addlinespace[3pt]
\hline\hline
\addlinespace[3pt]
\multicolumn{6}{l}{\emph{KM Subset}}\\
\addlinespace[2pt]
\ageet{} ($\Theta^{\tau}_3$)    & \textbf{0.709} & \textbf{0.395} & \textbf{$+$0.021} & \textbf{27.8} & 73.2 \\
\agepfh{} ($P_{rot},f_H$) & 0.489 & 0.530 & $-$0.092 & 37.5 & \textbf{66.2} \\
\agep{} ($P_{rot}$)      & 0.348 & 0.567 & $-$0.138 & 42.5 & 75.2 \\
\bottomrule
\end{tabular}
\caption{Results from three different age recovery models: \ageet{} which uses the 3-dim $\Theta^{\tau}_3$ as its indicator, \agepfh{} which uses both $P_{rot}$ and $f_H$ as indicators, and \agep{} which uses only $P_{rot}$ as its age indicator. The models are compared using 5 different metrics: Pearson $r$, median absolute error (MAE), median residual, \% of residuals $>0.5$ dex, and \% of true ages covered by the p16--p84 posterior interval. This last metric should be close to 68\% for a well calibrated model. In 4/5 metrics, \ageet{} outperforms \agepfh{} which outperforms \agep. In p16-p84 coverage, \agepfh{} outperforms the other two models, which both exhibit overly conservative posteriors (although only by 4--7\%). We have measured these metrics over all stars in our catalog (upper block), and again over only the region of relatively young, active stars (we call this the \textit{KM} region for simplicity) as outlined in Figures~\ref{fig:logL_grid_P} and \ref{fig:logL_grid_PfH}. The difference between \ageet and the other 2 models is larger in this regime.}
\label{tab:observable_comparison}
\end{table*}

To determine whether the $\Theta$ is more informative than $P_{\rm rot}$ as an age indicator, we compare \ageet's age recovery performance against two baseline models:

\begin{enumerate}
    \item \agep: A baseline gyrochronology model that is architecturally identical to \texttt{ChronoFlow} \citepalias{ChronoFlow...2025ApJ...986...59V}.
    \item \agepfh: A model that uses both $P_{rot}$ and $f_H$ as observational indicators.
\end{enumerate}

All 3 model variants were trained on the 2,893 stars in the \et catalog with documented ages and rotation periods in the literature sources, to ensure that performance differences were purely based on the model. Exoplanet hosts were excluded from this due to the heterogeneity and typically high uncertainty of their age measurements. They all implement the same Bayesian inference framework, priors, and gridding strategy to evaluate the age posteriors as described in \citetalias{ChronoFlow...2025ApJ...986...59V}.

We used ``leave-one-sector-out'' (LOSO) cross validation to test each model. In this scheme, we constructed 10 folds of stars, with no overlap in sectors. We trained 10 different versions of each model using 9/10 folds for training with the 10th was held out to compute test statistics. Within the 9 training folds, an 85\%/15\% training/validation split was used to select the best model.

This entire procedure was repeated 5 times for each model with different random seeds, and the posterior age distributions from each were combined to get an ensemble age posterior for each star. 

There is some nuance to the LOSO mechanics, since many stars appear in multiple sectors. To ensure that there was absolutely no leakage between the training and test datasets, if sector X was part of the test dataset, any star that had a LC in sector X was excluded from the training and validation datasets. This is a conservative way to prevent \ageet from overfitting to any unique stellar signatures, as the purpose of this test is to evaluate \ageet's ability to characterize LCs from ``new'' sectors.

Preventing leakage between training and test sectors is important to consider because direct encoders like \et{} naturally include sector-specific systematics. This can be an advantage, as described in \S\ref{sec:data}, since this enables \et{} to separate systematic from astrophysical signatures. However it also requires more thoughtful testing since age information can be leaked into test datasets through these systematics. This leakage can allow ``cheating'', where \et learns to use the correlation between sector and age, since these stars are in localized open clusters and not randomly distributed across the sky. This is also important for \agepfh since the $f_H$ metric is sensitive to sector-specific noise. It is not as important for \agep since $P_{rot}$ measurements are more robust to sector-specific systematics, but we nevertheless train each model with LOSO for consistency.

\begin{figure*}
  \centering
  \includegraphics[width=\textwidth]{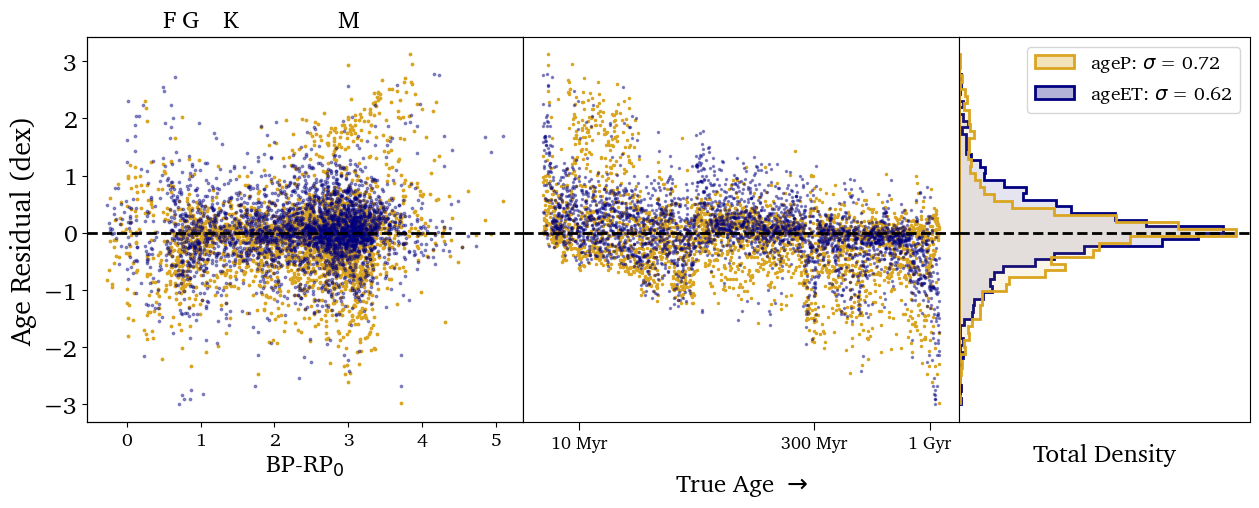}
  \caption{A comparison of ages inferred from \ageet and \agep. The left panel plots the residuals against $(BP-RP)_0$, with approximate spectral types shown on top. The middle panel plots them against true age; we place stars along the x-axis evenly here in order of increasing age for ease of interpretation. The right panel shows the overall distribution of residuals. \ageet is more precise for younger and redder stars, and less biased to the mean than \agep. From the histogram we also see that \ageet is less systematically biased and has a narrower distribution of residuals than \agep, indicating that its posterior age estimates are better calibrated.}
  \label{fig:residuals}
\end{figure*}

\begin{figure}
  \centering
  \includegraphics[width=\columnwidth]{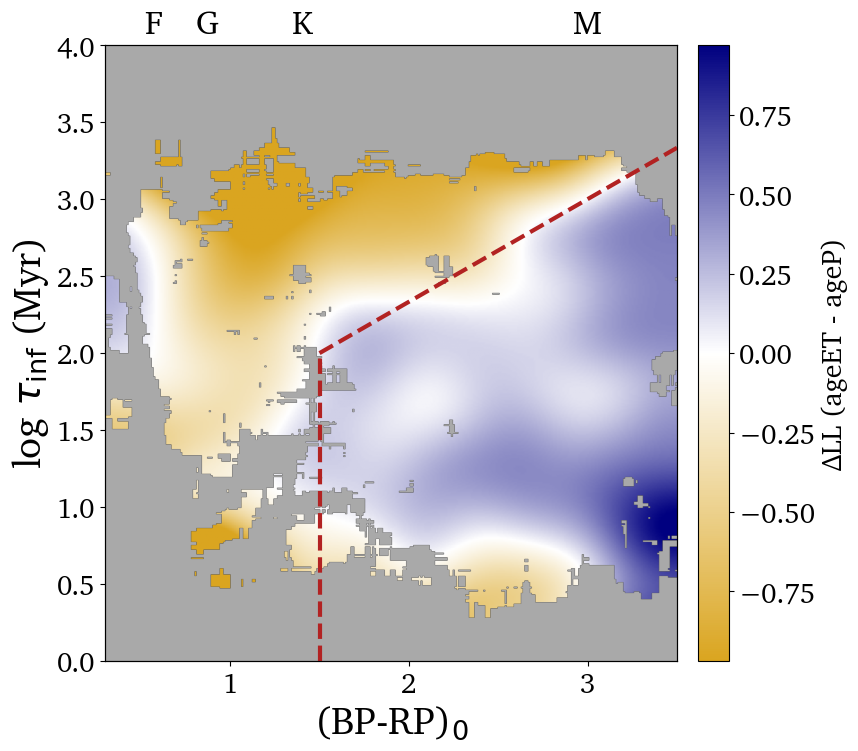}
  \caption{Difference in LL values from \ageet{} and \agep{} for stars in our catalog with measured rotation periods and ages. Blue regions indicate where \ageet{} is stronger, and gold regions indicate where \agep{} is stronger by this metric. LL values are binned with a resolution of 0.01 in $log (\tau_{inf})$ and 0.01 in $(BP-RP)_0$. Bins are smoothed over 0.2 in each direction to remove noise, and only bins with $\geq5$ stars are kept. Approximate spectral types are shown on top, as described in \ref{fig:residuals}. To the bottom right of the red dashed line is the \textit{KM} region described in Table~\ref{tab:observable_comparison}, where \ageet consistently outperforms \agep.}
  \label{fig:logL_grid_P}
\end{figure}

\begin{figure}
  \centering
  \includegraphics[width=\columnwidth]{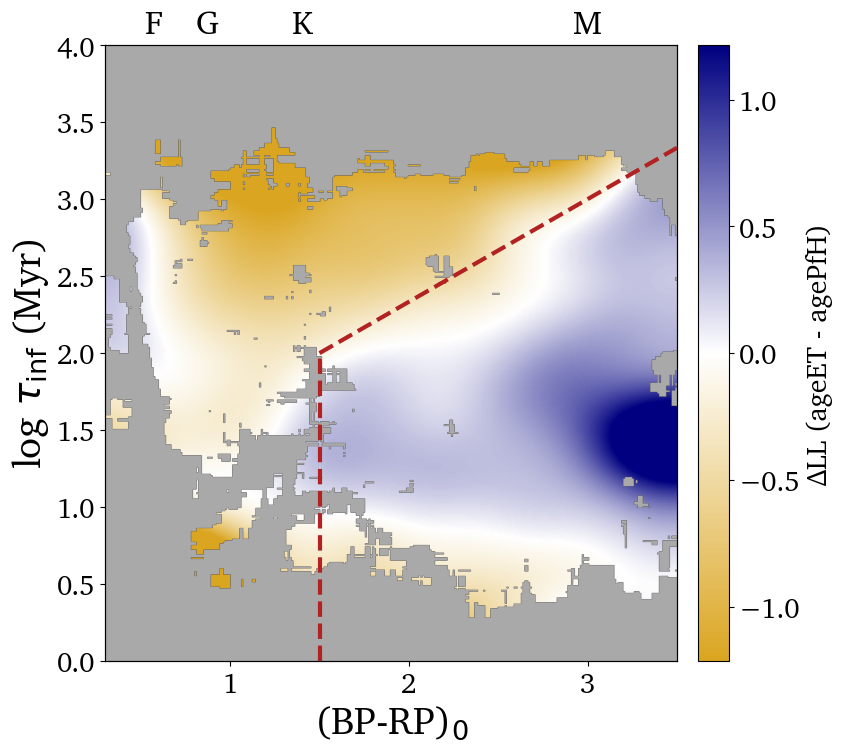}
  \caption{Analogous to Figure~\ref{fig:logL_grid_P} but comparing \ageet to \agepfh. \agepfh is stronger than \agep, but \ageet still outperforms \agepfh over the \textit{KM} region.}
  \label{fig:logL_grid_PfH}
\end{figure}

Table \ref{tab:observable_comparison} presents the results from these tests. In all metrics, \agepfh outperforms \agep. In 4/5 metrics, \ageet outperforms them both. We see by Pearson $r$ and median absolute error (MAE) that the \ageet age estimates are more accurate than the other models. It also has the lowest median residual, indicating that it has the least systematic bias. Finally, it has the lowest percentage of residuals $>0.5$ dex, meaning it has the fewest catastrophic failures. The only metric it does not beat \agepfh in is the percentage of true ages that lie with the p16--p84 posterior interval. This is not an indication of poor age estimates, just that the width of its posteriors are overly conservative.

We computed these metrics for all stars, and then for a subset of $\sim$KM stars that lie in the region highlighted in Figures~\ref{fig:logL_grid_P} and \ref{fig:logL_grid_PfH}. This is not an unbiased subset: we chose this region specifically because this is where \ageet performs the best. In this parameter space, it outperforms the other models by even wider margins.

In Figure~\ref{fig:residuals} we compare the distribution of residuals ($\tau_{\rm{inf}} - \tau_{\rm{lit}}$) from \agep{} and the \ageet{} LOSO test, where $\tau_{inf}$ is the median of the age posterior. The approximate spectral types shown in the left panel are based on the updated table of \citet{PecautMamajek...sptype...2013ApJS..208....9P}\footnote{While the original paper does not include $(BP-RP)_0$ colors, we use the table that is maintained \href{https://github.com/emamajek/SpectralType/blob/master/EEM_dwarf_UBVIJHK_colors_Teff.txt}{online}, which does include $(BP-RP)_0$.}. Overall, \ageet{} residuals are narrower and centered closer to 0, indicating that $\Theta$ is a better age calibrator overall than $P_{rot}$. This is particularly true for young red stars, as shown in the left and middle panels of Figure~\ref{fig:residuals}.

In Figures~\ref{fig:logL_grid_P} and \ref{fig:logL_grid_PfH}, we compare the difference in log likelihood (LL) between \ageet{}/\agep and \ageet/\agepfh respectively. Using LL as a metric is more informative than residuals because it considers both the shape and width of $p(\tau)$. A larger LL corresponds to a better match, so \ageet{} outperforms \agep{} in positive (blue) regions, and the inverse is true in negative (gold) regions. It is worth mentioning that the age axis in this figure is $\tau_{\mathrm{inf}}$, not $\tau_{\mathrm{lit}}$, so the interpretation is \textit{which age indicator is more reliable in a specific region of this parameter space} instead of \textit{which types of stars are better characterized by each age indicator}. However, these questions are highly correlated, especially in regions where the models do well.

In these figures we include all stars in our catalog with ages and rotation periods from the literature sources (excluding the exoplanet hosts and thick disk stars from \citet{Kim...thickdisk...2022MNRAS.510.4308K}). We bin results in a grid size of 0.2 in $(BP-RP)_0$ and $\log(\tau)$ and only show $\Delta LL$ for bins that have $\geq 5$ stars. We also apply Gaussian smoothing for ease of visualization.

There are clear broad trends that emerge across this parameter space. \agep{} is more reliable for $\sim$F and G type stars, and for K and M stars past an age that increases with color. This generally follows where we expect FGKM stars to converge onto a slowly-rotating sequence that has minimal scatter \citep[e.g.,][]{GodoyRivera...2021...rotcatalogue,ChronoFlow...2025ApJ...986...59V}, so it is not surprisingly that $P_{rot}$ is very useful in this regime. However, \ageet{} outperforms \agep{} for stars that are younger/redder than this threshold, which includes late K stars younger than $\sim100$ Myr, M stars $\lesssim1$ Gyr, where stellar spindown has not narrowed the $P_{rot}$ distribution from the initial periods yet \citep[e.g.,][]{BoyleBouma...gyro.Aper...2022arXiv221109822B,ChronoFlow...2025ApJ...986...59V}. Generally, younger and redder stars are also more magnetically active and therefore exhibit stronger variability in other ways, so this is evidence that other variability signatures captured by $\Theta$ do provide age information that is \textit{complementary to rotation}.

Since \ageet outperforms \agepfh in the same parameter space that it outperforms \agep, although by a smaller margin, \et is clearly extracting variability signals relevant to age that is complementary to both $P_{rot}$ and the amplitude of variability.

We note here that there is another regime at very young inferred ages ($\log \tau_{inf} \lesssim 0.75$ where \agep{} beats \ageet{}, even for young red stars. This is because there is a small population of stars across a range of older true ages that \ageet{} estimates as extremely young. This indicates that any stars with very young \ageet{} ages ($\lesssim$ a few Myr) should be treated with caution, but otherwise the broad trends described previously are consistent.

\section{Application to Field Stars}
\label{sec:field_ages}

We also examined whether $\Theta$ contains age information for field stars, since these stars are typically older and less variable than open cluster stars. To test this, we used the subset of 2,430 exoplanet host stars from our catalog having an estimated stellar age in their default record in the NASA Exoplanet Archive. Most of these are older than the oldest cluster stars in our catalog. We note that these age estimates are extremely heterogeneous from a wide variety of sources and methods, and often with large uncertainty, so there are inherent systematic bias in using these as ``true'' ages that we hope to recover. A detailed analysis of these systematics are outside the scope of this paper, so we take these age estimates at face value to see qualitatively whether our $\Theta$ can provide value for field star age estimates.

We tested age recovery with k-fold validation on the 2,430 exoplanet hosts. For each fold, we balanced sampling so that there was uniform representation across 10 different age bins spanning 0--14 Gyr in the training and validation samples. We did not perform LOSO tests since the exoplanet hosts do not experience the same spatial biases as cluster stars. Instead of using Bayesian inference, we train the age inference model (which we call \ageetxh) to learn $p(\tau)$ directly:
\begin{equation}
    \label{eq:NPE_field_ages}
    p(\tau\;|\;\Theta,(BP-RP)_0,\sigma_{BR0})
\end{equation}
which can be done in this case because we have a relatively uniform distribution of host ages, in contrast to \ageet which would be biased to the clustered ages seen in its training dataset. Here we do not need to compress the dimensionality of $\Theta$ because predicting a single variable from the high-dimensional $\Theta$ is more stable than trying to do the reverse.

\begin{figure*}
  \centering
  \includegraphics[width=\textwidth]{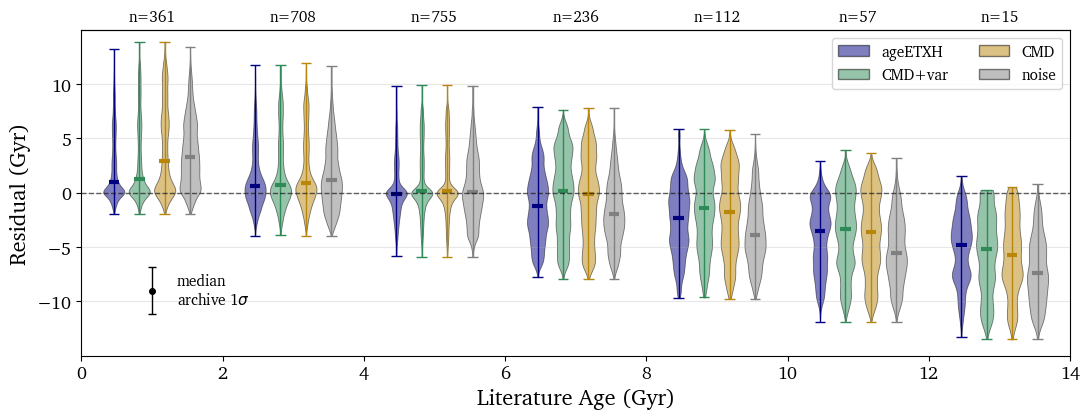}
  \caption{Violin plot showing age recovery for the 2,430 exoplanet hosts in our catalog with ages in the default NASA Exoplanet Archive records, grouped into 2 Gyr age bins. \et{} performance is shown in blue, the noise baseline in grey, and the CMD-only and CMD+variability variants in gold and green respectively. The majority of \ageetxh's predictive power for these stars comes from CMD indicators, with high level variability contributing in the 0-2 Gyr bin. The number of stars in each bin is shown on top of the plot. Past $\approx6$ Gyr, all models struggle.}
  \label{fig:field_age_recovery}
\end{figure*}

Architecturally, \ageetxh consists of an MLP with 2 hidden layers to compress $\Theta$ into 4 dimensions, then a CNF with 6 transforms and 2 hidden layers to predict age from these 4 dimensions.

We compare against three baselines:

\begin{enumerate}
    \item A pure noise model in which we replace $\Theta$ with Gaussian noise of the same scale as $\Theta$; this is analogous to the noise model in \S\ref{subsec:var_state_recovery_perf_comp}.
    \item A model that only includes CMD-related metadata fields as input instead of $\Theta$: $(BP-RP)_0$, $T_{mag}$, $G_0$, $\pi$, and associated errors.
    \item A model that includes the above CMD fields plus the 2 LC normalization metadata fields as input: $f_H$ and med($f$).
\end{enumerate}

The noise baseline represents a performance floor that is achieved by a model that only receives color as an input; practically it learns to predict the mean age conditioned on color for every star.

The second variant represents a CMD-based estimate of the stellar age, which for many of the older exoplanet hosts is the primary way that they are constrained.

The third model represents age estimates based on CMD properties and high level summary statistics of variability: this is the performance that we would expect from \ageetxh if the addition of $H_t$ in $\Theta$ was noninformative for these stars.

Figure~\ref{fig:field_age_recovery} shows the results from comparing these three variants, with residuals binned into seven different age groups where each spans 2 Gyr. It is clear that the majority of \ageetxh's predictive power comes from CMD information, as those variants are statistically indistinguishable past $\approx2$ Gyr. In the 0-2 Gyr bin, $f_H$ and med$(f)$ do provide additional useful information.

Overall, the performance of \ageetxh is comparable to the variant with CMD and variability metadata inputs, indicating \ageetxh is \textit{not} gaining any additional age information from the RNN architecture for these stars.

\section{Discussion and Future Work}
\label{sec:future}

So far, we have developed \et{} using a small subset of all of the \textit{TESS} data that are available, but in the future this could be expanded to all \textit{TESS} LCs at all cadences from all sectors. Since we developed \et{} to be time-aware, functionally the architecture can already accommodate this, however it will require testing. It could be valuable to make \et ``live'' by finetuning it on new sectors as more LCs become available, and updating a public library of the encodings $\Theta$ for every LC as the model receives more training data. As part of this work we provide $\Theta$ for all LCs processed by this version of \et, 

This architecture should be applicable to other survey data as well, such as \textit{Kepler} and \textit{K2} LCs. The upcoming \textit{PLATO} \citep[][]{PLATO...2025ExA....59...26R} mission will provide LCs over a much longer baseline than \textit{TESS} (2 years per hemisphere compared to 27 days per sector), and over a much larger field of view than \textit{Kepler} ($\approx2100$ deg$^2$ compared to $\approx115$ deg$^2$). Those characteristics make it a prime opportunity to test \et's ability to capture longer term variability. This type of framework could also conceivably be used for other astronomical time series products such as cluster-integrated LCs \citep[e.g.,][]{Wainer...ClusterLCs...2023AJ....166..106W}.

We also note here that machine learning and artificial intelligence capabilities are changing rapidly, and within the last few years there have been significant advances in TSFMs such as TimesFM, Moirai-MoE, Moirai 2.0, Chronos-2, and Timer-S1 \citep[][]{TimesFM...2023arXiv231010688D,MoiraiMoE...2024arXiv241010469L,Moirai2...2025arXiv251111698L,Chronos2...2025arXiv251015821F,TimerS1...2026arXiv260304791L}, and in particular SSMs \citep[e.g.,][]{Mamba...2023arXiv231200752G}. Many of these aim to address the scalability and heterogeneity concerns of existing TSFMs, which were two of the key motivations for developing \et{} from scratch. While most literature TSFMs still do not incorporate measurement uncertainty, it is quite conceivable that state-of-the-art industry TSFMs will soon be able to handle astronomy LCs well. Also, since general TSFMs \citep[e.g.,][]{TimesFM...2023arXiv231010688D,MoiraiMoE...2024arXiv241010469L} can now achieve performance comparable to fully supervised task-specific models in forecasting, it may be more efficient in the near future to fine-tune general TSFMs to astronomy data instead of pre-training an astronomy-specific TSFM from scratch. In addition to considering pre-built TSFMs, it may be useful to also test the performance of off-the-shelf tabular models such as \texttt{TabPFN} \citep[][]{TabPFN...2022arXiv220701848H} or Google's \texttt{TabFM} for downstream tasks.

Regardless whether we maintain the core architecture of \et{} as is going forward, or adopt components that have been proven in other contexts, this work shows that a stellar TSFM trained on multi-survey, multi-cadence, LCs has tremendous potential.

\subsection{Systematics}
\label{subsec:disc_systematics}

\citet{ChronoFlow...2025ApJ...986...59V} provide a detailed discussion of systematics that affect age inference in their gyrochronology model \texttt{ChronoFlow}; specifically the choice of dust map used for de-reddening, cluster membership criteria, and choice of cluster age used for calibration. Together, these factors contribute $\approx0.08$ dex of uncertainty to age estimates. This is an important source of error when posterior age predictions are narrow, but posteriors from \et typically have a width of $\approx0.4$ dex, so that error dominates. Nevertheless, we suggest adding this 0.08 dex systematic error to any reported age estimates from \et.

Since \et uses raw LCs an input, there are other systematics to consider related to the data acquisition, such as the instrumental noise for any specific observation. However, \et is designed to explicitly use observational parameters as input to account for this. We include the sector, camera, and CCD of each LC explicitly as metadata, and also $f_H$, med($f$), and $T_{mag}$, which carry both signal and noise properties. \et therefore has this observational context that can help separate observational artifacts from astrophysical signal.

This could be further improved with a contrastive learning scheme similar to that implemented by \citet{Ding...StarCLR...2026ApJ..1003..141D}. While our training scheme was purely self-supervised flux reconstruction, and $\Theta$ was calculated as a byproduct, contrastive learning would aim to make $\Theta$ from LCs of the same star (and similar stars) as equal as possible, and maximize the difference in $\Theta$ between very different types of stars. This would help remove observational artifacts from $\Theta$ while keeping the important astrophysical signal.

\subsection{Downstream Astrophysics}
\label{subsec:disc_astro}

We have deeply investigated the potential of \et for age inference using its encodings, however this is one of many potentially useful applications of \et. Another is a variability classifier, which \citet{Poznanski...TESSLCs...2026arXiv260505324P} and \citet{Ding...StarCLR...2026ApJ..1003..141D} both presented as an application of their frameworks. We could also identify regions in $\Theta$ space that correspond to known distinct populations of stars, and find other candidates for the population by proximity in $\Theta$.

Parameter inference is also possible, as the \ageet architecture could be easily modified to predict properties such as mass or temperature instead of age.

In the near future, expanding our training sample to include asteroseismic calibrators is well attainable and could provide additional value for age inference and these other applications.

\section{Conclusion}
\label{sec:conclusion}

In this work we developed \et: a Time Series Foundation Model (TSFM) based on a Recurrent Neural Network (RNN) architecture, trained on 69,345 2-minute \textit{TESS} light curves from 21,507 main sequence stars of spectral type $\sim$~FGKM. We built \et{} to be robust to high levels of noise, and to handle measurement uncertainty, irregular sampling, and large data gaps. \et{} uses raw light curves, stellar properties, and data acquisition parameters as input, to distinguish between astrophysical effects and observational systematics, and produces a meaningful fixed-dimension light curve encoding that can be used for downstream tasks such as classification or age inference.

The combination of the RNN architecture and probabilistic prediction head enables \et{} to outperform more naive models at predicting flux and recovering summary statistics. Interestingly, the forwards and backwards training components naturally encode different LC features, with the forward model being more sensitive to local flux variations but both contributing equally to the encoding. 

Importantly,

\begin{itemize}
    \item \et{} is only $\approx$64,000 parameters, which is at least two orders of magnitude smaller than typical literature TSFMs, and can trivially be run on most modern laptops.
    
    \item We showed that \et{} is able to characterize stars using their raw light curves, and encode that information into a latent representation that tracks different types of variability such as rotation, flaring, and higher order moments of the flux distribution.

    \item We found that \et{} outperforms $P_{rot}$ and variability amplitude as age indicators for younger, redder stars that have not converged onto the slowly-rotating sequence. This confirms that \et{} is able to extract age-dependent variability signatures that are complementary to $P_{rot}$.

    \item These results indicate that \et{} is useful for age inference in regimes where gyrochronology is less precise.
\end{itemize}

While \et{} is specifically trained on \textit{TESS} 2-minute light curves, expanding its scope to multi-cadence data, as well as \textit{Kepler}, \textit{K2}, and \textit{PLATO} light curves, has the potential to be incredibly fruitful for characterizing stellar variability, age inference, and therefore for downstream science. With this architecture, it would also be possible to train an \et-type model live as light curves are observed, and maintain a library of encodings that get updates as new data is acquired.

Overall, \et{} provides a promising conceptual foundation for a data-driven approach to characterizing stellar variability, and generating useful data products for downstream science.

%% Please use the acknowledgment and contribution environments. This will 
%% be anonomyized when the "anonymous" style option is used. 
\begin{acknowledgments}

P.R.V. was supported by funding from the Dunlap Institute, and a Postgraduate Scholarships - Doctoral (PGS-D) Award [PGS D - 589175 - 2024] from the Natural Sciences and Engineering Research Council of Canada (NSERC). PV would also like to thank Jonathan Gagné at the Montreal Planetarium and Université de Montréal for helping to extract data from MocaDB \citep[][]{MOCAdb...2026arXiv260215695G}, Andrea Zonca and Javier Hernandez Nicolau at SDSC for helping to optimize the \et{} code, and Ting Li at University of Toronto for providing local compute resourcing.

J.S.S. acknowledges the support of the Data Science Institute at the University of Toronto. J.S.S. was also supported by NSERC Discovery Grant RGPIN-2023-04849 and a University of Toronto Connaught New Researcher Award.

R.C. acknowledges the support of the Natural Sciences and Engineering Research Council of Canada (NSERC) and the Canada Research Chairs (CRC) Program [Award ID: CRC-2024-00372].

Gaia DR3 and TIC IDs of exoplanet hosts were retrieved from the Planetary Systems Composite Table~\citep[][]{NEA...PlanetSystemsTable...2020ipac.data..N13N} of the NASA Exoplanet Archive \citep[][]{Akeson...NASAExoplanetArchive...2013PASP..125..989A} on March 16, 2026. The NASA Exoplanet Archive is operated by the California Institute of Technology under contract with NASA under the Exoplanet Exploration Program.

This project made use of GPU resources from the \texttt{Bridges-2} supercomputer at the Pittsburgh Supercomputing Center (PSC), and the \texttt{Expanse} system at the San Diego Supercomputer Center (SDSC), allocated under the NSF Access project \textbf{PHY240253}.

This research has made use of the Astrophysics Data System, funded by NASA under Cooperative Agreement 80NSSC21M00561.

The authors disclose the use of Claude AI in this work for the purposes of developing and testing \et. Specifically, while all architectural and technical decisions were made by the authors, Claude Opus 4.6--5 was used to create and modify code. All AI-developed code was reviewed and validated by the authors, for which the authors assume complete ownership and responsibility.

All \et source code is available on Zenodo at: [insert here] and on GitHub at \href{https://github.com/philvanlane/encotess}{https://github.com/philvanlane/encotess}. These repositories also contain our catalog of light curve encodings.

\end{acknowledgments}

\facilities{Transiting Exoplanet Survey Satellite (\textit{TESS}), \texttt{Bridges-2} at Pittsburgh Supercomputing Center (PSC), \texttt{Expanse} at San Diego Supercomputer Center (SDSC).}

\software{}
\smallskip

This work made use of the following software packages: \texttt{Jupyter} \citep{2007CSE.....9c..21P,kluyver2016jupyter}, \texttt{matplotlib} \citep{Hunter:2007}, \texttt{numpy} \citep{numpy}, \texttt{pandas} \citep{mckinney-proc-scipy-2010,pandas_17229934}, \texttt{python} \citep{python}, \texttt{scipy} \citep{2020SciPy-NMeth,scipy_17467817}, \texttt{h5py} \citep{collette_python_hdf5_2014,h5py_7560547}, \texttt{PyTorch} v2.9.1\footnote{\href{https://github.com/pytorch/pytorch}{https://github.com/pytorch/pytorch}}, \texttt{umap-learn} \citep{mcinnes2018umap-software,2018arXivUMAP}, and \texttt{Zuko} \citep{rozet2022zuko,Zuko_10571785}.

This research made use of \texttt{Lightkurve}, a Python package for Kepler and TESS data analysis \citep[][]{Lightkurve...2018ascl.soft12013L}.

Software citation information aggregated using \texttt{\href{https://www.tomwagg.com/software-citation-station/}{The Software Citation Station}} \citep{software-citation-station-paper,software-citation-station-zenodo}.

%% Figure~and Table counter will not reset.

\bibliography{refs}{}

@ARTICLE{ChronoFlow...2025ApJ...986...59V,
       author = {{Van-Lane}, Phil R. and {Speagle}, Joshua S. and {Eadie}, Gwendolyn M. and {Douglas}, Stephanie T. and {Cargile}, Phillip A. and {Zucker}, Catherine and {Lu}, Yuxi (Lucy) and {Angus}, Ruth},
        title = "{ChronoFlow: A Data-driven Model for Gyrochronology}",
      journal = {\apj},
         year = 2025,
        month = jun,
       volume = {986},
       number = {1},
          eid = {59},
        pages = {59},
          doi = {10.3847/1538-4357/adcd73},
archivePrefix = {arXiv},
       eprint = {2412.12244},
 primaryClass = {astro-ph.SR},
       adsurl = {https://ui.adsabs.harvard.edu/abs/2025ApJ...986...59V}
}

@ARTICLE{TimesFM...2023arXiv231010688D,
       author = {{Das}, Abhimanyu and {Kong}, Weihao and {Sen}, Rajat and {Zhou}, Yichen},
        title = "{A decoder-only foundation model for time-series forecasting}",
      journal = {arXiv e-prints},
         year = 2023,
        month = oct,
          eid = {arXiv:2310.10688},
        pages = {arXiv:2310.10688},
          doi = {10.48550/arXiv.2310.10688},
archivePrefix = {arXiv},
       eprint = {2310.10688},
 primaryClass = {cs.CL},
       adsurl = {https://ui.adsabs.harvard.edu/abs/2023arXiv231010688D}
}

@ARTICLE{GaiaDR3...2023A&A...674A...1G,
       author = {{Gaia Collaboration} and {Vallenari}, A. and {Brown}, A.~G.~A. and {Prusti}, T. and {de Bruijne}, J.~H.~J. and {Arenou}, F. and {Babusiaux}, C. and {Biermann}, M. and {Creevey}, O.~L. and {Ducourant}, C. and {Evans}, D.~W. and {Eyer}, L. and {Guerra}, R. and {Hutton}, A. and {Jordi}, C. and {Klioner}, S.~A. and {Lammers}, U.~L. and {Lindegren}, L. and {Luri}, X. and {Mignard}, F. and {Panem}, C. and {Pourbaix}, D. and {Randich}, S. and {Sartoretti}, P. and {Soubiran}, C. and {Tanga}, P. and {Walton}, N.~A. and {Bailer-Jones}, C.~A.~L. and {Bastian}, U. and {Drimmel}, R. and {Jansen}, F. and {Katz}, D. and {Lattanzi}, M.~G. and {van Leeuwen}, F. and {Bakker}, J. and {Cacciari}, C. and {Casta{\~n}eda}, J. and {De Angeli}, F. and {Fabricius}, C. and {Fouesneau}, M. and {Fr{\'e}mat}, Y. and {Galluccio}, L. and {Guerrier}, A. and {Heiter}, U. and {Masana}, E. and {Messineo}, R. and {Mowlavi}, N. and {Nicolas}, C. and {Nienartowicz}, K. and {Pailler}, F. and {Panuzzo}, P. and {Riclet}, F. and {Roux}, W. and {Seabroke}, G.~M. and {Sordo}, R. and {Th{\'e}venin}, F. and {Gracia-Abril}, G. and {Portell}, J. and {Teyssier}, D. and {Altmann}, M. and {Andrae}, R. and {Audard}, M. and {Bellas-Velidis}, I. and {Benson}, K. and {Berthier}, J. and {Blomme}, R. and {Burgess}, P.~W. and {Busonero}, D. and {Busso}, G. and {C{\'a}novas}, H. and {Carry}, B. and {Cellino}, A. and {Cheek}, N. and {Clementini}, G. and {Damerdji}, Y. and {Davidson}, M. and {de Teodoro}, P. and {Nu{\~n}ez Campos}, M. and {Delchambre}, L. and {Dell'Oro}, A. and {Esquej}, P. and {Fern{\'a}ndez-Hern{\'a}ndez}, J. and {Fraile}, E. and {Garabato}, D. and {Garc{\'\i}a-Lario}, P. and {Gosset}, E. and {Haigron}, R. and {Halbwachs}, J. -L. and {Hambly}, N.~C. and {Harrison}, D.~L. and {Hern{\'a}ndez}, J. and {Hestroffer}, D. and {Hodgkin}, S.~T. and {Holl}, B. and {Jan{\ss}en}, K. and {Jevardat de Fombelle}, G. and {Jordan}, S. and {Krone-Martins}, A. and {Lanzafame}, A.~C. and {L{\"o}ffler}, W. and {Marchal}, O. and {Marrese}, P.~M. and {Moitinho}, A. and {Muinonen}, K. and {Osborne}, P. and {Pancino}, E. and {Pauwels}, T. and {Recio-Blanco}, A. and {Reyl{\'e}}, C. and {Riello}, M. and {Rimoldini}, L. and {Roegiers}, T. and {Rybizki}, J. and {Sarro}, L.~M. and {Siopis}, C. and {Smith}, M. and {Sozzetti}, A. and {Utrilla}, E. and {van Leeuwen}, M. and {Abbas}, U. and {{\'A}brah{\'a}m}, P. and {Abreu Aramburu}, A. and {Aerts}, C. and {Aguado}, J.~J. and {Ajaj}, M. and {Aldea-Montero}, F. and {Altavilla}, G. and {{\'A}lvarez}, M.~A. and {Alves}, J. and {Anders}, F. and {Anderson}, R.~I. and {Anglada Varela}, E. and {Antoja}, T. and {Baines}, D. and {Baker}, S.~G. and {Balaguer-N{\'u}{\~n}ez}, L. and {Balbinot}, E. and {Balog}, Z. and {Barache}, C. and {Barbato}, D. and {Barros}, M. and {Barstow}, M.~A. and {Bartolom{\'e}}, S. and {Bassilana}, J. -L. and {Bauchet}, N. and {Becciani}, U. and {Bellazzini}, M. and {Berihuete}, A. and {Bernet}, M. and {Bertone}, S. and {Bianchi}, L. and {Binnenfeld}, A. and {Blanco-Cuaresma}, S. and {Blazere}, A. and {Boch}, T. and {Bombrun}, A. and {Bossini}, D. and {Bouquillon}, S. and {Bragaglia}, A. and {Bramante}, L. and {Breedt}, E. and {Bressan}, A. and {Brouillet}, N. and {Brugaletta}, E. and {Bucciarelli}, B. and {Burlacu}, A. and {Butkevich}, A.~G. and {Buzzi}, R. and {Caffau}, E. and {Cancelliere}, R. and {Cantat-Gaudin}, T. and {Carballo}, R. and {Carlucci}, T. and {Carnerero}, M.~I. and {Carrasco}, J.~M. and {Casamiquela}, L. and {Castellani}, M. and {Castro-Ginard}, A. and {Chaoul}, L. and {Charlot}, P. and {Chemin}, L. and {Chiaramida}, V. and {Chiavassa}, A. and {Chornay}, N. and {Comoretto}, G. and {Contursi}, G. and {Cooper}, W.~J. and {Cornez}, T. and {Cowell}, S. and {Crifo}, F. and {Cropper}, M. and {Crosta}, M. and {Crowley}, C. and {Dafonte}, C. and {Dapergolas}, A. and {David}, M. and {David}, P. and {de Laverny}, P. and {De Luise}, F. and {De March}, R. and {De Ridder}, J. and {de Souza}, R. and {de Torres}, A. and {del Peloso}, E.~F. and {del Pozo}, E. and {Delbo}, M. and {Delgado}, A. and {Delisle}, J. -B. and {Demouchy}, C. and {Dharmawardena}, T.~E. and {Di Matteo}, P. and {Diakite}, S. and {Diener}, C. and {Distefano}, E. and {Dolding}, C. and {Edvardsson}, B. and {Enke}, H. and {Fabre}, C. and {Fabrizio}, M. and {Faigler}, S. and {Fedorets}, G. and {Fernique}, P. and {Fienga}, A. and {Figueras}, F. and {Fournier}, Y. and {Fouron}, C. and {Fragkoudi}, F. and {Gai}, M. and {Garcia-Gutierrez}, A. and {Garcia-Reinaldos}, M. and {Garc{\'\i}a-Torres}, M. and {Garofalo}, A. and {Gavel}, A. and {Gavras}, P. and {Gerlach}, E. and {Geyer}, R. and {Giacobbe}, P. and {Gilmore}, G. and {Girona}, S. and {Giuffrida}, G. and {Gomel}, R. and {Gomez}, A. and {Gonz{\'a}lez-N{\'u}{\~n}ez}, J. and {Gonz{\'a}lez-Santamar{\'\i}a}, I. and {Gonz{\'a}lez-Vidal}, J.~J. and {Granvik}, M. and {Guillout}, P. and {Guiraud}, J. and {Guti{\'e}rrez-S{\'a}nchez}, R. and {Guy}, L.~P. and {Hatzidimitriou}, D. and {Hauser}, M. and {Haywood}, M. and {Helmer}, A. and {Helmi}, A. and {Sarmiento}, M.~H. and {Hidalgo}, S.~L. and {Hilger}, T. and {H{\l}adczuk}, N. and {Hobbs}, D. and {Holland}, G. and {Huckle}, H.~E. and {Jardine}, K. and {Jasniewicz}, G. and {Jean-Antoine Piccolo}, A. and {Jim{\'e}nez-Arranz}, {\'O}. and {Jorissen}, A. and {Juaristi Campillo}, J. and {Julbe}, F. and {Karbevska}, L. and {Kervella}, P. and {Khanna}, S. and {Kontizas}, M. and {Kordopatis}, G. and {Korn}, A.~J. and {K{\'o}sp{\'a}l}, {\'A}. and {Kostrzewa-Rutkowska}, Z. and {Kruszy{\'n}ska}, K. and {Kun}, M. and {Laizeau}, P. and {Lambert}, S. and {Lanza}, A.~F. and {Lasne}, Y. and {Le Campion}, J. -F. and {Lebreton}, Y. and {Lebzelter}, T. and {Leccia}, S. and {Leclerc}, N. and {Lecoeur-Taibi}, I. and {Liao}, S. and {Licata}, E.~L. and {Lindstr{\o}m}, H.~E.~P. and {Lister}, T.~A. and {Livanou}, E. and {Lobel}, A. and {Lorca}, A. and {Loup}, C. and {Madrero Pardo}, P. and {Magdaleno Romeo}, A. and {Managau}, S. and {Mann}, R.~G. and {Manteiga}, M. and {Marchant}, J.~M. and {Marconi}, M. and {Marcos}, J. and {Marcos Santos}, M.~M.~S. and {Mar{\'\i}n Pina}, D. and {Marinoni}, S. and {Marocco}, F. and {Marshall}, D.~J. and {Martin Polo}, L. and {Mart{\'\i}n-Fleitas}, J.~M. and {Marton}, G. and {Mary}, N. and {Masip}, A. and {Massari}, D. and {Mastrobuono-Battisti}, A. and {Mazeh}, T. and {McMillan}, P.~J. and {Messina}, S. and {Michalik}, D. and {Millar}, N.~R. and {Mints}, A. and {Molina}, D. and {Molinaro}, R. and {Moln{\'a}r}, L. and {Monari}, G. and {Mongui{\'o}}, M. and {Montegriffo}, P. and {Montero}, A. and {Mor}, R. and {Mora}, A. and {Morbidelli}, R. and {Morel}, T. and {Morris}, D. and {Muraveva}, T. and {Murphy}, C.~P. and {Musella}, I. and {Nagy}, Z. and {Noval}, L. and {Oca{\~n}a}, F. and {Ogden}, A. and {Ordenovic}, C. and {Osinde}, J.~O. and {Pagani}, C. and {Pagano}, I. and {Palaversa}, L. and {Palicio}, P.~A. and {Pallas-Quintela}, L. and {Panahi}, A. and {Payne-Wardenaar}, S. and {Pe{\~n}alosa Esteller}, X. and {Penttil{\"a}}, A. and {Pichon}, B. and {Piersimoni}, A.~M. and {Pineau}, F. -X. and {Plachy}, E. and {Plum}, G. and {Poggio}, E. and {Pr{\v{s}}a}, A. and {Pulone}, L. and {Racero}, E. and {Ragaini}, S. and {Rainer}, M. and {Raiteri}, C.~M. and {Rambaux}, N. and {Ramos}, P. and {Ramos-Lerate}, M. and {Re Fiorentin}, P. and {Regibo}, S. and {Richards}, P.~J. and {Rios Diaz}, C. and {Ripepi}, V. and {Riva}, A. and {Rix}, H. -W. and {Rixon}, G. and {Robichon}, N. and {Robin}, A.~C. and {Robin}, C. and {Roelens}, M. and {Rogues}, H.~R.~O. and {Rohrbasser}, L. and {Romero-G{\'o}mez}, M. and {Rowell}, N. and {Royer}, F. and {Ruz Mieres}, D. and {Rybicki}, K.~A. and {Sadowski}, G. and {S{\'a}ez N{\'u}{\~n}ez}, A. and {Sagrist{\`a} Sell{\'e}s}, A. and {Sahlmann}, J. and {Salguero}, E. and {Samaras}, N. and {Sanchez Gimenez}, V. and {Sanna}, N. and {Santove{\~n}a}, R. and {Sarasso}, M. and {Schultheis}, M. and {Sciacca}, E. and {Segol}, M. and {Segovia}, J.~C. and {S{\'e}gransan}, D. and {Semeux}, D. and {Shahaf}, S. and {Siddiqui}, H.~I. and {Siebert}, A. and {Siltala}, L. and {Silvelo}, A. and {Slezak}, E. and {Slezak}, I. and {Smart}, R.~L. and {Snaith}, O.~N. and {Solano}, E. and {Solitro}, F. and {Souami}, D. and {Souchay}, J. and {Spagna}, A. and {Spina}, L. and {Spoto}, F. and {Steele}, I.~A. and {Steidelm{\"u}ller}, H. and {Stephenson}, C.~A. and {S{\"u}veges}, M. and {Surdej}, J. and {Szabados}, L. and {Szegedi-Elek}, E. and {Taris}, F. and {Taylor}, M.~B. and {Teixeira}, R. and {Tolomei}, L. and {Tonello}, N. and {Torra}, F. and {Torra}, J. and {Torralba Elipe}, G. and {Trabucchi}, M. and {Tsounis}, A.~T. and {Turon}, C. and {Ulla}, A. and {Unger}, N. and {Vaillant}, M.~V. and {van Dillen}, E. and {van Reeven}, W. and {Vanel}, O. and {Vecchiato}, A. and {Viala}, Y. and {Vicente}, D. and {Voutsinas}, S. and {Weiler}, M. and {Wevers}, T. and {Wyrzykowski}, {\L}. and {Yoldas}, A. and {Yvard}, P. and {Zhao}, H. and {Zorec}, J. and {Zucker}, S. and {Zwitter}, T.},
        title = "{Gaia Data Release 3. Summary of the content and survey properties}",
      journal = {\aap},
         year = 2023,
        month = jun,
       volume = {674},
          eid = {A1},
        pages = {A1},
          doi = {10.1051/0004-6361/202243940},
archivePrefix = {arXiv},
       eprint = {2208.00211},
 primaryClass = {astro-ph.GA},
       adsurl = {https://ui.adsabs.harvard.edu/abs/2023A&A...674A...1G}
}

@ARTICLE{AION1...2025arXiv251017960P,
       author = {{Parker}, Liam and {Lanusse}, Francois and {Shen}, Jeff and {Liu}, Ollie and {Hehir}, Tom and {Sarra}, Leopoldo and {Meyer}, Lucas and {Bowles}, Micah and {Wagner-Carena}, Sebastian and {Qu}, Helen and {Golkar}, Siavash and {Bietti}, Alberto and {Bourfoune}, Hatim and {Casserau}, Nathan and {Cornette}, Pierre and {Hirashima}, Keiya and {Krawezik}, Geraud and {Ohana}, Ruben and {Lourie}, Nicholas and {McCabe}, Michael and {Morel}, Rudy and {Mukhopadhyay}, Payel and {Pettee}, Mariel and {Regaldo-Saint Blancard}, Bruno and {Cho}, Kyunghyun and {Cranmer}, Miles and {Ho}, Shirley},
        title = "{AION-1: Omnimodal Foundation Model for Astronomical Sciences}",
      journal = {arXiv e-prints},
         year = 2025,
        month = oct,
          eid = {arXiv:2510.17960},
        pages = {arXiv:2510.17960},
          doi = {10.48550/arXiv.2510.17960},
archivePrefix = {arXiv},
       eprint = {2510.17960},
 primaryClass = {astro-ph.IM},
       adsurl = {https://ui.adsabs.harvard.edu/abs/2025arXiv251017960P}
}

@ARTICLE{Kraft...spindown...1967ApJ...150..551K,
       author = {{Kraft}, Robert P.},
        title = "{Studies of Stellar Rotation. V. The Dependence of Rotation on Age among Solar-Type Stars}",
      journal = {\apj},
         year = 1967,
        month = nov,
       volume = {150},
        pages = {551},
          doi = {10.1086/149359},
       adsurl = {https://ui.adsabs.harvard.edu/abs/1967ApJ...150..551K}
}

@ARTICLE{TimerS1...2026arXiv260304791L,
       author = {{Liu}, Yong and {Su}, Xingjian and {Wang}, Shiyu and {Zhang}, Haoran and {Liu}, Haixuan and {Wang}, Yuxuan and {Ye}, Zhou and {Xiang}, Yang and {Wang}, Jianmin and {Long}, Mingsheng},
        title = "{Timer-S1: A Billion-Scale Time Series Foundation Model with Serial Scaling}",
      journal = {arXiv e-prints},
         year = 2026,
        month = mar,
          eid = {arXiv:2603.04791},
        pages = {arXiv:2603.04791},
          doi = {10.48550/arXiv.2603.04791},
archivePrefix = {arXiv},
       eprint = {2603.04791},
 primaryClass = {cs.AI},
       adsurl = {https://ui.adsabs.harvard.edu/abs/2026arXiv260304791L}
}

@ARTICLE{Chronos2...2025arXiv251015821F,
       author = {{Fatir Ansari}, Abdul and {Shchur}, Oleksandr and {K{\"u}ken}, Jaris and {Auer}, Andreas and {Han}, Boran and {Mercado}, Pedro and {Sundar Rangapuram}, Syama and {Shen}, Huibin and {Stella}, Lorenzo and {Zhang}, Xiyuan and {Goswami}, Mononito and {Kapoor}, Shubham and {Maddix}, Danielle C. and {Guerron}, Pablo and {Hu}, Tony and {Yin}, Junming and {Erickson}, Nick and {Mutalik Desai}, Prateek and {Wang}, Hao and {Rangwala}, Huzefa and {Karypis}, George and {Wang}, Yuyang and {Bohlke-Schneider}, Michael},
        title = "{Chronos-2: From Univariate to Universal Forecasting}",
      journal = {arXiv e-prints},
         year = 2025,
        month = oct,
          eid = {arXiv:2510.15821},
        pages = {arXiv:2510.15821},
          doi = {10.48550/arXiv.2510.15821},
archivePrefix = {arXiv},
       eprint = {2510.15821},
 primaryClass = {stat.ML},
       adsurl = {https://ui.adsabs.harvard.edu/abs/2025arXiv251015821F}
}

@ARTICLE{Moirai2...2025arXiv251111698L,
       author = {{Liu}, Chenghao and {Aksu}, Taha and {Liu}, Juncheng and {Liu}, Xu and {Yan}, Hanshu and {Pham}, Quang and {Savarese}, Silvio and {Sahoo}, Doyen and {Xiong}, Caiming and {Li}, Junnan},
        title = "{Moirai 2.0: When Less Is More for Time Series Forecasting}",
      journal = {arXiv e-prints},
         year = 2025,
        month = nov,
          eid = {arXiv:2511.11698},
        pages = {arXiv:2511.11698},
          doi = {10.48550/arXiv.2511.11698},
archivePrefix = {arXiv},
       eprint = {2511.11698},
 primaryClass = {stat.ML},
       adsurl = {https://ui.adsabs.harvard.edu/abs/2025arXiv251111698L}
}

@ARTICLE{MoiraiMoE...2024arXiv241010469L,
       author = {{Liu}, Xu and {Liu}, Juncheng and {Woo}, Gerald and {Aksu}, Taha and {Liang}, Yuxuan and {Zimmermann}, Roger and {Liu}, Chenghao and {Savarese}, Silvio and {Xiong}, Caiming and {Sahoo}, Doyen},
        title = "{Moirai-MoE: Empowering Time Series Foundation Models with Sparse Mixture of Experts}",
      journal = {arXiv e-prints},
         year = 2024,
        month = oct,
          eid = {arXiv:2410.10469},
        pages = {arXiv:2410.10469},
          doi = {10.48550/arXiv.2410.10469},
archivePrefix = {arXiv},
       eprint = {2410.10469},
 primaryClass = {cs.LG},
       adsurl = {https://ui.adsabs.harvard.edu/abs/2024arXiv241010469L}
}

@ARTICLE{Kepler...2010Sci...327..977B,
       author = {{Borucki}, William J. and {Koch}, David and {Basri}, Gibor and {Batalha}, Natalie and {Brown}, Timothy and {Caldwell}, Douglas and {Caldwell}, John and {Christensen-Dalsgaard}, J{\o}rgen and {Cochran}, William D. and {DeVore}, Edna and {Dunham}, Edward W. and {Dupree}, Andrea K. and {Gautier}, Thomas N. and {Geary}, John C. and {Gilliland}, Ronald and {Gould}, Alan and {Howell}, Steve B. and {Jenkins}, Jon M. and {Kondo}, Yoji and {Latham}, David W. and {Marcy}, Geoffrey W. and {Meibom}, S{\o}ren and {Kjeldsen}, Hans and {Lissauer}, Jack J. and {Monet}, David G. and {Morrison}, David and {Sasselov}, Dimitar and {Tarter}, Jill and {Boss}, Alan and {Brownlee}, Don and {Owen}, Toby and {Buzasi}, Derek and {Charbonneau}, David and {Doyle}, Laurance and {Fortney}, Jonathan and {Ford}, Eric B. and {Holman}, Matthew J. and {Seager}, Sara and {Steffen}, Jason H. and {Welsh}, William F. and {Rowe}, Jason and {Anderson}, Howard and {Buchhave}, Lars and {Ciardi}, David and {Walkowicz}, Lucianne and {Sherry}, William and {Horch}, Elliott and {Isaacson}, Howard and {Everett}, Mark E. and {Fischer}, Debra and {Torres}, Guillermo and {Johnson}, John Asher and {Endl}, Michael and {MacQueen}, Phillip and {Bryson}, Stephen T. and {Dotson}, Jessie and {Haas}, Michael and {Kolodziejczak}, Jeffrey and {Van Cleve}, Jeffrey and {Chandrasekaran}, Hema and {Twicken}, Joseph D. and {Quintana}, Elisa V. and {Clarke}, Bruce D. and {Allen}, Christopher and {Li}, Jie and {Wu}, Haley and {Tenenbaum}, Peter and {Verner}, Ekaterina and {Bruhweiler}, Frederick and {Barnes}, Jason and {Prsa}, Andrej},
        title = "{Kepler Planet-Detection Mission: Introduction and First Results}",
      journal = {Science},
         year = 2010,
        month = feb,
       volume = {327},
       number = {5968},
        pages = {977},
          doi = {10.1126/science.1185402},
       adsurl = {https://ui.adsabs.harvard.edu/abs/2010Sci...327..977B}
}

@ARTICLE{Esquivel...HMMs...2025ApJ...979..141E,
       author = {{Esquivel}, J. Arturo and {Shen}, Yunyi and {Leos-Barajas}, Vianey and {Eadie}, Gwendolyn and {Speagle}, Joshua S. and {Craiu}, Radu V. and {Medina}, Amber and {Davenport}, James R.~A.},
        title = "{Detecting Stellar Flares in Photometric Data Using Hidden Markov Models}",
      journal = {\apj},
         year = 2025,
        month = feb,
       volume = {979},
       number = {2},
          eid = {141},
        pages = {141},
          doi = {10.3847/1538-4357/ad95f6},
archivePrefix = {arXiv},
       eprint = {2404.13145},
 primaryClass = {astro-ph.SR},
       adsurl = {https://ui.adsabs.harvard.edu/abs/2025ApJ...979..141E}
}

@ARTICLE{TabPFN...2022arXiv220701848H,
       author = {{Hollmann}, Noah and {M{\"u}ller}, Samuel and {Eggensperger}, Katharina and {Hutter}, Frank},
        title = "{TabPFN: A Transformer That Solves Small Tabular Classification Problems in a Second}",
      journal = {arXiv e-prints},
         year = 2022,
        month = jul,
          eid = {arXiv:2207.01848},
        pages = {arXiv:2207.01848},
          doi = {10.48550/arXiv.2207.01848},
archivePrefix = {arXiv},
       eprint = {2207.01848},
 primaryClass = {cs.LG},
       adsurl = {https://ui.adsabs.harvard.edu/abs/2022arXiv220701848H}
}

@ARTICLE{Messina...PhotAmp...2021A&A...645A.144M,
       author = {{Messina}, S.},
        title = "{Constraining the age of young stellar clusters via the amplitude of photometric variability}",
      journal = {\aap},
         year = 2021,
        month = jan,
       volume = {645},
          eid = {A144},
        pages = {A144},
          doi = {10.1051/0004-6361/202038739},
archivePrefix = {arXiv},
       eprint = {2011.13640},
 primaryClass = {astro-ph.SR},
       adsurl = {https://ui.adsabs.harvard.edu/abs/2021A&A...645A.144M}
}

@ARTICLE{Ding...StarCLR...2026ApJ..1003..141D,
       author = {{Ding}, Junyao and {Chen}, Xiaodian and {Gao}, Xinyi and {Tang}, Xiaoyu and {Wang}, Shu and {Huang}, Yang and {Qi}, Xinyu and {Xue}, Guirong and {Luo}, Ali and {Liu}, Jifeng},
        title = "{StarCLR: Contrastive Learning Representation for Astronomical Light Curves}",
      journal = {\apj},
         year = 2026,
        month = jun,
       volume = {1003},
       number = {2},
          eid = {141},
        pages = {141},
          doi = {10.3847/1538-4357/ae64ef},
archivePrefix = {arXiv},
       eprint = {2604.24516},
 primaryClass = {astro-ph.SR},
       adsurl = {https://ui.adsabs.harvard.edu/abs/2026ApJ..1003..141D}
}

@ARTICLE{Poznanski...TESSLCs...2026arXiv260505324P,
       author = {{Poznanski}, Dovi},
        title = "{A useful representation of TESS light curves}",
      journal = {arXiv e-prints},
         year = 2026,
        month = may,
          eid = {arXiv:2605.05324},
        pages = {arXiv:2605.05324},
          doi = {10.48550/arXiv.2605.05324},
archivePrefix = {arXiv},
       eprint = {2605.05324},
 primaryClass = {astro-ph.IM},
       adsurl = {https://ui.adsabs.harvard.edu/abs/2026arXiv260505324P}
}

@ARTICLE{Zimmerman...HMM...2024MNRAS.534.2142Z,
       author = {{Zimmerman}, Robert and {van Dyk}, David A. and {Kashyap}, Vinay L. and {Siemiginowska}, Aneta},
        title = "{Separating states in astronomical sources using hidden Markov models: with a case study of flaring and quiescence on EV Lac}",
      journal = {\mnras},
         year = 2024,
        month = nov,
       volume = {534},
       number = {3},
        pages = {2142-2167},
          doi = {10.1093/mnras/stae2082},
archivePrefix = {arXiv},
       eprint = {2405.06540},
 primaryClass = {astro-ph.SR},
       adsurl = {https://ui.adsabs.harvard.edu/abs/2024MNRAS.534.2142Z}
}

@ARTICLE{Wainer...ClusterLCs...2023AJ....166..106W,
       author = {{Wainer}, Tobin M. and {Zasowski}, Gail and {Pepper}, Joshua and {Wagg}, Tom and {Hedges}, Christina L. and {Poovelil}, Vijith Jacob and {Fetherolf}, Tara and {Davenport}, James R.~A. and {Christodoulou}, P. Marios and {Dinsmore}, Jack T. and {Patel}, Avi and {Goold}, Kameron and {Gibson}, Benjamin J.},
        title = "{Catalog of Integrated-light Star Cluster Light Curves in TESS}",
      journal = {\aj},
         year = 2023,
        month = sep,
       volume = {166},
       number = {3},
          eid = {106},
        pages = {106},
          doi = {10.3847/1538-3881/ace960},
archivePrefix = {arXiv},
       eprint = {2307.09510},
 primaryClass = {astro-ph.GA},
       adsurl = {https://ui.adsabs.harvard.edu/abs/2023AJ....166..106W}
}

@ARTICLE{Herrerra...flaring...2026arXiv260622601H,
       author = {{Herrera}, Rodrigo and {Leos-Barajas}, Vianey and {Eadie}, Gwendolyn and {Semenova}, Elizaveta and {Davenport}, James},
        title = "{Scalable Bayesian Additive Models for Stellar Flare Detection via Amortized Gaussian Process Inference and Hidden Markov Models}",
      journal = {arXiv e-prints},
         year = 2026,
        month = jun,
          eid = {arXiv:2606.22601},
        pages = {arXiv:2606.22601},
          doi = {10.48550/arXiv.2606.22601},
archivePrefix = {arXiv},
       eprint = {2606.22601},
 primaryClass = {stat.ML},
       adsurl = {https://ui.adsabs.harvard.edu/abs/2026arXiv260622601H}
}

@ARTICLE{Bouma...gyrointerp...2023ApJ...947L...3B,
       author = {{Bouma}, Luke G. and {Palumbo}, Elsa K. and {Hillenbrand}, Lynne A.},
        title = "{The Empirical Limits of Gyrochronology}",
      journal = {\apjl},
         year = 2023,
        month = apr,
       volume = {947},
       number = {1},
          eid = {L3},
        pages = {L3},
          doi = {10.3847/2041-8213/acc589},
archivePrefix = {arXiv},
       eprint = {2303.08830},
 primaryClass = {astro-ph.SR},
       adsurl = {https://ui.adsabs.harvard.edu/abs/2023ApJ...947L...3B}
}

@ARTICLE{BoyleBouma...gyro.Aper...2022arXiv221109822B,
       author = {{Boyle}, Andrew W. and {Bouma}, Luke G.},
        title = "{Stellar Rotation and Structure of the $\alpha$ Persei Complex: When Does Gyrochronology Start to Work?}",
      journal = {arXiv e-prints},
         year = 2022,
        month = nov,
          eid = {arXiv:2211.09822},
        pages = {arXiv:2211.09822},
          doi = {10.48550/arXiv.2211.09822},
archivePrefix = {arXiv},
       eprint = {2211.09822},
 primaryClass = {astro-ph.SR},
       adsurl = {https://ui.adsabs.harvard.edu/abs/2022arXiv221109822B}
}

@ARTICLE{DonosoOliva...Astromer2...2026A&A...707A.170D,
       author = {{Donoso-Oliva}, Cristobal and {Becker}, Ignacio and {Protopapas}, Pavlos and {Cabrera-Vives}, Guillermo and {C{\'a}diz-Leyton}, Martina and {Moreno-Cartagena}, Daniel},
        title = "{Generalizing across astronomical surveys: Few-shot light curve classification with Astromer 2}",
      journal = {\aap},
         year = 2026,
        month = mar,
       volume = {707},
          eid = {A170},
        pages = {A170},
          doi = {10.1051/0004-6361/202554026},
archivePrefix = {arXiv},
       eprint = {2502.02717},
 primaryClass = {astro-ph.IM},
       adsurl = {https://ui.adsabs.harvard.edu/abs/2026A&A...707A.170D}
}

@ARTICLE{Ye...TSFM...2024arXiv240502358Y,
       author = {{Ye}, Jiexia and {Yu}, Yongzi and {Zhang}, Weiqi and {Wang}, Le and {Li}, Jia and {Tsung}, Fugee},
        title = "{Empowering Time Series Analysis with Foundation Models: A Comprehensive Survey}",
      journal = {arXiv e-prints},
         year = 2024,
        month = may,
          eid = {arXiv:2405.02358},
        pages = {arXiv:2405.02358},
          doi = {10.48550/arXiv.2405.02358},
archivePrefix = {arXiv},
       eprint = {2405.02358},
 primaryClass = {stat.ML},
       adsurl = {https://ui.adsabs.harvard.edu/abs/2024arXiv240502358Y}
}

@dataset{NEA...PlanetSystemsTable...2020ipac.data..N13N,
       author = {{NASA Exoplanet Science Institute}},
        title = "{Planetary Systems Composite Table}",
 howpublished = {NASA IPAC DataSet, NEA13},
         year = 2020,
        month = jan,
          doi = {10.26133/NEA13},
       adsurl = {https://ui.adsabs.harvard.edu/abs/2020ipac.data..N13N}
}

@article{2007CSE.....9c..21P,
  author        = {{Perez}, Fernando and {Granger}, Brian E.},
  title         = "{IPython: A System for Interactive Scientific Computing}",
  journal       = {Computing in Science and Engineering},
  year          = "2007",
  month         = "Jan",
  volume        = {9},
  number        = {3},
  pages         = {21--29},
  doi           = {10.1109/MCSE.2007.53},
  adsurl        = {https://ui.adsabs.harvard.edu/abs/2007CSE.....9c..21P}
}

@conference{kluyver2016jupyter,
  title         = {Jupyter Notebooks -- a publishing format for reproducible computational workflows},
  author        = {Thomas Kluyver and Benjamin Ragan-Kelley and Fernando P{\'e}rez and Brian Granger and Matthias Bussonnier and Jonathan Frederic and Kyle Kelley and Jessica Hamrick and Jason Grout and Sylvain Corlay and Paul Ivanov and Dami{\'a}n Avila and Safia Abdalla and Carol Willing},
  booktitle     = {Positioning and Power in Academic Publishing: Players, Agents and Agendas},
  editor        = {F. Loizides and B. Schmidt},
  organization  = {IOS Press},
  pages         = {87--90},
  year          = {2016}
}

@article{Hunter:2007,
  author        = {Hunter, J. D.},
  title         = {Matplotlib: A 2D graphics environment},
  journal       = {Computing in Science \& Engineering},
  volume        = {9},
  number        = {3},
  pages         = {90--95},
  publisher     = {IEEE COMPUTER SOC},
  doi           = {10.1109/MCSE.2007.55},
  year          = 2007
}

@article{numpy,
  title         = {Array programming with {NumPy}},
  author        = {Charles R. Harris and K. Jarrod Millman and St{\'{e}}fan J. van der Walt and Ralf Gommers and Pauli Virtanen and David Cournapeau and Eric Wieser and Julian Taylor and Sebastian Berg and Nathaniel J. Smith and Robert Kern and Matti Picus and Stephan Hoyer and Marten H. van Kerkwijk and Matthew Brett and Allan Haldane and Jaime Fern{\'{a}}ndez del R{\'{i}}o and Mark Wiebe and Pearu Peterson and Pierre G{\'{e}}rard-Marchant and Kevin Sheppard and Tyler Reddy and Warren Weckesser and Hameer Abbasi and Christoph Gohlke and Travis E. Oliphant},
  year          = {2020},
  month         = sep,
  journal       = {Nature},
  volume        = {585},
  number        = {7825},
  pages         = {357--362},
  doi           = {10.1038/s41586-020-2649-2},
  publisher     = {Springer Science and Business Media {LLC}},
  url           = {https://doi.org/10.1038/s41586-020-2649-2}
}

@software{pandas_17229934,
  author       = {The pandas development team},
  title        = {pandas-dev/pandas: Pandas},
  month        = sep,
  year         = 2025,
  publisher    = {Zenodo},
  version      = {v2.3.3},
  doi          = {10.5281/zenodo.17229934},
  url          = {https://doi.org/10.5281/zenodo.17229934},
  swhid        = {swh:1:dir:5bc73e541e362cf8e7068ebb434b9d7bc1194c14
                   ;origin=https://doi.org/10.5281/zenodo.3509134;vis
                   it=swh:1:snp:8c6e563b81f60a3d62fec309b38f470a0eacf
                   596;anchor=swh:1:rel:b0f463d37766850a89633bd05643d
                   45342b9241d;path=pandas-dev-pandas-1efe649
                  },
}

@inproceedings{mckinney-proc-scipy-2010,
  author        = {{W}es {M}c{K}inney},
  title         = {{D}ata {S}tructures for {S}tatistical {C}omputing in {P}ython},
  booktitle     = {{P}roceedings of the 9th {P}ython in {S}cience {C}onference},
  pages         = {56--61},
  year          = {2010},
  editor        = {{S}t\'efan van der {W}alt and {J}arrod {M}illman},
  doi           = {10.25080/Majora-92bf1922-00a}
}

@book{python,
  author        = {Van Rossum, Guido and Drake, Fred L.},
  title         = {Python 3 Reference Manual},
  year          = {2009},
  isbn          = {1441412697},
  publisher     = {CreateSpace},
  address       = {Scotts Valley, CA}
}

@software{scipy_17467817,
  author       = {Ralf Gommers and
                  Pauli Virtanen and
                  Matt Haberland and
                  Evgeni Burovski and
                  Tyler Reddy and
                  Warren Weckesser and
                  Travis E. Oliphant and
                  Andrew Nelson and
                  David Cournapeau and
                  alexbrc and
                  Pamphile Roy and
                  Ilhan Polat and
                  Pearu Peterson and
                  Josh Wilson and
                  Lucas Colley and
                  endolith and
                  Nikolay Mayorov and
                  Stefan van der Walt and
                  Jake Bowhay and
                  Matthew Brett and
                  Denis Laxalde and
                  Albert Steppi and
                  Eric Larson and
                  Atsushi Sakai and
                  Jarrod Millman and
                  Lars and
                  peterbell10 and
                  CJ Carey and
                  Paul van Mulbregt and
                  eric-jones},
  title        = {scipy/scipy: SciPy 1.16.3},
  month        = oct,
  year         = 2025,
  publisher    = {Zenodo},
  version      = {v1.16.3},
  doi          = {10.5281/zenodo.17467817},
  url          = {https://doi.org/10.5281/zenodo.17467817},
  swhid        = {swh:1:dir:69c06d8cf0d2614216b02c9525123a28184c300a
                   ;origin=https://doi.org/10.5281/zenodo.595738;visi
                   t=swh:1:snp:91bcaa77109605ecbc37265041dc57ca5c3695
                   11;anchor=swh:1:rel:4946397cdc012ff0f73bb27005b937
                   5ebef4ad52;path=scipy-scipy-56e567e
                  },
}

@article{2020SciPy-NMeth,
  author        = {Virtanen, Pauli and Gommers, Ralf and Oliphant, Travis E. and Haberland, Matt and Reddy, Tyler and Cournapeau, David and Burovski, Evgeni and Peterson, Pearu and Weckesser, Warren and Bright, Jonathan and {van der Walt}, St{\'e}fan J. and Brett, Matthew and Wilson, Joshua and Millman, K. Jarrod and Mayorov, Nikolay and Nelson, Andrew R. J. and Jones, Eric and Kern, Robert and Larson, Eric and Carey, C J and Polat, {\.I}lhan and Feng, Yu and Moore, Eric W. and {VanderPlas}, Jake and Laxalde, Denis and Perktold, Josef and Cimrman, Robert and Henriksen, Ian and Quintero, E. A. and Harris, Charles R. and Archibald, Anne M. and Ribeiro, Ant{\^o}nio H. and Pedregosa, Fabian and {van Mulbregt}, Paul and {SciPy 1.0 Contributors}},
  title         = {{{SciPy} 1.0: Fundamental Algorithms for Scientific Computing in Python}},
  journal       = {Nature Methods},
  year          = {2020},
  volume        = {17},
  pages         = {261--272},
  adsurl        = {https://rdcu.be/b08Wh},
  doi           = {10.1038/s41592-019-0686-2}
}

@software{h5py_7560547,
  author       = {Andrew Collette and
                  Thomas Kluyver and
                  Thomas A Caswell and
                  James Tocknell and
                  Jerome Kieffer and
                  Aleksandar Jelenak and
                  Anthony Scopatz and
                  Darren Dale and
                  Chen and
                  Thomas VINCENT and
                  Matt Einhorn and
                  payno and
                  juliagarriga and
                  Pierlauro Sciarelli and
                  Valentin Valls and
                  Satrajit Ghosh and
                  Ulrik Kofoed Pedersen and
                  Mark Kittisopikul and
                  jakirkham and
                  Martin Raspaud and
                  Cyril Danilevski and
                  Hameer Abbasi and
                  John Readey and
                  Kai Mühlbauer and
                  Andrey Paramonov and
                  Lawrence Chan and
                  Robin De Schepper and
                  V. Armando Solé and
                  jialin and
                  Daniel Hay Guest},
  title        = {h5py/h5py: 3.8.0},
  month        = jan,
  year         = 2023,
  publisher    = {Zenodo},
  version      = {3.8.0},
  doi          = {10.5281/zenodo.7560547},
  url          = {https://doi.org/10.5281/zenodo.7560547},
}

@book{collette_python_hdf5_2014,
  year          = {2013},
  publisher     = {O'Reilly},
  title         = {Python and HDF5},
  author        = {Andrew Collette}
}

@article{mcinnes2018umap-software,
  title         = {UMAP: Uniform Manifold Approximation and Projection},
  author        = {McInnes, Leland and Healy, John and Saul, Nathaniel and Grossberger, Lukas},
  journal       = {The Journal of Open Source Software},
  volume        = {3},
  number        = {29},
  pages         = {861},
  year          = {2018}
}

@article{2018arXivUMAP,
  author        = {{McInnes}, L. and {Healy}, J. and {Melville}, J.},
  title         = "{UMAP: Uniform Manifold Approximation and Projection for Dimension Reduction}",
  journal       = {ArXiv e-prints},
  archiveprefix = "arXiv",
  eprint        = {1802.03426},
  primaryclass  = "stat.ML",
  year          = 2018,
  month         = feb
}

@software{Zuko_10571785,
  author       = {François Rozet and
                  Felix Divo and
                  Simon Schnake},
  title        = {probabilists/zuko: Zuko 1.1.0},
  month        = jan,
  year         = 2024,
  publisher    = {Zenodo},
  version      = {1.1.0},
  doi          = {10.5281/zenodo.10571785},
  url          = {https://doi.org/10.5281/zenodo.10571785},
}

@software{rozet2022zuko,
  title         = {{Zuko}: Normalizing flows in PyTorch},
  author        = {Rozet, Fran\c{c}ois and others},
  year          = {2022},
  doi           = {10.5281/zenodo.7625672},
  license       = {MIT},
  url           = {https://pypi.org/project/zuko}
}

@article{software-citation-station-paper,
  author        = {{Wagg}, Tom and {Broekgaarden}, Floor S.},
  title         = "{Streamlining and standardizing software citations with The Software Citation Station}",
  journal       = {arXiv e-prints},
  year          = 2024,
  month         = jun,
  eid           = {arXiv:2406.04405},
  pages         = {arXiv:2406.04405},
  archiveprefix = {arXiv},
  eprint        = {2406.04405},
  primaryclass  = {astro-ph.IM},
  adsurl        = {https://ui.adsabs.harvard.edu/abs/2024arXiv240604405W}
}

@ARTICLE{Chaplin...Asteroseismology...K...2026A&A...710A.361C,
       author = {{Chaplin}, William J. and {Campante}, Tiago L. and {Lund}, Mikkel N. and {Nielsen}, Martin B. and {Davies}, Guy R. and {Hatt}, Emily and {Howe}, Rachel and {Stokholm}, Amalie},
        title = "{Ensemble asteroseismology: An ensemble approach to detecting signatures of solar-like oscillations in K dwarfs}",
      journal = {\aap},
         year = 2026,
        month = jun,
       volume = {710},
          eid = {A361},
        pages = {A361},
          doi = {10.1051/0004-6361/202659600},
archivePrefix = {arXiv},
       eprint = {2605.23515},
 primaryClass = {astro-ph.SR},
       adsurl = {https://ui.adsabs.harvard.edu/abs/2026A&A...710A.361C}
}

@ARTICLE{UNet...2015arXiv150504597R,
       author = {{Ronneberger}, Olaf and {Fischer}, Philipp and {Brox}, Thomas},
        title = "{U-Net: Convolutional Networks for Biomedical Image Segmentation}",
      journal = {arXiv e-prints},
         year = 2015,
        month = may,
          eid = {arXiv:1505.04597},
        pages = {arXiv:1505.04597},
          doi = {10.48550/arXiv.1505.04597},
archivePrefix = {arXiv},
       eprint = {1505.04597},
 primaryClass = {cs.CV},
       adsurl = {https://ui.adsabs.harvard.edu/abs/2015arXiv150504597R}
}

@article{UMAP...McInnes2018, doi = {10.21105/joss.00861}, url = {https://doi.org/10.21105/joss.00861}, year = {2018}, publisher = {The Open Journal}, volume = {3}, number = {29}, pages = {861}, author = {McInnes, Leland and Healy, John and Saul, Nathaniel and Großberger, Lukas}, title = {UMAP: Uniform Manifold Approximation and Projection}, journal = {Journal of Open Source Software} }

@ARTICLE{Mamba...2023arXiv231200752G,
       author = {{Gu}, Albert and {Dao}, Tri},
        title = "{Mamba: Linear-Time Sequence Modeling with Selective State Spaces}",
      journal = {arXiv e-prints},
         year = 2023,
        month = dec,
          eid = {arXiv:2312.00752},
        pages = {arXiv:2312.00752},
          doi = {10.48550/arXiv.2312.00752},
archivePrefix = {arXiv},
       eprint = {2312.00752},
 primaryClass = {cs.LG},
       adsurl = {https://ui.adsabs.harvard.edu/abs/2023arXiv231200752G}
}

@ARTICLE{MLP...1986Natur.323..533R,
       author = {{Rumelhart}, David E. and {Hinton}, Geoffrey E. and {Williams}, Ronald J.},
        title = "{Learning representations by back-propagating errors}",
      journal = {\nat},
         year = 1986,
        month = oct,
       volume = {323},
       number = {6088},
        pages = {533-536},
          doi = {10.1038/323533a0},
       adsurl = {https://ui.adsabs.harvard.edu/abs/1986Natur.323..533R}
}

@techreport{tess_sdpdd,
  author      = {Tenenbaum, Peter and Jenkins, Jon M.},
  title       = {{TESS} Science Data Products Description Document},
  institution = {NASA Ames Research Center},
  number      = {EXP-TESS-ARC-ICD-TM-0014, Rev.~D},
  year        = {2018},
  note        = {\url{https://archive.stsci.edu/missions/tess/doc/EXP-TESS-ARC-ICD-TM-0014.pdf}}
}

@MISC{Lightkurve...2018ascl.soft12013L,
   author = {{Lightkurve Collaboration} and {Cardoso}, J.~V.~d.~M. and
             {Hedges}, C. and {Gully-Santiago}, M. and {Saunders}, N. and
             {Cody}, A.~M. and {Barclay}, T. and {Hall}, O. and
             {Sagear}, S. and {Turtelboom}, E. and {Zhang}, J. and
             {Tzanidakis}, A. and {Mighell}, K. and {Coughlin}, J. and
             {Bell}, K. and {Berta-Thompson}, Z. and {Williams}, P. and
             {Dotson}, J. and {Barentsen}, G.},
    title = "{Lightkurve: Kepler and TESS time series analysis in Python}",
howpublished = {Astrophysics Source Code Library},
     year = 2018,
    month = dec,
archivePrefix = "ascl",
   eprint = {1812.013},
   adsurl = {http://adsabs.harvard.edu/abs/2018ascl.soft12013L},
}

@ARTICLE{Stumpe...PDCSAP...2014PASP..126..100S,
       author = {{Stumpe}, Martin C. and {Smith}, Jeffrey C. and {Catanzarite}, Joseph H. and {Van Cleve}, Jeffrey E. and {Jenkins}, Jon M. and {Twicken}, Joseph D. and {Girouard}, Forrest R.},
        title = "{Multiscale Systematic Error Correction via Wavelet-Based Bandsplitting in Kepler Data}",
      journal = {\pasp},
         year = 2014,
        month = jan,
       volume = {126},
       number = {935},
        pages = {100},
          doi = {10.1086/674989},
       adsurl = {https://ui.adsabs.harvard.edu/abs/2014PASP..126..100S}
}

@ARTICLE{Stumpe...PDCSAP...2012PASP..124..985S,
       author = {{Stumpe}, Martin C. and {Smith}, Jeffrey C. and {Van Cleve}, Jeffrey E. and {Twicken}, Joseph D. and {Barclay}, Thomas S. and {Fanelli}, Michael N. and {Girouard}, Forrest R. and {Jenkins}, Jon M. and {Kolodziejczak}, Jeffery J. and {McCauliff}, Sean D. and {Morris}, Robert L.},
        title = "{Kepler Presearch Data Conditioning I{\textemdash}Architecture and Algorithms for Error Correction in Kepler Light Curves}",
      journal = {\pasp},
         year = 2012,
        month = sep,
       volume = {124},
       number = {919},
        pages = {985},
          doi = {10.1086/667698},
archivePrefix = {arXiv},
       eprint = {1203.1382},
 primaryClass = {astro-ph.IM},
       adsurl = {https://ui.adsabs.harvard.edu/abs/2012PASP..124..985S}
}

@ARTICLE{Smith...PDCSAP...2012PASP..124.1000S,
       author = {{Smith}, Jeffrey C. and {Stumpe}, Martin C. and {Van Cleve}, Jeffrey E. and {Jenkins}, Jon M. and {Barclay}, Thomas S. and {Fanelli}, Michael N. and {Girouard}, Forrest R. and {Kolodziejczak}, Jeffery J. and {McCauliff}, Sean D. and {Morris}, Robert L. and {Twicken}, Joseph D.},
        title = "{Kepler Presearch Data Conditioning II - A Bayesian Approach to Systematic Error Correction}",
      journal = {\pasp},
         year = 2012,
        month = sep,
       volume = {124},
       number = {919},
        pages = {1000},
          doi = {10.1086/667697},
archivePrefix = {arXiv},
       eprint = {1203.1383},
 primaryClass = {astro-ph.IM},
       adsurl = {https://ui.adsabs.harvard.edu/abs/2012PASP..124.1000S}
}

@INPROCEEDINGS{Jenkins...SPOC...2016SPIE.9913E..3EJ,
       author = {{Jenkins}, Jon M. and {Twicken}, Joseph D. and {McCauliff}, Sean and {Campbell}, Jennifer and {Sanderfer}, Dwight and {Lung}, David and {Mansouri-Samani}, Masoud and {Girouard}, Forrest and {Tenenbaum}, Peter and {Klaus}, Todd and {Smith}, Jeffrey C. and {Caldwell}, Douglas A. and {Chacon}, A.~D. and {Henze}, Christopher and {Heiges}, Cory and {Latham}, David W. and {Morgan}, Edward and {Swade}, Daryl and {Rinehart}, Stephen and {Vanderspek}, Roland},
        title = "{The TESS science processing operations center}",
    booktitle = {Software and Cyberinfrastructure for Astronomy IV},
         year = 2016,
       editor = {{Chiozzi}, Gianluca and {Guzman}, Juan C.},
       series = {Society of Photo-Optical Instrumentation Engineers (SPIE) Conference Series},
       volume = {9913},
        month = aug,
          eid = {99133E},
        pages = {99133E},
          doi = {10.1117/12.2233418},
       adsurl = {https://ui.adsabs.harvard.edu/abs/2016SPIE.9913E..3EJ}
}

@ARTICLE{Angus...GP.Prot...2018MNRAS.474.2094A,
       author = {{Angus}, Ruth and {Morton}, Timothy and {Aigrain}, Suzanne and {Foreman-Mackey}, Daniel and {Rajpaul}, Vinesh},
        title = "{Inferring probabilistic stellar rotation periods using Gaussian processes}",
      journal = {\mnras},
         year = 2018,
        month = feb,
       volume = {474},
       number = {2},
        pages = {2094-2108},
          doi = {10.1093/mnras/stx2109},
archivePrefix = {arXiv},
       eprint = {1706.05459},
 primaryClass = {astro-ph.SR},
       adsurl = {https://ui.adsabs.harvard.edu/abs/2018MNRAS.474.2094A}
}

@ARTICLE{Garcia...Wavelet...2014A&A...572A..34G,
       author = {{Garc{\'\i}a}, R.~A. and {Ceillier}, T. and {Salabert}, D. and {Mathur}, S. and {van Saders}, J.~L. and {Pinsonneault}, M. and {Ballot}, J. and {Beck}, P.~G. and {Bloemen}, S. and {Campante}, T.~L. and {Davies}, G.~R. and {do Nascimento}, Jr., J.-D. and {Mathis}, S. and {Metcalfe}, T.~S. and {Nielsen}, M.~B. and {Su{\'a}rez}, J.~C. and {Chaplin}, W.~J. and {Jim{\'e}nez}, A. and {Karoff}, C.},
        title = "{Rotation and magnetism of Kepler pulsating solar-like stars. Towards asteroseismically calibrated age-rotation relations}",
      journal = {\aap},
         year = 2014,
        month = dec,
       volume = {572},
          eid = {A34},
        pages = {A34},
          doi = {10.1051/0004-6361/201423888},
archivePrefix = {arXiv},
       eprint = {1403.7155},
 primaryClass = {astro-ph.SR},
       adsurl = {https://ui.adsabs.harvard.edu/abs/2014A&A...572A..34G}
}

@ARTICLE{McQuillan...ACF...2013ApJ...775L..11M,
       author = {{McQuillan}, A. and {Mazeh}, T. and {Aigrain}, S.},
        title = "{Stellar Rotation Periods of the Kepler Objects of Interest: A Dearth of Close-in Planets around Fast Rotators}",
      journal = {\apjl},
         year = 2013,
        month = sep,
       volume = {775},
       number = {1},
          eid = {L11},
        pages = {L11},
          doi = {10.1088/2041-8205/775/1/L11},
archivePrefix = {arXiv},
       eprint = {1308.1845},
 primaryClass = {astro-ph.EP},
       adsurl = {https://ui.adsabs.harvard.edu/abs/2013ApJ...775L..11M}
}

@ARTICLE{VanCleve...PhotTTPrecision...2016PASP..128g5002V,
       author = {{Van Cleve}, Jeffrey E. and {Howell}, Steve B. and {Smith}, Jeffrey C. and {Clarke}, Bruce D. and {Thompson}, Susan E. and {Bryson}, Stephen T. and {Lund}, Mikkel N. and {Handberg}, Rasmus and {Chaplin}, William J.},
        title = "{That's How We Roll: The NASA K2 Mission Science Products and Their Performance Metrics}",
      journal = {\pasp},
         year = 2016,
        month = jul,
       volume = {128},
       number = {965},
        pages = {075002},
          doi = {10.1088/1538-3873/128/965/075002},
archivePrefix = {arXiv},
       eprint = {1512.06162},
 primaryClass = {astro-ph.IM},
       adsurl = {https://ui.adsabs.harvard.edu/abs/2016PASP..128g5002V}
}

@ARTICLE{Stassun...TESS.CTL...2019AJ....158..138S,
       author = {{Stassun}, Keivan G. and {Oelkers}, Ryan J. and {Paegert}, Martin and {Torres}, Guillermo and {Pepper}, Joshua and {De Lee}, Nathan and {Collins}, Kevin and {Latham}, David W. and {Muirhead}, Philip S. and {Chittidi}, Jay and {Rojas-Ayala}, B{\'a}rbara and {Fleming}, Scott W. and {Rose}, Mark E. and {Tenenbaum}, Peter and {Ting}, Eric B. and {Kane}, Stephen R. and {Barclay}, Thomas and {Bean}, Jacob L. and {Brassuer}, C.~E. and {Charbonneau}, David and {Ge}, Jian and {Lissauer}, Jack J. and {Mann}, Andrew W. and {McLean}, Brian and {Mullally}, Susan and {Narita}, Norio and {Plavchan}, Peter and {Ricker}, George R. and {Sasselov}, Dimitar and {Seager}, S. and {Sharma}, Sanjib and {Shiao}, Bernie and {Sozzetti}, Alessandro and {Stello}, Dennis and {Vanderspek}, Roland and {Wallace}, Geoff and {Winn}, Joshua N.},
        title = "{The Revised TESS Input Catalog and Candidate Target List}",
      journal = {\aj},
         year = 2019,
        month = oct,
       volume = {158},
       number = {4},
          eid = {138},
        pages = {138},
          doi = {10.3847/1538-3881/ab3467},
archivePrefix = {arXiv},
       eprint = {1905.10694},
 primaryClass = {astro-ph.SR},
       adsurl = {https://ui.adsabs.harvard.edu/abs/2019AJ....158..138S}
}

@ARTICLE{Eschen...TESS...M...2024MNRAS.531.5053E,
       author = {{Eschen}, Yoshi Nike Emilia and {Kunimoto}, Michelle},
        title = "{Nine new M dwarf planet candidates from TESS including five gas giants}",
      journal = {\mnras},
         year = 2024,
        month = jul,
       volume = {531},
       number = {4},
        pages = {5053-5060},
          doi = {10.1093/mnras/stae1496},
archivePrefix = {arXiv},
       eprint = {2406.06688},
 primaryClass = {astro-ph.EP},
       adsurl = {https://ui.adsabs.harvard.edu/abs/2024MNRAS.531.5053E}
}

@ARTICLE{Gaidos...Kepler..M...2016MNRAS.457.2877G,
       author = {{Gaidos}, E. and {Mann}, A.~W. and {Kraus}, A.~L. and {Ireland}, M.},
        title = "{They are small worlds after all: revised properties of Kepler M dwarf stars and their planets}",
      journal = {\mnras},
         year = 2016,
        month = apr,
       volume = {457},
       number = {3},
        pages = {2877-2899},
          doi = {10.1093/mnras/stw097},
archivePrefix = {arXiv},
       eprint = {1512.04437},
 primaryClass = {astro-ph.EP},
       adsurl = {https://ui.adsabs.harvard.edu/abs/2016MNRAS.457.2877G}
}

@ARTICLE{Soderblom...ages...2010ARA&A..48..581S,
       author = {{Soderblom}, David R.},
        title = "{The Ages of Stars}",
      journal = {\araa},
         year = 2010,
        month = sep,
       volume = {48},
        pages = {581-629},
          doi = {10.1146/annurev-astro-081309-130806},
archivePrefix = {arXiv},
       eprint = {1003.6074},
 primaryClass = {astro-ph.SR},
       adsurl = {https://ui.adsabs.harvard.edu/abs/2010ARA&A..48..581S}
}

@software{software-citation-station-zenodo,
  author       = {Tom Wagg and
                  Floor Broekgaarden and
                  Phil Van-Lane and
                  Kai Wu and
                  Kayhan Gültekin},
  title        = {TomWagg/software-citation-station: v1.4},
  month        = nov,
  year         = 2025,
  publisher    = {Zenodo},
  version      = {v1.4},
  doi          = {10.5281/zenodo.17654855},
  url          = {https://doi.org/10.5281/zenodo.17654855},
  swhid        = {swh:1:dir:9a009430037c791424a572f542e9a5d5c1fb44ff
                   ;origin=https://doi.org/10.5281/zenodo.13225526;vi
                   sit=swh:1:snp:ef11f058d718d691f0661c9445c2251328cb
                   ac95;anchor=swh:1:rel:84cde4e532032537b7f014c4467c
                   102430a53fa2;path=TomWagg-software-citation-
                   station-61a588a
                  },
}

@ARTICLE{Akeson...NASAExoplanetArchive...2013PASP..125..989A,
       author = {{Akeson}, R.~L. and {Chen}, X. and {Ciardi}, D. and {Crane}, M. and {Good}, J. and {Harbut}, M. and {Jackson}, E. and {Kane}, S.~R. and {Laity}, A.~C. and {Leifer}, S. and {Lynn}, M. and {McElroy}, D.~L. and {Papin}, M. and {Plavchan}, P. and {Ram{\'\i}rez}, S.~V. and {Rey}, R. and {von Braun}, K. and {Wittman}, M. and {Abajian}, M. and {Ali}, B. and {Beichman}, C. and {Beekley}, A. and {Berriman}, G.~B. and {Berukoff}, S. and {Bryden}, G. and {Chan}, B. and {Groom}, S. and {Lau}, C. and {Payne}, A.~N. and {Regelson}, M. and {Saucedo}, M. and {Schmitz}, M. and {Stauffer}, J. and {Wyatt}, P. and {Zhang}, A.},
        title = "{The NASA Exoplanet Archive: Data and Tools for Exoplanet Research}",
      journal = {\pasp},
         year = 2013,
        month = aug,
       volume = {125},
       number = {930},
        pages = {989},
          doi = {10.1086/672273},
archivePrefix = {arXiv},
       eprint = {1307.2944},
 primaryClass = {astro-ph.IM},
       adsurl = {https://ui.adsabs.harvard.edu/abs/2013PASP..125..989A}
}

@ARTICLE{IsmailFawaz...TimeSeriesClassification...2020arXiv201000567I,
       author = {{Ismail Fawaz}, Hassan},
        title = "{Deep learning for time series classification}",
      journal = {arXiv e-prints},
         year = 2020,
        month = oct,
          eid = {arXiv:2010.00567},
        pages = {arXiv:2010.00567},
          doi = {10.48550/arXiv.2010.00567},
archivePrefix = {arXiv},
       eprint = {2010.00567},
 primaryClass = {cs.LG},
       adsurl = {https://ui.adsabs.harvard.edu/abs/2020arXiv201000567I}
}

@ARTICLE{DonosoOliva...Astromer...2023A&A...670A..54D,
       author = {{Donoso-Oliva}, C. and {Becker}, I. and {Protopapas}, P. and {Cabrera-Vives}, G. and {Vishnu}, M. and {Vardhan}, H.},
        title = "{ASTROMER. A transformer-based embedding for the representation of light curves}",
      journal = {\aap},
         year = 2023,
        month = feb,
       volume = {670},
          eid = {A54},
        pages = {A54},
          doi = {10.1051/0004-6361/202243928},
archivePrefix = {arXiv},
       eprint = {2205.01677},
 primaryClass = {astro-ph.IM},
       adsurl = {https://ui.adsabs.harvard.edu/abs/2023A&A...670A..54D}
}

@ARTICLE{Mathur...Sph...2023ApJ...952..131M,
       author = {{Mathur}, Savita and {Claytor}, Zachary R. and {Santos}, {\^A}ngela R.~G. and {Garc{\'\i}a}, Rafael A. and {Amard}, Louis and {Bugnet}, Lisa and {Corsaro}, Enrico and {Bonanno}, Alfio and {Breton}, Sylvain N. and {Godoy-Rivera}, Diego and {Pinsonneault}, Marc H. and {van Saders}, Jennifer},
        title = "{Magnetic Activity Evolution of Solar-like Stars. I. S $_{ph}$-Age Relation Derived from Kepler Observations}",
      journal = {\apj},
         year = 2023,
        month = aug,
       volume = {952},
       number = {2},
          eid = {131},
        pages = {131},
          doi = {10.3847/1538-4357/acd118},
archivePrefix = {arXiv},
       eprint = {2306.11657},
 primaryClass = {astro-ph.SR},
       adsurl = {https://ui.adsabs.harvard.edu/abs/2023ApJ...952..131M}
}

@ARTICLE{PLATO...2025ExA....59...26R,
       author = {{Rauer}, Heike and {Aerts}, Conny and {Cabrera}, Juan and {Deleuil}, Magali and {Erikson}, Anders and {Gizon}, Laurent and {Goupil}, Mariejo and {Heras}, Ana and {Walloschek}, Thomas and {Lorenzo-Alvarez}, Jose and {Marliani}, Filippo and {Martin-Garcia}, C{\'e}sar and {Mas-Hesse}, J. Miguel and {O'Rourke}, Laurence and {Osborn}, Hugh and {Pagano}, Isabella and {Piotto}, Giampaolo and {Pollacco}, Don and {Ragazzoni}, Roberto and {Ramsay}, Gavin and {Udry}, St{\'e}phane and {Appourchaux}, Thierry and {Benz}, Willy and {Brandeker}, Alexis and {G{\"u}del}, Manuel and {Janot-Pacheco}, Eduardo and {Kabath}, Petr and {Kjeldsen}, Hans and {Min}, Michiel and {Santos}, Nuno and {Smith}, Alan and {Suarez}, Juan-Carlos and {Werner}, Stephanie C. and {Aboudan}, Alessio and {Abreu}, Manuel and {Acu{\~n}a}, Lorena and {Adams}, Moritz and {Adibekyan}, Vardan and {Affer}, Laura and {Agneray}, Fran{\c{c}}ois and {Agnor}, Craig and {Aguirre B{\o}rsen-Koch}, Victor and {Ahmed}, Saad and {Aigrain}, Suzanne and {Al-Bahlawan}, Ashraf and {Alcacera Gil}, Ma de los Angeles and {Alei}, Eleonora and {Alencar}, Silvia and {Alexander}, Richard and {Alfonso-Garz{\'o}n}, Julia and {Alibert}, Yann and {Allende Prieto}, Carlos and {Almeida}, Leonardo and {Alonso Sobrino}, Roi and {Altavilla}, Giuseppe and {Althaus}, Christian and {Alvarez Trujillo}, Luis Alonso and {Amarsi}, Anish and {Ammler-von Eiff}, Matthias and {Am{\^o}res}, Eduardo and {Andrade}, Laerte and {Antoniadis-Karnavas}, Alexandros and {Ant{\'o}nio}, Carlos and {Aparicio del Moral}, Beatriz and {Appolloni}, Matteo and {Arena}, Claudio and {Armstrong}, David and {Aroca Aliaga}, Jose and {Asplund}, Martin and {Audenaert}, Jeroen and {Auricchio}, Natalia and {Avelino}, Pedro and {Baeke}, Ann and {Bailli{\'e}}, Kevin and {Balado}, Ana and {Ballber Balaguer{\'o}}, Pau and {Balestra}, Andrea and {Ball}, Warrick and {Ballans}, Herve and {Ballot}, Jerome and {Barban}, Caroline and {Barbary}, Ga{\"e}le and {Barbieri}, Mauro and {Barcel{\'o} Forteza}, Sebasti{\`a} and {Barker}, Adrian and {Barklem}, Paul and {Barnes}, Sydney and {Barrado Navascues}, David and {Barragan}, Oscar and {Baruteau}, Cl{\'e}ment and {Basu}, Sarbani and {Baudin}, Frederic and {Baumeister}, Philipp and {Bayliss}, Daniel and {Bazot}, Michael and {Beck}, Paul G. and {Belkacem}, Kevin and {Bellinger}, Earl and {Benatti}, Serena and {Benomar}, Othman and {B{\'e}rard}, Diane and {Bergemann}, Maria and {Bergomi}, Maria and {Bernardo}, Pierre and {Biazzo}, Katia and {Bignamini}, Andrea and {Bigot}, Lionel and {Billot}, Nicolas and {Binet}, Martin and {Biondi}, David and {Biondi}, Federico and {Birch}, Aaron C. and {Bitsch}, Bertram and {Bluhm Ceballos}, Paz Victoria and {B{\'o}di}, Attila and {Bogn{\'a}r}, Zs{\'o}fia and {Boisse}, Isabelle and {Bolmont}, Emeline and {Bonanno}, Alfio and {Bonavita}, Mariangela and {Bonfanti}, Andrea and {Bonfils}, Xavier and {Bonito}, Rosaria and {Bonomo}, Aldo Stefano and {B{\"o}rner}, Anko and {Boro Saikia}, Sudeshna and {Borreguero Mart{\'\i}n}, Elisa and {Borsa}, Francesco and {Borsato}, Luca and {Bossini}, Diego and {Bouchy}, Francois and {Bou{\'e}}, Gwena{\"e}l and {Boufleur}, Rodrigo and {Boumier}, Patrick and {Bourrier}, Vincent and {Bowman}, Dominic M. and {Bozzo}, Enrico and {Bradley}, Louisa and {Bray}, John and {Bressan}, Alessandro and {Breton}, Sylvain and {Brienza}, Daniele and {Brito}, Ana and {Brogi}, Matteo and {Brown}, Beverly and {Brown}, David J.~A. and {Brun}, Allan Sacha and {Bruno}, Giovanni and {Bruns}, Michael and {Buchhave}, Lars A. and {Bugnet}, Lisa and {Buldgen}, Ga{\"e}l and {Burgess}, Patrick and {Busatta}, Andrea and {Busso}, Giorgia and {Buzasi}, Derek and {Caballero}, Jos{\'e} A. and {Cabral}, Alexandre and {Cabrero Gomez}, Juan-Francisco and {Calderone}, Flavia and {Cameron}, Robert and {Cameron}, Andrew and {Campante}, Tiago and {Campos Gestal}, N{\'e}stor and {Canto Martins}, Bruno Leonardo and {Cara}, Christophe and {Carone}, Ludmila and {Carrasco}, Josep Manel and {Casagrande}, Luca and {Casewell}, Sarah L. and {Cassisi}, Santi and {Castellani}, Marco and {Castro}, Matthieu and {Catala}, Claude and {Catal{\'a}n Fern{\'a}ndez}, Irene and {Catelan}, M{\'a}rcio and {Cegla}, Heather and {Cerruti}, Chiara and {Cessa}, Virginie and {Chadid}, Merieme and {Chaplin}, William and {Charpinet}, Stephane and {Chiappini}, Cristina and {Chiarucci}, Simone and {Chiavassa}, Andrea and {Chinellato}, Simonetta and {Chirulli}, Giovanni and {Christensen-Dalsgaard}, J{\o}rgen and {Church}, Ross and {Claret}, Antonio and {Clarke}, Cathie and {Claudi}, Riccardo and {Clermont}, Lionel and {Coelho}, Hugo and {Coelho}, Joao and {Cogato}, Fabrizio and {Colom{\'e}}, Josep and {Condamin}, Mathieu and {Conde Garc{\'\i}a}, Fernando and {Conseil}, Simon},
        title = "{The PLATO mission}",
      journal = {Experimental Astronomy},
         year = 2025,
        month = jun,
       volume = {59},
       number = {3},
          eid = {26},
        pages = {26},
          doi = {10.1007/s10686-025-09985-9},
archivePrefix = {arXiv},
       eprint = {2406.05447},
 primaryClass = {astro-ph.IM},
       adsurl = {https://ui.adsabs.harvard.edu/abs/2025ExA....59...26R}
}

@ARTICLE{GodoyRivera...2021...rotcatalogue,
       author = {{Godoy-Rivera}, Diego and {Pinsonneault}, Marc H. and {Rebull}, Luisa M.},
        title = "{Stellar Rotation in the Gaia Era: Revised Open Clusters' Sequences}",
      journal = {\apjs},
         year = 2021,
        month = dec,
       volume = {257},
       number = {2},
          eid = {46},
        pages = {46},
          doi = {10.3847/1538-4365/ac2058},
archivePrefix = {arXiv},
       eprint = {2101.01183},
 primaryClass = {astro-ph.SR},
       adsurl = {https://ui.adsabs.harvard.edu/abs/2021ApJS..257...46G}
}

@ARTICLE{Tran...flaring...2026arXiv260220402T,
       author = {{Tran}, Andrew and {Song}, Inseok},
        title = "{Stellar flare study of nearby young moving group members with TESS Data}",
      journal = {arXiv e-prints},
         year = 2026,
        month = feb,
          eid = {arXiv:2602.20402},
        pages = {arXiv:2602.20402},
          doi = {10.48550/arXiv.2602.20402},
archivePrefix = {arXiv},
       eprint = {2602.20402},
 primaryClass = {astro-ph.SR},
       adsurl = {https://ui.adsabs.harvard.edu/abs/2026arXiv260220402T}
}

@ARTICLE{Medina...2022ApJ...935..104M,
       author = {{Medina}, Amber A. and {Winters}, Jennifer G. and {Irwin}, Jonathan M. and {Charbonneau}, David},
        title = "{Galactic Kinematics and Observed Flare Rates of a Volume-complete Sample of Mid-to-late M Dwarfs: Constraints on the History of the Stellar Radiation Environment of Planets Orbiting Low-mass Stars}",
      journal = {\apj},
         year = 2022,
        month = aug,
       volume = {935},
       number = {2},
          eid = {104},
        pages = {104},
          doi = {10.3847/1538-4357/ac77f9},
archivePrefix = {arXiv},
       eprint = {2205.02331},
 primaryClass = {astro-ph.SR},
       adsurl = {https://ui.adsabs.harvard.edu/abs/2022ApJ...935..104M}
}

@ARTICLE{Feinstein...2020AJ....160..219F,
       author = {{Feinstein}, Adina D. and {Montet}, Benjamin T. and {Ansdell}, Megan and {Nord}, Brian and {Bean}, Jacob L. and {G{\"u}nther}, Maximilian N. and {Gully-Santiago}, Michael A. and {Schlieder}, Joshua E.},
        title = "{Flare Statistics for Young Stars from a Convolutional Neural Network Analysis of TESS Data}",
      journal = {\aj},
         year = 2020,
        month = nov,
       volume = {160},
       number = {5},
          eid = {219},
        pages = {219},
          doi = {10.3847/1538-3881/abac0a},
archivePrefix = {arXiv},
       eprint = {2005.07710},
 primaryClass = {astro-ph.SR},
       adsurl = {https://ui.adsabs.harvard.edu/abs/2020AJ....160..219F}
}

@ARTICLE{Davenport...flaring...2019ApJ...871..241D,
       author = {{Davenport}, James R.~A. and {Covey}, Kevin R. and {Clarke}, Riley W. and {Boeck}, Austin C. and {Cornet}, Jonathan and {Hawley}, Suzanne L.},
        title = "{The Evolution of Flare Activity with Stellar Age}",
      journal = {\apj},
         year = 2019,
        month = feb,
       volume = {871},
       number = {2},
          eid = {241},
        pages = {241},
          doi = {10.3847/1538-4357/aafb76},
archivePrefix = {arXiv},
       eprint = {1901.00890},
 primaryClass = {astro-ph.SR},
       adsurl = {https://ui.adsabs.harvard.edu/abs/2019ApJ...871..241D}
}

@ARTICLE{Lu...Day&Age...2024AJ....167..159L,
       author = {{Lu}, Yuxi(Lucy) and {Angus}, Ruth and {Foreman-Mackey}, Daniel and {Hattori}, Soichiro},
        title = "{In This Day and Age: An Empirical Gyrochronology Relation for Partially and Fully Convective Single Field Stars}",
      journal = {\aj},
         year = 2024,
        month = apr,
       volume = {167},
       number = {4},
          eid = {159},
        pages = {159},
          doi = {10.3847/1538-3881/ad28b9},
archivePrefix = {arXiv},
       eprint = {2310.14990},
 primaryClass = {astro-ph.SR},
       adsurl = {https://ui.adsabs.harvard.edu/abs/2024AJ....167..159L}
}

@ARTICLE{BailerJones...dist...2021AJ....161..147B,
       author = {{Bailer-Jones}, C.~A.~L. and {Rybizki}, J. and {Fouesneau}, M. and {Demleitner}, M. and {Andrae}, R.},
        title = "{Estimating Distances from Parallaxes. V. Geometric and Photogeometric Distances to 1.47 Billion Stars in Gaia Early Data Release 3}",
      journal = {\aj},
         year = 2021,
        month = mar,
       volume = {161},
       number = {3},
          eid = {147},
        pages = {147},
          doi = {10.3847/1538-3881/abd806},
archivePrefix = {arXiv},
       eprint = {2012.05220},
 primaryClass = {astro-ph.SR},
       adsurl = {https://ui.adsabs.harvard.edu/abs/2021AJ....161..147B}
}

@ARTICLE{Barnes...2003ApJ...586..464B,
       author = {{Barnes}, Sydney A.},
        title = "{On the Rotational Evolution of Solar- and Late-Type Stars, Its Magnetic Origins, and the Possibility of Stellar Gyrochronology}",
      journal = {\apj},
         year = 2003,
        month = mar,
       volume = {586},
       number = {1},
        pages = {464-479},
          doi = {10.1086/367639},
archivePrefix = {arXiv},
       eprint = {astro-ph/0303631},
 primaryClass = {astro-ph},
       adsurl = {https://ui.adsabs.harvard.edu/abs/2003ApJ...586..464B}
}

@ARTICLE{TESS...2015JATIS...1a4003R,
       author = {{Ricker}, George R. and {Winn}, Joshua N. and {Vanderspek}, Roland and {Latham}, David W. and {Bakos}, G{\'a}sp{\'a}r {\'A}. and {Bean}, Jacob L. and {Berta-Thompson}, Zachory K. and {Brown}, Timothy M. and {Buchhave}, Lars and {Butler}, Nathaniel R. and {Butler}, R. Paul and {Chaplin}, William J. and {Charbonneau}, David and {Christensen-Dalsgaard}, J{\o}rgen and {Clampin}, Mark and {Deming}, Drake and {Doty}, John and {De Lee}, Nathan and {Dressing}, Courtney and {Dunham}, Edward W. and {Endl}, Michael and {Fressin}, Francois and {Ge}, Jian and {Henning}, Thomas and {Holman}, Matthew J. and {Howard}, Andrew W. and {Ida}, Shigeru and {Jenkins}, Jon M. and {Jernigan}, Garrett and {Johnson}, John Asher and {Kaltenegger}, Lisa and {Kawai}, Nobuyuki and {Kjeldsen}, Hans and {Laughlin}, Gregory and {Levine}, Alan M. and {Lin}, Douglas and {Lissauer}, Jack J. and {MacQueen}, Phillip and {Marcy}, Geoffrey and {McCullough}, Peter R. and {Morton}, Timothy D. and {Narita}, Norio and {Paegert}, Martin and {Palle}, Enric and {Pepe}, Francesco and {Pepper}, Joshua and {Quirrenbach}, Andreas and {Rinehart}, Stephen A. and {Sasselov}, Dimitar and {Sato}, Bun'ei and {Seager}, Sara and {Sozzetti}, Alessandro and {Stassun}, Keivan G. and {Sullivan}, Peter and {Szentgyorgyi}, Andrew and {Torres}, Guillermo and {Udry}, Stephane and {Villasenor}, Joel},
        title = "{Transiting Exoplanet Survey Satellite (TESS)}",
      journal = {Journal of Astronomical Telescopes, Instruments, and Systems},
         year = 2015,
        month = jan,
       volume = {1},
          eid = {014003},
        pages = {014003},
          doi = {10.1117/1.JATIS.1.1.014003},
       adsurl = {https://ui.adsabs.harvard.edu/abs/2015JATIS...1a4003R}
}

@ARTICLE{Jeffries...LiDepAges...2023MNRAS.523..802J,
       author = {{Jeffries}, R.~D. and {Jackson}, R.~J. and {Wright}, Nicholas J. and {Weaver}, G. and {Gilmore}, G. and {Randich}, S. and {Bragaglia}, A. and {Korn}, A.~J. and {Smiljanic}, R. and {Biazzo}, K. and {Casey}, A.~R. and {Frasca}, A. and {Gonneau}, A. and {Guiglion}, G. and {Morbidelli}, L. and {Prisinzano}, L. and {Sacco}, G.~G. and {Tautvai{\v{s}}ien{\.{e}}}, G. and {Worley}, C.~C. and {Zaggia}, S.},
        title = "{The Gaia-ESO Survey: empirical estimates of stellar ages from lithium equivalent widths (EAGLES)}",
      journal = {\mnras},
         year = 2023,
        month = jul,
       volume = {523},
       number = {1},
        pages = {802-824},
          doi = {10.1093/mnras/stad1293},
archivePrefix = {arXiv},
       eprint = {2304.12197},
 primaryClass = {astro-ph.SR},
       adsurl = {https://ui.adsabs.harvard.edu/abs/2023MNRAS.523..802J}
}

@ARTICLE{K2...Howell...2014PASP..126..398H,
       author = {{Howell}, Steve B. and {Sobeck}, Charlie and {Haas}, Michael and {Still}, Martin and {Barclay}, Thomas and {Mullally}, Fergal and {Troeltzsch}, John and {Aigrain}, Suzanne and {Bryson}, Stephen T. and {Caldwell}, Doug and {Chaplin}, William J. and {Cochran}, William D. and {Huber}, Daniel and {Marcy}, Geoffrey W. and {Miglio}, Andrea and {Najita}, Joan R. and {Smith}, Marcie and {Twicken}, J.~D. and {Fortney}, Jonathan J.},
        title = "{The K2 Mission: Characterization and Early Results}",
      journal = {\pasp},
         year = 2014,
        month = apr,
       volume = {126},
       number = {938},
        pages = {398},
          doi = {10.1086/676406},
archivePrefix = {arXiv},
       eprint = {1402.5163},
 primaryClass = {astro-ph.IM},
       adsurl = {https://ui.adsabs.harvard.edu/abs/2014PASP..126..398H}
}

@ARTICLE{Wilson...1968ApJ...153..221W,
       author = {{Wilson}, O.~C.},
        title = "{Flux Measurements at the Centers of Stellar H- and K-Lines}",
      journal = {\apj},
         year = 1968,
        month = jul,
       volume = {153},
        pages = {221},
          doi = {10.1086/149652},
       adsurl = {https://ui.adsabs.harvard.edu/abs/1968ApJ...153..221W}
}

@ARTICLE{NeuralSplineFlows...2019arXiv190604032D,
       author = {{Durkan}, Conor and {Bekasov}, Artur and {Murray}, Iain and {Papamakarios}, George},
        title = "{Neural Spline Flows}",
      journal = {arXiv e-prints},
         year = 2019,
        month = jun,
          eid = {arXiv:1906.04032},
        pages = {arXiv:1906.04032},
          doi = {10.48550/arXiv.1906.04032},
archivePrefix = {arXiv},
       eprint = {1906.04032},
 primaryClass = {stat.ML},
       adsurl = {https://ui.adsabs.harvard.edu/abs/2019arXiv190604032D}
}

@article{PCA...Hotelling,
author = {Hotelling, H.},
year = {1932},
month = {11},
pages = {417-441},
title = {Analysis of a complex of statistical variables into principal components},
volume = {24},
journal = {Journal of Educational Psychology},
doi = {10.1037/h0071325}
}

@ARTICLE{Edenhofer...2024A&A...685A..82E,
       author = {{Edenhofer}, Gordian and {Zucker}, Catherine and {Frank}, Philipp and {Saydjari}, Andrew K. and {Speagle}, Joshua S. and {Finkbeiner}, Douglas and {En{\ss}lin}, Torsten A.},
        title = "{A parsec-scale Galactic 3D dust map out to 1.25 kpc from the Sun}",
      journal = {\aap},
         year = 2024,
        month = may,
       volume = {685},
          eid = {A82},
        pages = {A82},
          doi = {10.1051/0004-6361/202347628},
archivePrefix = {arXiv},
       eprint = {2308.01295},
 primaryClass = {astro-ph.GA},
       adsurl = {https://ui.adsabs.harvard.edu/abs/2024A&A...685A..82E}
}

@ARTICLE{PecautMamajek...sptype...2013ApJS..208....9P,
       author = {{Pecaut}, Mark J. and {Mamajek}, Eric E.},
        title = "{Intrinsic Colors, Temperatures, and Bolometric Corrections of Pre-main-sequence Stars}",
      journal = {\apjs},
         year = 2013,
        month = sep,
       volume = {208},
       number = {1},
          eid = {9},
        pages = {9},
          doi = {10.1088/0067-0049/208/1/9},
archivePrefix = {arXiv},
       eprint = {1307.2657},
 primaryClass = {astro-ph.SR},
       adsurl = {https://ui.adsabs.harvard.edu/abs/2013ApJS..208....9P}
}

@ARTICLE{MOCAdb...2026arXiv260215695G,
       author = {{Gagn{\'e}}, Jonathan and {Moranta}, Leslie and {Faherty}, Jacqueline K. and {Curtis}, Jason Lee and {Bickle}, Thomas P. and {Couture}, Dominic and {Chiasson David}, Am{\'e}lie and {Christie}, Katie and {Lambier}, Samantha and {Leclerc}, Elise and {Poliquin}, Livia and {Belzile}, Danika and {Mamajek}, Eric E.},
        title = "{The Montreal Open Clusters and Associations (MOCA) Database: A Census of Nearby Associations, Open Clusters, and Young Substellar Objects within 500 pc of the Sun}",
      journal = {arXiv e-prints},
         year = 2026,
        month = feb,
          eid = {arXiv:2602.15695},
        pages = {arXiv:2602.15695},
          doi = {10.48550/arXiv.2602.15695},
archivePrefix = {arXiv},
       eprint = {2602.15695},
 primaryClass = {astro-ph.SR},
       adsurl = {https://ui.adsabs.harvard.edu/abs/2026arXiv260215695G}
}

@ARTICLE{Lomb...1976Ap&SS..39..447L,
       author = {{Lomb}, N.~R.},
        title = "{Least-Squares Frequency Analysis of Unequally Spaced Data}",
      journal = {\apss},
         year = 1976,
        month = feb,
       volume = {39},
       number = {2},
        pages = {447-462},
          doi = {10.1007/BF00648343},
       adsurl = {https://ui.adsabs.harvard.edu/abs/1976Ap&SS..39..447L}
}

@ARTICLE{Thomson...multitaper...1982IEEEP..70.1055T,
       author = {{Thomson}, D.~J.},
        title = "{Spectrum Estimation and Harmonic Analysis}",
      journal = {IEEE Proceedings},
         year = 1982,
        month = jan,
       volume = {70},
        pages = {1055-1096},
          doi = {10.1109/PROC.1982.12433},
       adsurl = {https://ui.adsabs.harvard.edu/abs/1982IEEEP..70.1055T}
}

@ARTICLE{Scargle...1982ApJ...263..835S,
       author = {{Scargle}, J.~D.},
        title = "{Studies in astronomical time series analysis. II. Statistical aspects of spectral analysis of unevenly spaced data.}",
      journal = {\apj},
         year = 1982,
        month = dec,
       volume = {263},
        pages = {835-853},
          doi = {10.1086/160554},
       adsurl = {https://ui.adsabs.harvard.edu/abs/1982ApJ...263..835S}
}

@ARTICLE{Pass...2024ApJ...966..231P,
       author = {{Pass}, Emily K. and {Charbonneau}, David and {Latham}, David W. and {Berlind}, Perry and {Calkins}, Michael L. and {Esquerdo}, Gilbert A. and {Mink}, Jessica},
        title = "{The Mass Dependence of H{\ensuremath{\alpha}} Emission and Stellar Spindown for Fully Convective M Dwarfs}",
      journal = {\apj},
         year = 2024,
        month = may,
       volume = {966},
       number = {2},
          eid = {231},
        pages = {231},
          doi = {10.3847/1538-4357/ad3631},
archivePrefix = {arXiv},
       eprint = {2401.10167},
 primaryClass = {astro-ph.SR},
       adsurl = {https://ui.adsabs.harvard.edu/abs/2024ApJ...966..231P}
}

@ARTICLE{Pass...2023AJ....166...16P,
       author = {{Pass}, Emily K. and {Winters}, Jennifer G. and {Charbonneau}, David and {Irwin}, Jonathan M. and {Medina}, Amber A.},
        title = "{Active Stars in the Spectroscopic Survey of Mid-to-late M Dwarfs within 15 pc}",
      journal = {\aj},
         year = 2023,
        month = jul,
       volume = {166},
       number = {1},
          eid = {16},
        pages = {16},
          doi = {10.3847/1538-3881/acd6a2},
archivePrefix = {arXiv},
       eprint = {2306.00799},
 primaryClass = {astro-ph.SR},
       adsurl = {https://ui.adsabs.harvard.edu/abs/2023AJ....166...16P}
}

@ARTICLE{LWRD...Engle...2024ApJ...960...62E,
       author = {{Engle}, Scott G.},
        title = "{Living with a Red Dwarf: X-Ray, UV, and Ca II Activity-Age Relationships of M Dwarfs}",
      journal = {\apj},
         year = 2024,
        month = jan,
       volume = {960},
       number = {1},
          eid = {62},
        pages = {62},
          doi = {10.3847/1538-4357/ad0840},
archivePrefix = {arXiv},
       eprint = {2310.04302},
 primaryClass = {astro-ph.SR},
       adsurl = {https://ui.adsabs.harvard.edu/abs/2024ApJ...960...62E}
}

@ARTICLE{Mamonova...2025AA...700A..53M,
       author = {{Mamonova}, E. and {Shan}, Y. and {Kowalski}, A.~F. and {Wedemeyer}, S. and {Werner}, S.~C.},
        title = "{Flare frequency in M dwarfs belonging to young moving groups}",
      journal = {\aap},
         year = 2025,
        month = aug,
       volume = {700},
          eid = {A53},
        pages = {A53},
          doi = {10.1051/0004-6361/202554614},
archivePrefix = {arXiv},
       eprint = {2506.04465},
 primaryClass = {astro-ph.SR},
       adsurl = {https://ui.adsabs.harvard.edu/abs/2025A&A...700A..53M}
}

@ARTICLE{Kiman...2021AJ....161..277K,
       author = {{Kiman}, Rocio and {Faherty}, Jacqueline K. and {Cruz}, Kelle L. and {Gagn{\'e}}, Jonathan and {Angus}, Ruth and {Schmidt}, Sarah J. and {Mann}, Andrew W. and {Bardalez Gagliuffi}, Daniella C. and {Rice}, Emily},
        title = "{Calibration of the H{\ensuremath{\alpha}} Age-Activity Relation for M Dwarfs}",
      journal = {\aj},
         year = 2021,
        month = jun,
       volume = {161},
       number = {6},
          eid = {277},
        pages = {277},
          doi = {10.3847/1538-3881/abf561},
archivePrefix = {arXiv},
       eprint = {2104.01232},
 primaryClass = {astro-ph.SR},
       adsurl = {https://ui.adsabs.harvard.edu/abs/2021AJ....161..277K}
}

@ARTICLE{Facco...TwoNN...2017NatSR...712140F,
       author = {{Facco}, Elena and {d'Errico}, Maria and {Rodriguez}, Alex and {Laio}, Alessandro},
        title = "{Estimating the intrinsic dimension of datasets by a minimal neighborhood information}",
      journal = {Scientific Reports},
         year = 2017,
        month = sep,
       volume = {7},
          eid = {12140},
        pages = {12140},
          doi = {10.1038/s41598-017-11873-y},
archivePrefix = {arXiv},
       eprint = {1803.06992},
 primaryClass = {stat.ML},
       adsurl = {https://ui.adsabs.harvard.edu/abs/2017NatSR...712140F}
}

@ARTICLE{TARS...Boyle...2026arXiv260305586B,
       author = {{Boyle}, Andrew W. and {Bouma}, Luke G. and {Mann}, Andrew W.},
        title = "{The TESS All-Sky Rotation Survey: Periods for 1,046,317 Stars Within 500 pc}",
      journal = {arXiv e-prints},
         year = 2026,
        month = mar,
          eid = {arXiv:2603.05586},
        pages = {arXiv:2603.05586},
          doi = {10.48550/arXiv.2603.05586},
archivePrefix = {arXiv},
       eprint = {2603.05586},
 primaryClass = {astro-ph.SR},
       adsurl = {https://ui.adsabs.harvard.edu/abs/2026arXiv260305586B}
}

@article{Time2Vec...DBLP:journals/corr/abs-1907-05321,
  author       = {Seyed Mehran Kazemi and
                  Rishab Goel and
                  Sepehr Eghbali and
                  Janahan Ramanan and
                  Jaspreet Sahota and
                  Sanjay Thakur and
                  Stella Wu and
                  Cathal Smyth and
                  Pascal Poupart and
                  Marcus A. Brubaker},
  title        = {Time2Vec: Learning a Vector Representation of Time},
  journal      = {CoRR},
  volume       = {abs/1907.05321},
  year         = {2019},
  url          = {http://arxiv.org/abs/1907.05321},
  eprinttype   = {arXiv},
  eprint       = {1907.05321},
  bibsource    = {dblp computer science bibliography, https://dblp.org}
}

@article{GRU...DBLP:journals/corr/ChoMGBSB14,
  author       = {Kyunghyun Cho and
                  Bart van Merrienboer and
                  {\c{C}}aglar G{\"{u}}l{\c{c}}ehre and
                  Fethi Bougares and
                  Holger Schwenk and
                  Yoshua Bengio},
  title        = {Learning Phrase Representations using {RNN} Encoder-Decoder for Statistical
                  Machine Translation},
  journal      = {CoRR},
  volume       = {abs/1406.1078},
  year         = {2014},
  url          = {http://arxiv.org/abs/1406.1078},
  eprinttype   = {arXiv},
  eprint       = {1406.1078},
  bibsource    = {dblp computer science bibliography, https://dblp.org}
}

@article{LSTM,
author = {Hochreiter, Sepp and Schmidhuber, Jürgen},
year = {1997},
month = {11},
pages = {1735-1780},
title = {Long Short-Term Memory},
volume = {9},
journal = {Neural Computation},
doi = {10.1162/neco.1997.9.8.1735}
}

@ARTICLE{Kim...thickdisk...2022MNRAS.510.4308K,
       author = {{Kim}, Bokyoung and {L{\'e}pine}, Sebastien},
        title = "{Stars in the local galactic thick disc and halo in Gaia EDR3: a catalogue of half a million local main-sequence stars with photometric metallicities}",
      journal = {\mnras},
         year = 2022,
        month = mar,
       volume = {510},
       number = {3},
        pages = {4308-4329},
          doi = {10.1093/mnras/stab3671},
archivePrefix = {arXiv},
       eprint = {2112.07419},
 primaryClass = {astro-ph.GA},
       adsurl = {https://ui.adsabs.harvard.edu/abs/2022MNRAS.510.4308K}
}

@ARTICLE{Zuo...FALCO...2026AJ....171...10Z,
       author = {{Zuo}, Xiaoxiong and {Tao}, Yihan and {Huang}, Yang and {Kang}, Zhixuan and {Chen}, Huaxi and {Cui}, Chenzhou and {Pan}, Jiashu and {Kong}, Xiao and {Ting}, Yuan-Sen and {Tang}, Xiaoyu and {Han}, Henggeng and {Mu}, Haiyang and {Xu}, Yunfei and {Fan}, Dongwei and {Xue}, Guirong and {Luo}, Ali and {Liu}, Jifeng},
        title = "{FALCO: Foundation Model of Astronomical Light Curves for Time Domain Astronomy. Implementation and Applications on Kepler Data}",
      journal = {\aj},
         year = 2026,
        month = jan,
       volume = {171},
       number = {1},
          eid = {10},
        pages = {10},
          doi = {10.3847/1538-3881/ae1467},
archivePrefix = {arXiv},
       eprint = {2504.20290},
 primaryClass = {astro-ph.IM},
       adsurl = {https://ui.adsabs.harvard.edu/abs/2026AJ....171...10Z}
}

@ARTICLE{DESA...2025ApJ...994..110K,
       author = {{Kamai}, Ilay and {Bronstein}, Alex M. and {Perets}, Hagai B.},
        title = "{Machine Learning Inference of Stellar Properties Using Integrated Photometric and Spectroscopic Data}",
      journal = {\apj},
         year = 2025,
        month = nov,
       volume = {994},
       number = {1},
          eid = {110},
        pages = {110},
          doi = {10.3847/1538-4357/ae0cbc},
archivePrefix = {arXiv},
       eprint = {2507.10666},
 primaryClass = {astro-ph.SR},
       adsurl = {https://ui.adsabs.harvard.edu/abs/2025ApJ...994..110K}
}

@ARTICLE{Bayestar19...dustmap...2019ApJ...887...93G,
       author = {{Green}, Gregory M. and {Schlafly}, Edward and {Zucker}, Catherine and {Speagle}, Joshua S. and {Finkbeiner}, Douglas},
        title = "{A 3D Dust Map Based on Gaia, Pan-STARRS 1, and 2MASS}",
      journal = {\apj},
         year = 2019,
        month = dec,
       volume = {887},
       number = {1},
          eid = {93},
        pages = {93},
          doi = {10.3847/1538-4357/ab5362},
archivePrefix = {arXiv},
       eprint = {1905.02734},
 primaryClass = {astro-ph.GA},
       adsurl = {https://ui.adsabs.harvard.edu/abs/2019ApJ...887...93G}
}

@ARTICLE{Lindegren...zeropoint...2021A&A...649A...4L,
       author = {{Lindegren}, L. and {Bastian}, U. and {Biermann}, M. and {Bombrun}, A. and {de Torres}, A. and {Gerlach}, E. and {Geyer}, R. and {Hern{\'a}ndez}, J. and {Hilger}, T. and {Hobbs}, D. and {Klioner}, S.~A. and {Lammers}, U. and {McMillan}, P.~J. and {Ramos-Lerate}, M. and {Steidelm{\"u}ller}, H. and {Stephenson}, C.~A. and {van Leeuwen}, F.},
        title = "{Gaia Early Data Release 3. Parallax bias versus magnitude, colour, and position}",
      journal = {\aap},
         year = 2021,
        month = may,
       volume = {649},
          eid = {A4},
        pages = {A4},
          doi = {10.1051/0004-6361/202039653},
archivePrefix = {arXiv},
       eprint = {2012.01742},
 primaryClass = {astro-ph.IM},
       adsurl = {https://ui.adsabs.harvard.edu/abs/2021A&A...649A...4L}
}

@ARTICLE{Newton...SouthernMdwarfs...2018AJ....156..217N,
       author = {{Newton}, Elisabeth R. and {Mondrik}, Nicholas and {Irwin}, Jonathan and {Winters}, Jennifer G. and {Charbonneau}, David},
        title = "{New Rotation Period Measurements for M Dwarfs in the Southern Hemisphere: An Abundance of Slowly Rotating, Fully Convective Stars}",
      journal = {\aj},
         year = 2018,
        month = nov,
       volume = {156},
       number = {5},
          eid = {217},
        pages = {217},
          doi = {10.3847/1538-3881/aad73b},
archivePrefix = {arXiv},
       eprint = {1807.09365},
 primaryClass = {astro-ph.SR},
       adsurl = {https://ui.adsabs.harvard.edu/abs/2018AJ....156..217N}
}

@ARTICLE{Newton...nearbyMdwarfs...2016ApJ...821...93N,
       author = {{Newton}, Elisabeth R. and {Irwin}, Jonathan and {Charbonneau}, David and {Berta-Thompson}, Zachory K. and {Dittmann}, Jason A. and {West}, Andrew A.},
        title = "{The Rotation and Galactic Kinematics of Mid M Dwarfs in the Solar Neighborhood}",
      journal = {\apj},
         year = 2016,
        month = apr,
       volume = {821},
       number = {2},
          eid = {93},
        pages = {93},
          doi = {10.3847/0004-637X/821/2/93},
archivePrefix = {arXiv},
       eprint = {1511.00957},
 primaryClass = {astro-ph.SR},
       adsurl = {https://ui.adsabs.harvard.edu/abs/2016ApJ...821...93N}
}

@ARTICLE{Magaudda...Mdwarfs...2020AA...638A..20M,
       author = {{Magaudda}, E. and {Stelzer}, B. and {Covey}, K.~R. and {Raetz}, St. and {Matt}, S.~P. and {Scholz}, A.},
        title = "{Relation of X-ray activity and rotation in M dwarfs and predicted time-evolution of the X-ray luminosity}",
      journal = {\aap},
         year = 2020,
        month = jun,
       volume = {638},
          eid = {A20},
        pages = {A20},
          doi = {10.1051/0004-6361/201937408},
archivePrefix = {arXiv},
       eprint = {2004.02904},
 primaryClass = {astro-ph.SR},
       adsurl = {https://ui.adsabs.harvard.edu/abs/2020A&A...638A..20M}
}

@ARTICLE{Santos...Prot...2019ApJS..244...21S,
       author = {{Santos}, A.~R.~G. and {Garc{\'\i}a}, R.~A. and {Mathur}, S. and {Bugnet}, L. and {van Saders}, J.~L. and {Metcalfe}, T.~S. and {Simonian}, G.~V.~A. and {Pinsonneault}, M.~H.},
        title = "{Surface Rotation and Photometric Activity for Kepler Targets. I. M and K Main-sequence Stars}",
      journal = {\apjs},
         year = 2019,
        month = sep,
       volume = {244},
       number = {1},
          eid = {21},
        pages = {21},
          doi = {10.3847/1538-4365/ab3b56},
archivePrefix = {arXiv},
       eprint = {1908.05222},
 primaryClass = {astro-ph.SR},
       adsurl = {https://ui.adsabs.harvard.edu/abs/2019ApJS..244...21S}
}

@ARTICLE{Santos...Prot...2021ApJS..255...17S,
       author = {{Santos}, A.~R.~G. and {Breton}, S.~N. and {Mathur}, S. and {Garc{\'\i}a}, R.~A.},
        title = "{Surface Rotation and Photometric Activity for Kepler Targets. II. G and F Main-sequence Stars and Cool Subgiant Stars}",
      journal = {\apjs},
         year = 2021,
        month = jul,
       volume = {255},
       number = {1},
          eid = {17},
        pages = {17},
          doi = {10.3847/1538-4365/ac033f},
archivePrefix = {arXiv},
       eprint = {2107.02217},
 primaryClass = {astro-ph.SR},
       adsurl = {https://ui.adsabs.harvard.edu/abs/2021ApJS..255...17S}
}

@ARTICLE{McQuillan...Prot...2014ApJS..211...24M,
       author = {{McQuillan}, A. and {Mazeh}, T. and {Aigrain}, S.},
        title = "{Rotation Periods of 34,030 Kepler Main-sequence Stars: The Full Autocorrelation Sample}",
      journal = {\apjs},
         year = 2014,
        month = apr,
       volume = {211},
       number = {2},
          eid = {24},
        pages = {24},
          doi = {10.1088/0067-0049/211/2/24},
archivePrefix = {arXiv},
       eprint = {1402.5694},
 primaryClass = {astro-ph.SR},
       adsurl = {https://ui.adsabs.harvard.edu/abs/2014ApJS..211...24M}
}

@ARTICLE{Mathur...activity...2025ApJ...982..114M,
       author = {{Mathur}, Savita and {Santos}, {\^A}ngela R.~G. and {Claytor}, Zachary R. and {Garc{\'\i}a}, Rafael A. and {Strugarek}, Antoine and {Finley}, Adam J. and {Noraz}, Quentin and {Amard}, Louis and {Beck}, Paul G. and {Bonanno}, Alfio and {Breton}, Sylvain N. and {Brun}, Allan S. and {Cao}, Lyra and {Corsaro}, Enrico and {Godoy-Rivera}, Diego and {Mathis}, St{\'e}phane and {Palakkatharappil}, Dinil B. and {Pinsonneault}, Marc H. and {van Saders}, Jennifer},
        title = "{Magnetic Activity Evolution of Solar-like Stars. II. S$_{ph}${\textendash}Ro Evolution of Kepler Main-sequence Targets}",
      journal = {\apj},
         year = 2025,
        month = apr,
       volume = {982},
       number = {2},
          eid = {114},
        pages = {114},
          doi = {10.3847/1538-4357/adb8cc},
archivePrefix = {arXiv},
       eprint = {2502.10109},
 primaryClass = {astro-ph.SR},
       adsurl = {https://ui.adsabs.harvard.edu/abs/2025ApJ...982..114M}
}

@ARTICLE{RNNs...2024arXiv241001201F,
       author = {{Feng}, Leo and {Tung}, Frederick and {Osama Ahmed}, Mohamed and {Bengio}, Yoshua and {Hajimirsadeghi}, Hossein},
        title = "{Were RNNs All We Needed?}",
      journal = {arXiv e-prints},
         year = 2024,
        month = oct,
          eid = {arXiv:2410.01201},
        pages = {arXiv:2410.01201},
          doi = {10.48550/arXiv.2410.01201},
archivePrefix = {arXiv},
       eprint = {2410.01201},
 primaryClass = {cs.LG},
       adsurl = {https://ui.adsabs.harvard.edu/abs/2024arXiv241001201F}
}

@ARTICLE{tapify...2024AJ....168..193P,
       author = {{Patil}, Aarya A. and {Eadie}, Gwendolyn M. and {Speagle}, Joshua S. and {Thomson}, David J.},
        title = "{Improving Power Spectrum Estimation Using Multitapering: Efficient Asteroseismic Analyses for Understanding Stars, the Milky Way, and Beyond}",
      journal = {\aj},
         year = 2024,
        month = nov,
       volume = {168},
       number = {5},
          eid = {193},
        pages = {193},
          doi = {10.3847/1538-3881/ad7029},
archivePrefix = {arXiv},
       eprint = {2209.15027},
 primaryClass = {astro-ph.IM},
       adsurl = {https://ui.adsabs.harvard.edu/abs/2024AJ....168..193P}
}

@ARTICLE{Feinstein...2024AJ....168...60F,
       author = {{Feinstein}, Adina D. and {Seligman}, Darryl Z. and {France}, Kevin and {Gagn{\'e}}, Jonathan and {Kowalski}, Adam},
        title = "{Evolution of Flare Activity in GKM Stars Younger Than 300 Myr over Five Years of TESS Observations}",
      journal = {\aj},
         year = 2024,
        month = aug,
       volume = {168},
       number = {2},
          eid = {60},
        pages = {60},
          doi = {10.3847/1538-3881/ad4edf},
archivePrefix = {arXiv},
       eprint = {2405.00850},
 primaryClass = {astro-ph.SR},
       adsurl = {https://ui.adsabs.harvard.edu/abs/2024AJ....168...60F}
}

@ARTICLE{Engle...LWRD...2024ApJ...960...62E,
       author = {{Engle}, Scott G.},
        title = "{Living with a Red Dwarf: X-Ray, UV, and Ca II Activity-Age Relationships of M Dwarfs}",
      journal = {\apj},
         year = 2024,
        month = jan,
       volume = {960},
       number = {1},
          eid = {62},
        pages = {62},
          doi = {10.3847/1538-4357/ad0840},
archivePrefix = {arXiv},
       eprint = {2310.04302},
 primaryClass = {astro-ph.SR},
       adsurl = {https://ui.adsabs.harvard.edu/abs/2024ApJ...960...62E}
}

@ARTICLE{Pass...MM...2022ApJ...936..109P,
       author = {{Pass}, Emily K. and {Charbonneau}, David and {Irwin}, Jonathan M. and {Winters}, Jennifer G.},
        title = "{Constraints on the Spindown of Fully Convective M Dwarfs Using Wide Field Binaries}",
      journal = {\apj},
         year = 2022,
        month = sep,
       volume = {936},
       number = {2},
          eid = {109},
        pages = {109},
          doi = {10.3847/1538-4357/ac7da8},
archivePrefix = {arXiv},
       eprint = {2206.15318},
 primaryClass = {astro-ph.SR},
       adsurl = {https://ui.adsabs.harvard.edu/abs/2022ApJ...936..109P}
}

@ARTICLE{TIMEtable...2023MNRAS.520.5283G,
       author = {{Gaidos}, Eric and {Claytor}, Zachary and {Dungee}, Ryan and {Ali}, Aleezah and {Feiden}, Gregory A.},
        title = "{The TIME Table: rotation and ages of cool exoplanet host stars}",
      journal = {\mnras},
         year = 2023,
        month = apr,
       volume = {520},
       number = {4},
        pages = {5283-5304},
          doi = {10.1093/mnras/stad343},
archivePrefix = {arXiv},
       eprint = {2301.12109},
 primaryClass = {astro-ph.EP},
       adsurl = {https://ui.adsabs.harvard.edu/abs/2023MNRAS.520.5283G}
}
\bibliographystyle{aasjournalv7}

\appendix

\section{Model performance with additional input channels}
\label{app:extra_channels}

\begin{figure}[b]
  \centering
  \includegraphics[width=\columnwidth]{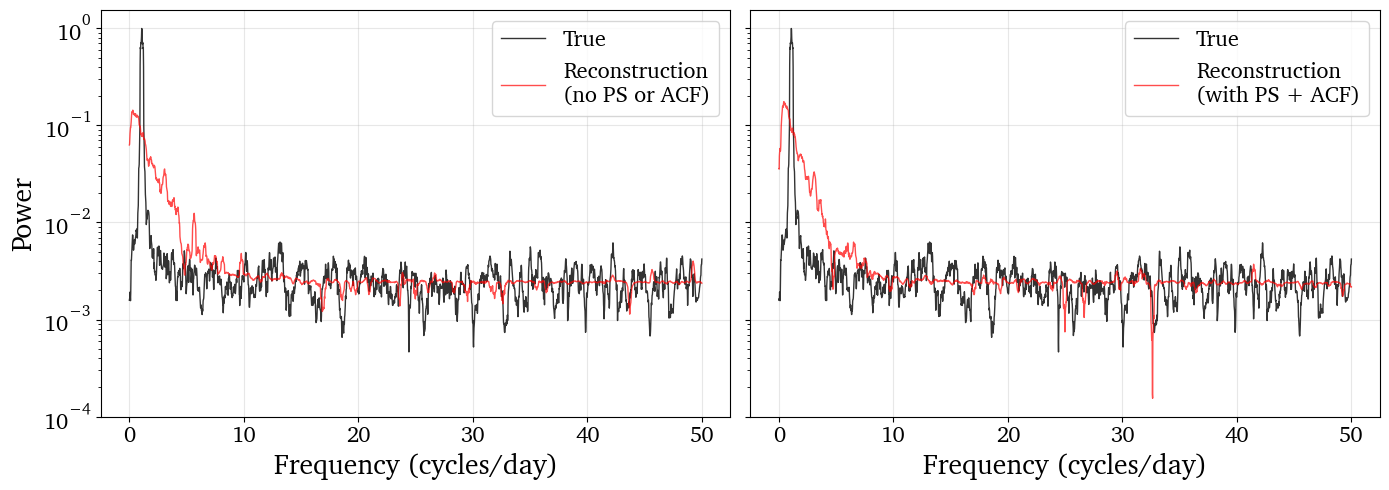}
  \caption{Reconstruction of the power spectrum of TIC 67787319 Sector 43. The left panel shows the reconstruction based on the core \et model, and the right panel shows the reconstruction when PS and ACF channels are included in training. Recovery of some features is marginally improved by including PS and ACF in training, but the overall shape is very similar.}
  \label{fig:ps_recon}
\end{figure}

We compared out model's ability to recover data both with and without extra input channels. The first extra input channel we considered for each LC is the power spectrum. This is a measure of how much information is contained in the LC at different frequencies. For signals that are strongly periodic at a single frequency, we would expect the power spectrum to have a strong peak at that frequency and smaller values at all other frequencies.

While Lomb-Scargle (LS) periodograms \citep[][]{Lomb...1976Ap&SS..39..447L,Scargle...1982ApJ...263..835S} are often used to compute these power spectra, they can suffer from spectral leakage, particularly in the case of irregularly sampled data. \citet{Scargle...1982ApJ...263..835S} suggested that tapering (reducing the amplitude at the edges of the time series data) can mitigate this spectral leakage and reduce the bias of the power spectrum estimate, but at the cost of increasing the variance. \citet{Thomson...multitaper...1982IEEEP..70.1055T} introduced the concept of multitapering, which combines the results from applying multiple orthogonal tapers to the same time series to produce a power spectrum estimate that reduces both bias and variance. \citet{tapify...2024AJ....168..193P} implement this concept in their \texttt{tapify} code, which we use to compute the power spectrum for each LC in our sample.

To homogenize our data and avoid including the frequencies of our power spectra as input, we re-sampled each power spectrum to a uniform grid of 16,000 frequency bins with a maximum of 360 cycles/day (which approximately corresponds to the Nyquist frequency of 2-minute cadence data. In addition to the power values, \texttt{tapify} provides the f-statistic for each frequency, which is a measure of how periodic the signal is. We combine the power values and the f-statistics in our first extra input channel.

\begin{figure}
\centering
\includegraphics[width=0.48\textwidth]{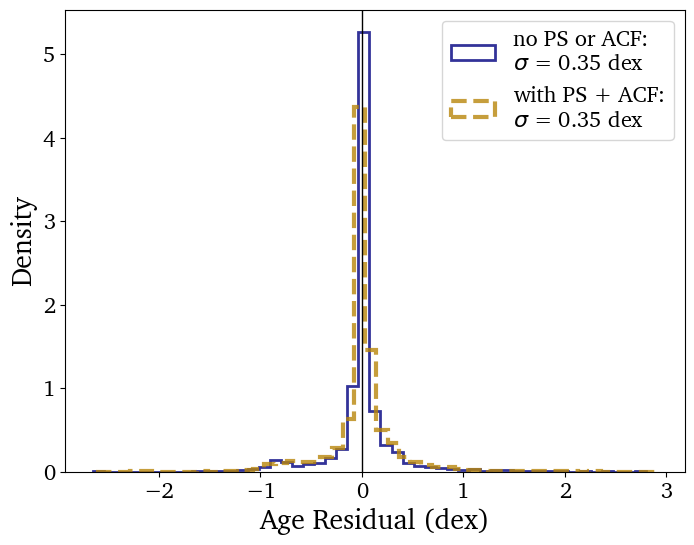}
\hfill
\includegraphics[width=0.4875\textwidth]{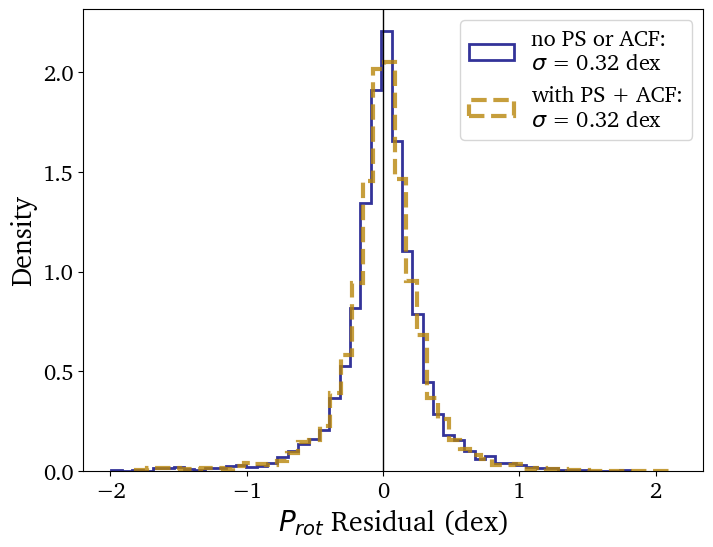}
\caption{Comparison of age (left) and $P_{rot}$ (right) inference tests using our $\Theta$, compared between the final \et model and a version that included the PS and ACF encoders in training.}
\label{fig:ps_acf_comp}
\end{figure}

The second extra input channel is the autocorrelation function (ACF) for each LC. We calculated this simply by applying the \texttt{scipy.fft.ifft} method to the re-sampled power spectrum, which provides an ``autocorrelation coefficient'' over a grid of time offsets $\Delta t$. At $\Delta t$ values that correspond to a strong periodic signal, the amplitude of the coefficient will be large.

We considered both of these channels because they are views of the data in different domains: the PS in the frequency domain and the ACF in the $\Delta t$ domain. Since RNNs can have difficulty learning patterns over long offsets, these channels could provide helpful additional context.

First, we trained autoencoders and decoders to reconstruct the PS and ACF. This was most successful using a UNet-style \citep[][]{UNet...2015arXiv150504597R} Convolutional Neural Network (CNN) architecture. We then trained two versions \et with and without these extra channels. They were included by concatenating the PS and ACF encodings into the input for the hidden state $H_t$ computation at each step in the RNN.

We compared the performance in several ways: (i) the reconstruction loss curve, (ii) ability to recover rotation periods, (iii), ability to recover ages, and (iv) ability to reconstruct the PS and ACFs from the latent space. All models had comparable performance across these tests, indicating that \et is able to capture the periodicity information from the flux channel equally as well as it can from the PS and ACFs equally well. Therefore it is not worth the extra model complexity and computational expense of including the PS and ACF channels in \et. Examples of the rotation period and age inference tests, and a random power spectrum reconstruction, are shown in Figures \ref{fig:ps_acf_comp} and \ref{fig:ps_recon}.

\section{Low-dimensional Encoding Representations}
\label{app:low_dim}

Here we discuss in detail three methods for reducing the dimensionality of the 1536-dim LC encodings $\Theta$.

\subsection{UMAP Representation}
\label{app:low_dim_umap}

\texttt{UMAP} learns how objects are clustered in a high dimensional space, and then places them in a lower dimensional space that is optimized to preserve that structure, so it is particularly useful for visualization. The actual values of the \texttt{UMAP} dimensions are not meaningful, so we do not show them any plots, but the relative positions of the LCs reveal important trends. 2-dimensional space is naturally intuitive for visualizations, so we use 2-dim \texttt{UMAP} projections of $\Theta$ frequently throughout the rest of this paper, but since it does not capture the intrinsic dimensionality of $\Theta$ well we do not use it for quantitative analysis.

\subsection{PCA Representation}
\label{app:low_dim_pca}

PCA re-orients a parameter space by iteratively defining axes that align with the directions of greatest variance. In the context of our 1536-dim $\Theta$ space, the first principal component (PC1) is a unit vector that maximizes the normalized variance of all $\Theta$ projected onto it. The next component PC2 is defined as the unit vector linearly orthogonal to PC1 that maximizes the projected variance of all $\Theta$. PC3 is the unit vector that maximizes the projected variance of $\Theta$ while being orthogonal to PC1 and PC2, and subsequent components are defined in the same manner.

This can be done for any number of components up to the dimensionality of the parameter space, but the marginal signal diminishes with each PC added. For example, PC1 explains 44\% of the variance in $\Theta$ on its own, PCs 2 through 4 explain an additional 28\% together, and PCs 5 through 1536 together explain the remaining 28\%.

Since PCA is a relatively intuitive way of choosing a low-dimensional representation of $\Theta$, we use it in most of the quantitative analysis in this work. PCA is also flexible; depending on the complexity of the downstream task that uses $\Theta$ the number of PCs can be chosen to balance performance with completeness. To this point, in this work we truncate to 4, 8, or 16 PCs for various tasks depending on the context and goal, referred to as $\Theta'_4$, $\Theta'_8$ and $\Theta'_{16}$.

It is also important to note that the number of useful PCs does not necessarily correlate with the intrinsic dimensionality of $\Theta$. While PCA is restricted to a \textit{linear} basis, \texttt{TwoNN} captures \textit{nonlinear} dimensionality, meaning that more than 13 PCs would be required to capture the nonlinear 13-dim space described by \texttt{TwoNN}.

\subsection{PLS Representation}
\label{app:low_dim_pls}

The PLS representation $\Theta^{\tau}$ is constructed using the same procedure as PCA, with one key difference. Instead of constructing axes to maximize the variance of the projected $\Theta$, PLS constructs axes to maximize the \textit{covariance} of the projected $\Theta$ with stellar age $\tau$.

This means that age information is concentrated by design in the top few PLS components, in contrast to PCA where its unsupervised nature might result in more age information occurring in lower components.

We specifically use the top 3 PLS dimensions for age inference analysis because they are very informative, and overfitting significantly increased at higher dimensions. We tested this by measuring $\Theta^{\tau}$'s correlation with both true ages and fabricated (inaccurate) ages at various dimensions, and found that the correlation with fabricated ages increased significantly past 3 dimensions, indicating that the marginal increase in signal was primarily due to noise.

\section{Detailed Model Architecture}
\label{app:detailed_model_architecture}

This Appendix describes the \et{} implementation in detail, which we have written with the goal of complete reproducibility. It is divided into four subsections, mirroring the brief descriptions provided in \S\ref{sec:model}:

\begin{enumerate}[itemsep=-1pt]
    \item A description of the \textbf{encoding layers} that transform the raw inputs into learned representations. 
    \item The \textbf{core RNN framework}.
    \item The \textbf{self-supervised training procedure} that uses a \textit{Conditional Normalizing Flow} (CNF) head to predict flux.
    \item The aggregation process that constructs the \textbf{fixed-size latent representation} ($\Theta$) from LCs.
\end{enumerate}

\subsection{Input Encoding}
\label{app:model_input_enc}

Our inputs for each light curve are the normalized time series channels
\begin{equation}
    [\hat{f},\hat{\sigma}_{f},t]
\end{equation}
and the normalized metadata fields
\begin{equation}
\begin{aligned}
    M_S = \{\;\;\;&\frac{G_0}{100},\;\log_{10}(\sigma_{G0}),\;(BP-RP)_0,\;\log_{10}(\sigma_{BR0}),\;\log_{10}(\pi),\;\log_{10}(\sigma_{\pi}),\\
    &\log_{10}(T_{mag}),\;\frac{S}{100},\;N_{cam},\;N_{ccd},\;\log_{10}(f_{med}),\;\log_{10}(f_{H})\;\;\;\}\\
\end{aligned}
\end{equation}

We encode the timestamps as $\tilde{t}=g(t)$, where $\tilde{t}$ is an 8-dim vector and $g$ is an 88-parameter \textit{Multilayer Perceptron} (MLP) having one 8-dim linear hidden layer with a \texttt{ReLU} activation and an 8-dim linear output layer with no activation. $g$ is trained in parallel with the rest of \et\footnote{MLPs are widely considered the canonical ``standard'' neural network \citep[][]{MLP...1986Natur.323..533R} and are often the building blocks of more complex models.}. The purpose of this encoding is to improve \et's ability to learn periodic and quasi-periodic temporal trends; this is a common encoding strategy \citep[e.g.][]{Time2Vec...DBLP:journals/corr/abs-1907-05321,DonosoOliva...Astromer2...2026A&A...707A.170D} so timestamps that are far apart in real time but closer in phase space become closer in the encoding space. This means that each observed data point consists of the 10-dimensional input $x_t = [\hat{f}_t, \hat{\sigma}_{f,t},\tilde{t}]$, so an entire LC with $L$ observations $X = [x_t,\;...,\;x_L]$ is described by a vector of size $10 \times L$.

We apply another MLP $q$ to process the metadata parameters into a 32-dim vector $\tilde{M}_S = q(M_S)$. $q$ has 41,376 parameters and consists of 3 128-dim hidden linear layers with layer normalization, \texttt{GELU} activation and 10\% dropout, and a 32-dim linear output layer with no activation. $q$'s first hidden layer has 26 input dimensions, corresponding to the 13 metadata fields and the 13-dim binary mask channel.

\subsubsection{Metadata Masking}
\label{app:input_enc_meta_mask}

The binary mask channel is used to indicate whether or not a specific metadata parameter is masked out or not, and is applied at two levels. First, a block mask is applied to all metadata based on:
\begin{equation}
    p_{\mathrm{meta\_block\_mask}} \sim \mathrm{Bernoulli}(0.15)
\end{equation}
meaning that every LC has a 15\% chance of having all metadata parameters masked. Then, for stars where that block mask is not applied, field level masking is applied according to:
\begin{equation}
    p_{\mathrm{meta\_field\_keep}} \sim \mathrm{Bernoulli}(U(0.3,1.0))
\end{equation}
for each field. Note that $p_{\mathrm{meta\_block\_mask}}$ is the probability of \textit{masking out} metadata completely, but $p_{\mathrm{meta\_field\_keep}}$ is the probability of \textit{keeping} the field. For stars that pass the block mask, a minimum of one metadata field is kept. 

The result of this procedure is that some stars have metadata completely masked from training, and others have varying numbers and combinations of individual metadata fields that are masked. Masking is functionally applied by setting the value of the field to 0 and the value of the corresponding mask channel field to 0. Any metadata fields that are unmasked keep their true value and the corresponding mask channel field is set to 1.

This masking scheme encourages \et{} to not over-rely on systematic metadata signatures during training, and the masking is recomputed every batch and every training epoch so that LCs are not masked in the same way every iteration.

When computing $\Theta$ for use in downstream age inference, we do not apply any metadata masking since we want $\Theta$ to have maximal information.

\subsubsection{Light Curve Masking}
\label{app:model_train_mask}

In addition to metadata masking, we mask a portion of each LC as a way to improve \et's robustness against data gaps. This also reduces overfitting as \et sees different portions of the same LC each iteration of training.

Practically, we do this by iteratively drawing mask segments of length
\begin{equation}
    L_{\mathrm{mask}} \sim \text{LogUniform}(5,2880)
\end{equation}
which we place at a starting point in the LC drawn from
\begin{equation}
    x_{\mathrm{start}} \sim U(0,L_{LC}-L_{\mathrm{mask}})
\end{equation}
so that the entire mask fits in the LC. We chose this distribution to prioritize learning short term variability while still being sensitive to offsets up to approximately the size of the largest gaps in \textit{TESS} LCs.

Masks are generate iteratively for each LC until the total length of masked segments reaches 50\% of the LC, and the last mask is truncated so that the total number of masked data points does not exceed 50\%. We ignore the effect of overlapping masks, so the typical LC ends up being $\approx 40\%$ masked.

Masking is functionally applied by excluding masked points from the hidden state computations described in Appendix~\ref{app:model_rnn_framework}. However, those points are still used as prediction targets in the training procedure described in Appendix~\ref{app:detailed_model_training}, since they provide an additional opportunity to train across data gaps.

Similar to the metadata masking, when computing $\Theta$ for use in downstream age inference we do not apply any masking to the LCs since we want $\Theta$ to have maximal information.

\subsection{Recurrent Neural Network Framework}
\label{app:model_rnn_framework}

The concept driving RNNs is that time series data can be encoded as ``hidden states'' $H_t$, where each $H_t$ represents the state of a process at some measurement after some measurement at time $t$. With this notation, $H_0$ is a state representing some process before any observations, $H_1$ is the state after the first observation, and $H_L$ is the final state after the $L$th observation. In this framework, these hidden states accumulate information as a set of observations is processed sequentially. $H_0$ is derived from initial conditions, $H_1$ is computed from $H_0$ and the 1st observation, $H_2$ is computed from $H_1$ and the 2nd observation, and so on. In this manner, each hidden state $H_t$ holds information about the first $t$ observations.

The downside to traditional RNNs is therefore that every $H_t$ must be computed sequentially. We do not provide equations here, but direct the reader to \citet{RNNs...2024arXiv241001201F}, who provide examples for two specific types of RNNs: \textit{Long Short-Term Memory} (LSTM) networks \citep[][]{LSTM} and \textit{Gate Recurrent Unit (GRU)} networks \citep[][]{GRU...DBLP:journals/corr/ChoMGBSB14}. However, they also demonstrated that it is possible to simplify LSTM and GRU networks in a way that allows \textit{parallel computation} of $H_t$ while achieving performance comparable to transformers and traditional LSTM and GRU networks. Critically, this lets us remove the computational bottleneck of sequential processing, while taking advantage of the other benefits of RNNs as described in \S\ref{subsec:model_rnn}. \citet{RNNs...2024arXiv241001201F} formalize both LSTM and GRU implementations of this parallel architecture, defined as \texttt{minLSTM} and \texttt{minGRU} models, which they provide example PyTorch implementations for. We use the \texttt{minGRU} implementation directly in \et{}, although we tested the \texttt{minLSTM} model as well and performance was comparable. As a benchmark, \texttt{minGRU} provided a $\approx50\times$ speedup in training over sequences of $\approx$16,000 data points compared to the sequential GRU implementation.

Specifically, the computation of hidden states within the \texttt{minGRU} network can be described as:

\begin{align}
H_t &= (1-z_t) \odot H_{t-1} + z_t \odot \tilde{H}_t, \\
z_t &= a(x_t), \\
\tilde{H}_t &= b(x_t),\\
x_t &= J(\hat{f}_t,\hat{\sigma}_{f,t},\tilde{t},\tilde{M}_S)
\end{align}
where $J$ is a 64-dim linear neural layer that transforms the 42-dim input for each point into $x_t$, $b$ is a 64-dim linear neural layer with a custom activation function defined by \citet{RNNs...2024arXiv241001201F}, and $a$ is a 64-dim linear neural layer with sigmoid activation.

While in this framing there still appears to be a dependency of $H_t$ on $H_{t-1}$, through clever refactoring \citet{RNNs...2024arXiv241001201F} show that this can be reduced to purely a dependence on ($x_0$...$x_t$), which allows all $H_t$ in a LC to be computed simultaneously. Compressing the above equations, we use the notation

\begin{equation}
    \label{eq:hidden_states}
    H_t = \mathtt{minGRU}(X_t)
\end{equation}
for simplicity, where $X_t$ represents all inputs from $x_0$ to $x_t$. The exact calculations for the \texttt{minGRU} ``cell'' can be found in our publicly available code, mirrors the architecture provided by \citet{RNNs...2024arXiv241001201F}. Including the input projection $J$, our \texttt{minGRU} cell is 11,072 parameters.

The result is that we are able to compute a 64-dim $H_t$ corresponding to every point in a LC. $H_t$ contains information from the entire context of the LC up to and including $t$, and also the LC metadata. Therefore through training, these $H_t$ should describe the variability that we see in each LC within the context of the stellar and LC metadata properties. These $H_t$ are therefore the core products of the \et architecture, and what we use for downstream analysis.

\subsubsection{Bidirectionality}
\label{app:model_rnn_bidirectionality}

While the description of our RNN architecture above has only considered modeling LCs forward through time, we can implement the same procedure for LCs in reverse. This is useful for capturing behavior that is unique in the forwards or backwards direction, which occurs in asymmetric LC features such as flares that typically exhibit an abrupt rise in flux but a smoother decay.

We implement distinct (but architecturally identical) \texttt{minGRU} cells in both the forwards and backwards directions, which we distinguish hereafter in this work as $\texttt{minGRU}_f$ and $\texttt{minGRU}_b$ respectively. There are therefore two different hidden states (the forward $H^f_t$ and the backward $H^b_t$) computed for every point in the LC:

\begin{align}
    H^f_t &= \texttt{minGRU}_f(X^f_t)\\
    H^b_t &= \texttt{minGRU}_b(X^b_t)
\end{align}
where now $X^f_t$ represents all inputs from $x_0$ to $x_t$ and $X^b_t$ represents all inputs from $x_t$ to $x_L$. At each point in the LC, we now have a 128-dim hidden state:
\begin{equation}
    H_t = (H^f_t,H^b_t)
\end{equation}
that characterizes the variability of both the preceding and following portions of the LC, with the context of the stellar and LC metadata.

Including the input projection $J$, each \texttt{minGRU} cell is 11,072 parameters, so the total size of our RNN architecture is 22,144 parameters.

\subsection{Model Training}
\label{app:detailed_model_training}

As described briefly in \S\ref{subsec:model_training}, we train \et to predict the flux at some time $t_y$, given both LC and metadata context, the encoded time of prediction $\tilde{t}_y$, and the measurement error of the prediction $\hat{\sigma_y}$. We never want to provide our flux predictor with information on the observation it is predicting, so the maximum LC context we provide is the forward hidden state at the preceding point $H^f_{t-1}$, and the backward hidden state at the following point $H^b_{t+1}$. To train \et to capture longer term patterns, we can extend this concept to arbitrary offsets of the hidden states, $H^f_{t-k}$ and $H^b_{t+k}$, for $k$ up to a practically useful limit.

Additionally, instead of training \et to predict a discrete flux value with some error, we use a \textit{Conditional Normalizing Flow} (CNF) which computes a full probabilistic likelihood of the target flux. We do this because individual LC flux measurements are highly noisy, so we do not consider the accuracy of any individual flux prediction as important as successfully predicting the general trends with a well-calibrated likelihood distribution. It also allows complete flexibility in the shape of the flux distribution.

Specifically, we use a 28,974-dim Neural Spline Flow \citep[][]{NeuralSplineFlows...2019arXiv190604032D} with 2 transforms. Each transform consists of a 2 64-dim hidden linear layers with \texttt{ReLU} activation, and a 23-dim linear output layer (corresponding to an 8-bin spline).

Formalizing all of the above, we train the CNF to predict:
\begin{equation}
    \label{eq:flux_prediction}
    p(\hat{f}_y\;|\;H^f_{y-k},\,H^b_{L-y+k},\;\hat{\sigma}_{f,y},\;\tilde{t_y})
\end{equation}

The bottom portion of Figure~\ref{fig:schematic} illustrates this architecture. We choose a different offset $k$ for each batch during training, which is drawn from:
\begin{equation}
    k\sim \text{LogUniform}(1,2880)
\end{equation}
$k=1$ is a 2-min offset and $k=2880$ is a 4-day offset assuming constant 2-min sampling. We choose this distribution to prioritize learning short term variability while still being sensitive to offsets up to approximately the size of the largest gaps in \textit{TESS} LCs, similar to the masking strategy described in Appendix~\ref{app:model_train_mask}.

To calculate loss we use the \textit{negative log-likelihood} (NLL; $\mathcal{L}_{NLL}$) of $p({\hat{f}_y})$, at the observed flux. NLL is widely used as a loss metric for non-Gaussian likelihoods, and in such cases is a better representation of the accuracy of $p(\hat{f}_y)$ than other metrics such as $\chi^2$. Conceptually, this is a measure of how well the true observed flux value is aligned with the predicted distribution. Predictions are scored based on the ``height'' of $p(\hat{f}_y)$ at the observed value; in other words the density of our predicted flux at that point. Our loss function is therefore:
\begin{equation}
    \label{eq:nll}
    \textrm{NLL} \equiv -\log p(\hat{f}_{y,obs}\;|\;H^f_{y-k},\,H^b_{L-y+k},\;\hat{\sigma}_{f,y},\;\tilde{t_y})
\end{equation}
This is calculated for every unmasked observation, and our training procedure minimizes the average NLL across the observations from all LCs in a batch.

\subsection{Computation of LC Encodings}
\label{app:model_output_computation}

Our goal is to generate a meaningful fixed-size encoding from every LC using \et. As described above, through training we iteratively optimize our model weights to maximize the LC reconstruction ability of \et. With the final optimized model, we can compute all hidden states \{$H_{t=0}$,...,$H_{t=L}$\} for any LC having $L$ observations, which contain characteristic information about each star is expected to behave. Naively, these could be used directly as $\Theta$, however each $H_t$ is a 128-dim vector that exist for each data point, resulting in a total size of $\mathcal{O}(10^7)$ features per LC. This is not practical for characterization or downstream analysis, and we also require a fixed size for our latent representations to be useful. So, we apply an aggregation strategy to try to reduce the encoding size by discarding redundant information while retaining important signal. Specifically, we save the following aggregations for each LC:

\begin{enumerate}
    \item The maximum, minimum, mean, and standard deviation of each dimension across all $H_t$ in the LC (512-dim).
    \item Hidden state at the first and last timestamps: $H_0$ and $H_L$ (256-dim).
    \item The mean and standard deviation of the rate of change of each dimension across all $H_t$, measured using absolute time (256-dim).
    \item The mean of each dimension of the hidden states, divided into four sequential bins with equal numbers of observations. This is essentially four versions of \#3, where each is calculated over 1/4 of the hidden states (512-dim).
\end{enumerate}

By concatenating these, we obtain a 1536-dim encoding $\Theta$ for each LC. These can be thought of as a series of stellar parameters. Although the physical meaning behind each of these numbers is obscured, thoughtful interpretation can help characterize this latent space, which we explore in \S\ref{sec:variability_stats}.

\section{Latent space corner plot}
\label{app:pc_corner_plot}

\begin{figure}[b]
  \centering
  \includegraphics[width=0.82\textwidth]{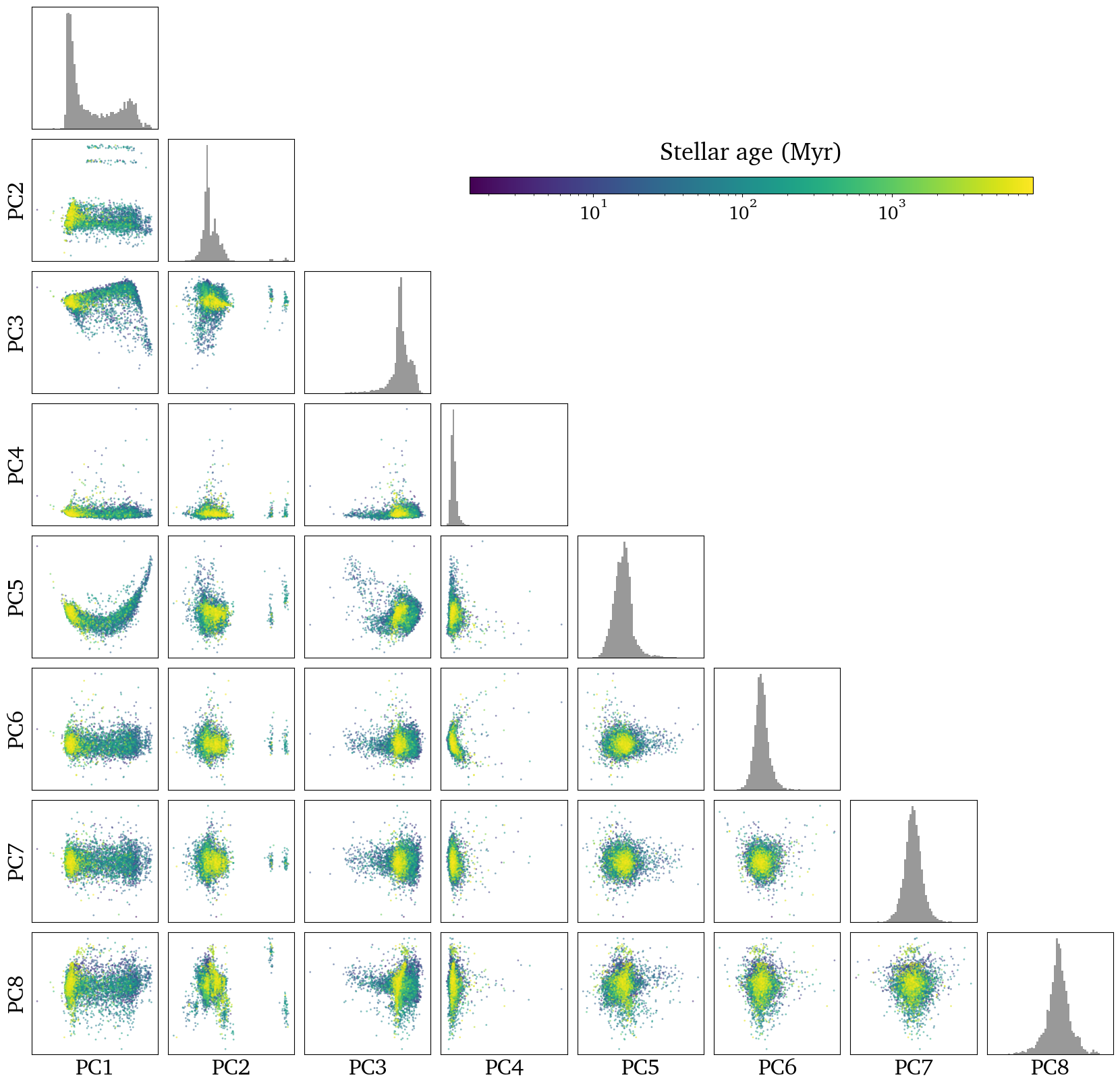}
  \caption{Age gradient over the top 8 PCs of $\Theta'$. It is apparent that the top component (PC1) carries strong age signal.}
  \label{fig:pc8_corner_plot}
\end{figure}

Figure~\ref{fig:pc8_corner_plot} presents a corner plot visualizing the correlations between age and the top 8 PCs in $\Theta'$. The intrinsic nonlinear dimensionality of $\Theta$ is $\approx13$ as measured by \texttt{TwoNN} (see \S\ref{subsec:model_output}), and PCA is linear, so therefore it takes $>13$ PC components to create a $\Theta$ subspace without losing any information. We show the top 8 PCs here simply for ease of visualization. 

It is clear that PC1 carries strong age signal, and to a lesser extent PC3. There appears to be relatively strong correlation between PC1/PC3 and PC1/PC5.

There are also distinct clustered subpopulations in PC2 that are not apparent in the other PCs, which could potentially be useful for classification.

\end{document}